\documentclass[12pt, notitlepage,aps,preprintnumbers,superscriptaddress,nofootinbib,aps,prd,floatfix]{revtex4-2}
\pdfoutput=1
\usepackage{graphicx} 
\usepackage{dcolumn} 
\usepackage{bm} 
\usepackage{slashed}
\usepackage{physics}
\usepackage{extarrows}
\usepackage{multirow,array}
\usepackage{amsmath,amssymb,physics}
\usepackage{subfigure}
\usepackage{enumitem}
\setlist{nolistsep}
\usepackage{xcolor}
\usepackage{url}
\usepackage{slashed}
\usepackage{hyperref}
\usepackage{float}
\definecolor{nicered}{rgb}{0.5,0.,0.}
\definecolor{nicegreen}{rgb}{0.,0.5,0.}
\definecolor{niceblue}{rgb}{0.,0.,0.5}
\hypersetup{colorlinks,citecolor=nicegreen,linkcolor=nicered,urlcolor=niceblue}
\usepackage{multirow}
\usepackage{braket}
\usepackage{colortbl}
\usepackage{gensymb}

\newcommand{\TeV}{\textrm{TeV}}

\newcommand{\fb}{\textrm{fb}}

\newcommand{\ifb}{\textrm{fb}$^{-1}$}

\usepackage{ulem}

\usepackage{comment}
\begin{document}
\preprint{MSUHEP-23-021, PITT-PACC-21XX}
\title{Exploring the impact of high-precision top-quark pair production data on the structure of the proton at the LHC}

\author{Alim Ablat}
\affiliation{School of Physics Science and Technology, Xinjiang University, Urumqi, Xinjiang 830046 China.}
\author{Marco Guzzi}
\email{mguzzi@kennesaw.edu}
\affiliation{Department of Physics, Kennesaw State University, Kennesaw, GA 30144, USA.}
\author{Keping Xie}
\email{xiekeping@pitt.edu}
\affiliation{
Pittsburgh Particle Physics Astrophysics and Cosmology Center, Department of Physics and Astronomy, University of Pittsburgh, Pittsburgh, PA 15260, USA.}
\author{Sayipjamal Dulat}
\affiliation{School of Physics Science and Technology, Xinjiang University, Urumqi, Xinjiang 830046 China.}
\author{Tie-Jiun Hou}
\affiliation{School of Nuclear Science and Technology, University of South China, Hengyang, Hunan 421001, China.}
\author{Ibrahim Sitiwaldi}
\affiliation{School of Physics Science and Technology, Xinjiang University, Urumqi, Xinjiang 830046 China.}
\author{C.-P. Yuan}
\affiliation{Department of Physics and Astronomy, Michigan State University, East Lansing, MI 48824, USA.}

\date{\today}
\begin{abstract}
The impact of recent LHC top-quark pair production single differential cross section measurements at 13 TeV collision energy on the structure of the proton is explored. In particular, the impact of these high-precision data on the gluon and other parton distribution functions (PDFs) of the proton at intermediate and large partonic momentum fraction $x$ is analyzed. This study extends the CT18 global analysis framework to include these new data. The interplay between top-quark pair and inclusive jet production as well as other processes at the LHC, is studied. 
In addition, a study of the impact of scale choice on the theory description of the new 13 TeV $t\bar t$ measurements is performed.

\end{abstract}

\maketitle
\tableofcontents

\section{Introduction}
\label{sec:intro}

The physics of the top quark, discovered by the D$0$ and CDF collaborations at the Tevatron proton-antiproton collider~\cite{CDF:1995wbb,D0:1995jca} in 1995, is central to a large number of precision programs for theory and experiments at the Large Hadron Collider (LHC) and its future upgrades.  

In proton-proton collisions at the LHC, properties of the top quark are being thoroughly investigated as they are critical to many endeavors in high-energy physics, {\it e.g.}, test the QCD theory and structure of the proton with unprecedented accuracy, unravel details of the electroweak (EW) sector of the Standard Model (SM) at high energies, and search for new physics interactions. 

The mass of the top quark ($m_t=172.69 \pm 0.30$ GeV from direct measurements~\cite{ParticleDataGroup:2022pth}) is very close to that of the recently discovered~\cite{ATLAS:2012yve,CMS:2012qbp} Higgs boson ($m_H = 125.25\pm 0.17$ GeV) and quantum corrections to these masses are deeply related to one another. In particular, the mass of the top quark is an important ingredient in the determination of absolute stability conditions for the EW vacuum as top-quark mass radiative corrections drive the couplings of the Higgs boson~\cite{Branchina:2018xdh,Bentivegna:2017qry,Alekhin:2012py,Bezrukov:2012sa,Degrassi:2012ry,Ellis:2009tp,Isidori:2001bm,Coleman:1977py,Callan:1977pt}. 
In addition, the top quark provides us with unique opportunities to search for signatures of new physics interactions at the TeV scale and beyond (see for instance refs.~\cite{Frederix:2007gi,Brivio:2019ius} and referenced therein).

The top quark mainly decays into a real $W$ boson and a $b$ quark before hadronization occurs. Therefore, precision measurements of top-quark pair production cross sections allow us to set stringent tests on perturbative QCD and to explore the structure of the proton with higher precision. In fact, more than 90\% of the production rate is due to the gluon-gluon fusion channel at the LHC with $\sqrt{S}=13~\TeV$ of center-of-mass collision energy. This makes top-quark pair production a unique probe of proton's parton content, especially the gluon, at intermediate and large partonic momentum fraction $x$, which motivates this work.

The ATLAS and CMS collaborations performed high-precision measurements of both total inclusive and differential top-quark pair production cross section at $\sqrt{S}=7$, $8$~\cite{ATLAS:2012vvt,ATLAS:2014ipf,ATLAS:2015lsn,CMS:2012hkm,CMS:2013hon,CMS:2017iqf} and 13 TeV~\cite{ATLAS:2019hxz,ATLAS:2020ccu,ATLAS:2023gsl,CMS:2018adi,CMS:2018htd,CMS:2021vhb} of center-of-mass collision energy, at parton and particle level, in different channels ({\it e.g.}, dilepton, semi-leptonic, all-hadronic, and lepton+jets) at different integrated luminosity (IL). 

The higher precision LHC data samples obtained at 13 TeV are at the core of this analysis which aims at studying their impact on parton distribution functions (PDFs) of the proton in global QCD analyses using the CTEQ framework. High-precision theoretical calculations to predict top-quark pair production cross sections for a variety of kinematic distributions obtained with different bin resolutions are therefore crucial for this task.    

{\bf Progress in theoretical calculations for $t\bar t$ cross sections.}
Radiative corrections to heavy-quark production at the next-to-leading order (NLO) ${\cal O}(\alpha_s^3)$ in QCD have been known for a long time~\cite{Nason:1987xz,Nason:1989zy,Beenakker:1988bq,Meng:1989rp,Beenakker:1990maa,Mangano:1991jk}, and NLO calculations for $t\bar{t}$ total and differential cross sections have been implemented in various parton-level Monte Carlo computer programs, {\it e.g.}, \texttt{MCFM}~\cite{Campbell:2015qma}, \texttt{MC@NLO}~\cite{Frixione:2003ei}, \texttt{POWHEG}~\cite{Frixione:2007nw}, 
\texttt{MadGraph/MadEvent}~\cite{Alwall:2007st,Frederix:2009yq}.

The progress made in the past two decades on both the theoretical and computational sides by many groups has been remarkable (see for instance refs.~\cite{Grozin:2004yc,Czakon:2007wk,Czakon:2007ej,Mitov:2006xs,Ferroglia:2009ep,Ferroglia:2009ii,Kidonakis:2009ev,Bierenbaum:2011gg,Barnreuther:2013qvf,Grozin:2022wse,Kidonakis:2023lgc} and references therein). The fixed-order calculation for $t\bar{t}$ production at next-to-next-to-leading order (NNLO) at ${\cal O}(\alpha_s^4)$ in QCD~\cite{Czakon:2013goa,Czakon:2012pz,Czakon:2012zr,Barnreuther:2012wtj,Czakon:2015owf,Czakon:2016ckf} has been accomplished based on the \texttt{STRIPPER} subtraction~\cite{Czakon:2010td,Czakon:2011ve,Czakon:2014oma} method.   
An independent calculation for the inclusive and differential $t\bar{t}$ production cross section has recently been accomplished based on the $q_T$-subtraction~\cite{Catani:2019iny,Catani:2019hip,Catani:2020tko} method and has been incorporated in the publicly available computer program \texttt{MATRIX}~\cite{Grazzini:2017mhc}. 
The inclusion of top-quark decay effects has recently been studied in refs.~\cite{Behring:2019iiv,Czakon:2020qbd}. 

The impact of logarithmic enhancements in threshold resummation on the cross section has also been extensively studied in the past decades, and relevant work can be found in refs.~\cite{Sterman:1986aj,Catani:1989ne,Kidonakis:1997gm,Laenen:1998qw,Bonciani:1998vc,Czakon:2009zw,Aliev:2010zk,Cacciari:2011hy,Czakon:2011xx} and in refs.~\cite{Kidonakis:2000ui,Kidonakis:2001nj,Kidonakis:2003qe,Kidonakis:2008mu,Kidonakis:2010dk,Kidonakis:2014pja,Kidonakis:2014isa,Kidonakis:2019yji,Kidonakis:2023juy}, while soft collinear effective field theory (SCET)~\cite{Neubert:1993mb,Bauer:2000yr, Bauer:2001yt,Beneke:2002ph} has been used in refs.~\cite{Ahrens:2009uz,Ahrens:2010zv,Ahrens:2011mw,Ahrens:2011px,Beneke:2011mq,Beneke:2012wb,Ferroglia:2012uy,Ferroglia:2012ku,Ferroglia:2013awa}.

The impact of electroweak (EW) corrections on $t\bar t$ production from both the weak and QED sectors has been studied in refs.~\cite{Beenakker:1993yr,Kuhn:2005it,Bernreuther:2005is,Kuhn:2006vh,Bernreuther:2006vg,Bernreuther:2008md,Manohar:2012rs,Kuhn:2013zoa,Campbell:2015vua,Hollik:2007sw,Bernreuther:2010ny,Hollik:2011ps,Kuhn:2011ri,Bernreuther:2012sx,Pagani:2016caq,Denner:2016jyo,Czakon:2017wor,Carrazza:2020gss,Campbell:2016dks}.  

Transverse momentum resummation for top-quark pair production has also been studied and recently documented in refs.~\cite{Ju:2022wia,Catani:2018mei,Angeles-Martinez:2018mqh,Bonciani:2015sha,Catani:2014qha,Li:2013mia,Zhu:2012ts}, while higher-order QCD corrections have been combined with Parton Shower (PS) simulations in 
\cite{Mazzitelli:2020jio,Mazzitelli:2021mmm,Alioli:2021ggd}, based on the general computer framework described in refs.~\cite{Nason:2004rx,Frixione:2007vw,Alioli:2010xd}.

{\bf Top-quark pair production in global PDF analyses.}
Top-quark pair production cross section measurements at the LHC and Tevatron are now a staple part of the data set baseline in modern global analyses at NNLO and beyond in QCD to determine proton PDFs. Examples of these analyses are ABMP~\cite{Alekhin:2017kpj}, CT18~\cite{Hou:2019efy}, MSHT20~\cite{Bailey:2020ooq,McGowan:2022nag}, and NNPDF~\cite{NNPDF:2017mvq,NNPDF:2021njg}. In particular, measurements of $t\bar t$ differential cross sections at the ATLAS and CMS experiments have been included in the most recent analyses~\cite{Hou:2019efy,Bailey:2020ooq,NNPDF:2017mvq,NNPDF:2021njg} at NNLO where, together with inclusive jet production measurements, they play an important role in constraining the gluon PDF in the intermediate to large-$x$ region as they complement each other. Though the $t\bar t$ and inclusive high-$p_T$ jet productions largely overlap in the $Q-x$ plane, their matrix elements and phase-space suppression are different so that their constraints on the gluon are placed at different values of $x$. 
However, the presence of tensions between experiments which can potentially result in different pulls on the gluon at intermediate/large $x$, and the strong correlation between the top-quark mass $m_t$, the strong coupling $\alpha_s$ and the gluon itself, make PDFs extraction particularly challenging~\cite{Alekhin:2017kpj,Cooper-Sarkar:2020twv,Cridge:2021qfd,Cridge:2023ztj}. 

Another complication arises from the amount of information on the statistical and systematical uncertainties published by the experimental collaborations. These uncertainties can be expressed in terms of either the covariance matrix or nuisance parameter representation and conversion from covariance matrix to nuisance parameters is not unique. Complete information on the statistical, uncorrelated and correlated systematic uncertainties (and their sources) is critical to maximize constraints from the data in PDF determinations.

A large number of studies appeared in the literature that have used total and differential $t\bar t$ cross section measurements at the LHC to constrain the gluon and other PDFs. Recent and less recent analyses can be found in refs.~\cite{Gauld:2013aja,Czakon:2013tha,Guzzi:2014wia,Czakon:2016olj,Czakon:2019yrx,Bailey:2019yze,Cooper-Sarkar:2020twv,Kadir:2020yml,Cridge:2021qfd,Kassabov:2023hbm}. 

{\bf Main goals of this analysis}. In this work, we shall study the impact of particular selections of 13 TeV $t\bar t$ single differential distributions at ATLAS (in the all-hadronic and lepton+jets channels)~\cite{ATLAS:2019hxz,ATLAS:2020ccu} and CMS (in the dilepton and lepton+jets channels)~\cite{CMS:2018adi,CMS:2021vhb}, on NNLO PDFs obtained by using the same framework ({\it i.e.}, strategy, and tolerance criteria definitions for the uncertainties) adopted in the CT18 analysis~\cite{Hou:2019efy}. In particular, PDFs are extracted by using optimal combinations of $t\bar t$ absolute differential cross sections measurements with different IL added on top of the CT18 baseline. The impact on the global fit from individual and combined kinematic distributions is analyzed by selecting different renormalization $\mu_R$ and factorization $\mu_F$ scale choices in the theory predictions. 
EW corrections are also considered, however their impact is found to be negligible.  

In addition, we analyze the impact of 13 TeV $t\bar{t}$ double differential distributions at ATLAS and CMS using the \texttt{ePump} (error PDF Updating Method Package) framework~\cite{Schmidt:2018hvu,Hou:2019gfw}. 
In Sec.~\ref{sec:1dvs2d} we discuss the observed impact from double differential distributions on PDFs, and find that this is comparable to that from single differential ones. However, the treatment and interpretation of correlated systematic uncertainties in the analysis with double differential distributions is more challenging and complicates the data vs theory description. 
This work mainly concerns the study of the impact of 13 TeV $t\bar{t}$ single differential distributions on PDFs extracted in CTEQ global QCD analyses. A thorough and more extensive investigation of $t\bar t$ double differential distributions in PDF determinations will be presented in the future, in a separate work. 
A previous study investigating $t\bar t$ double differential cross sections at the LHC at 8 TeV and their impact on PDFs with \texttt{ePump}, is discussed in ref.~\cite{Czakon:2019yrx}.  

{\bf Single top production.} Single (anti)top production cross sections in the $t$- and $s$-channel have also been measured at ATLAS and CMS at 7, 8, and 13 TeV collision energies. The impact of $t$-channel single (anti)top production has been explored in ref.~\cite{Nocera:2019wyk} where it is found that an optimal combination of single-top data constrains
the light quark and gluon PDFs with a reduction of their relative uncertainty by a fraction of a percent in the region $10^{-3} \leq x \leq 0.5$ with even more pronounced reduction on the ratio $u/d$
around $x \approx 0.1$. Part of these measurements are also included in the NNPDF4.0 global analysis~\cite{NNPDF:2021njg}. 
An investigation of the impact of single top-quark production cross section measurements at the LHC on CTEQ PDFs will also be presented in a separate work. 
 
The single and double differential cross sections of top-quark pair production as well as single (anti)top production cross sections will be ingredients of high importance in the next generation of PDF determinations which are going to use higher IL data.

The rest of this paper is organized as follows: in Sec.~\ref{recent-PDF-results} we summarize the findings from recent global QCD analyses using 7 and 8 TeV top-quark pair production measurements, while in Sec.~\ref{sec:13TeVdata} we describe the LHC measurements from ATLAS and CMS at 13 TeV considered in this study. In Sec.~\ref{sec:theory}, we describe the theoretical framework where details of the calculations such as NNLO QCD corrections, EW corrections, and scale dependence are discussed.    
In Sec.~\ref{sec:ATL-stat-corr}, we discuss the impact on PDFs from the bin-by-bin statistical correlations in the ATLAS 13 TeV lepton + jets channel measurements, while in Sec.~\ref{sec:1dvs2d} we discuss single vs double distributions at the LHC. In Sec.~\ref{sec:global-fit}, we describe the impact on PDFs from individual 13 TeV $t\bar t$ data sets considered in separate fits, and in Sec.~\ref{optimal-comb} we present the main results of this analysis obtained from two optimal combinations of 13 TeV $t\bar t$ measurements. We will conclude in Sec.~\ref{sec:conclusions}, while the details of theoretical calculations are presented in App.~\ref{app:theory-comp}, and the treatment of the correlated systematics are summarized in App. ~\ref{exp-unc-treatment}.

\section{Top-quark pair production at the LHC run I, and II}

We start with a brief overview of top-quark pair production measurements in recent global QCD analyses of PDFs which include LHC 7 and 8 TeV $t\bar t$ data from ATLAS and CMS, and summarize their findings. 
These measurements with respective analyses are also reported in Tab.~\ref{tab:LHCttbar-pdf-fits}. 
Next, we discuss the LHC 13 TeV data that are at the core of this work and will play an important role in all post-CT18 PDF determinations.

\subsection{Top-quark data in the CT18 era} 
\label{recent-PDF-results}

The CT18NNLO analysis~\cite{Hou:2019efy} includes 
two ATLAS absolute single-differential cross sections~\cite{ATLAS:2015lsn} $\dd\sigma/\dd p_{T}^{t}$ and $\dd\sigma/\dd m_{t\bar{t}}$ for the invariant mass with 20.3 fb$^{-1}$ of IL, and the CMS normalized double-differential cross section~\cite{CMS:2017iqf} $\dd^2\sigma/\dd p_{T,t}dy_t$, with 19.7 fb$^{-1}$. The two ATLAS measurements are combined into one single data set which includes the combination of the $e$+jets and $\mu$+jets channels for the $p_{T,t}$ and $m_{t\bar{t}}$ distributions with statistical correlations. These distributions are chosen according to their best compatibility within the global fit. In fact, the impact from the single-differential $y_t$ and $y_{t\bar{t}}$ rapidity distributions (absolute or normalized) at ATLAS~\cite{ATLAS:2015lsn} is also explored and tension is found with some other data sets. 
For example, the $y_t$ and $y_{t\bar{t}}$ rapidity distributions show agreement with HERA DIS data, but have opposite trend as compared to the CMS $d^2\sigma/dp_{T,t}dy_t$ and ATLAS $p_{T,t}$ and $m_{t\bar{t}}$ combined distributions. Their inclusion, either in the single-differential or double-differential form does not reduce PDF errors.
The resulting $t\bar{t}$ impact on the CT18 PDFs is found to be modest with a preference for a softer gluon at large $x$ in the $0.1 \lesssim x\lesssim 0.3$ range, and with changes that have no  statistically significant amount.

The MSHT20 analysis~\cite{Bailey:2020ooq}, in addition to the same CMS double-differential distribution considered in the CT18 study, includes ATLAS single-differential cross section measurements for the $m_{t\bar{t}},y_{t\bar{t}},p_{T,t}$, and $y_t$ distributions~\cite{ATLAS:2015lsn} in the lepton+jets channel combined with statistical correlations, ATLAS measurements of $y_{t\bar{t}}$ distribution~\cite{ATLAS:2016pal} in the dilepton channel, as well as the CMS normalized $y_{t\bar{t}}$ distribution in the lepton+jets channel~\cite{CMS:2015rld}. In addition, four total cross section measurements from ATLAS~\cite{ATLAS:2015xtk} and CMS~\cite{CMS:2013hon,CMS:2014btv,CMS:2015auz} are included.
For the ATLAS single-differential measurements in the lepton+jets channel, 
the correlated systematic parton-shower (PS) error across the four distributions has been decorrelated according to the procedure described in ref.~\cite{Bailey:2019yze}.
The impact on the MSHT20 gluon PDF from top-quark pair production data results in a suppressed high-$x$ gluon ($x\gtrsim 0.1$) with a complicated interplay/tension with the $Z$-$p_T$ and LHC jet data. PDF uncertainties at large $x$ are slightly reduced. Most of the impact is from the ATLAS combination of single differential distributions~\cite{ATLAS:2015lsn,ATLAS:2016pal} in the lepton+jets channel, and the treatment of systematic uncertainties plays a significant role in the description of data in terms of $\chi^2$.

The NNPDF4.0 analysis~\cite{NNPDF:2021njg} includes the ATLAS normalized  $y_t$ and $y_{t\bar{t}}$ distributions~\cite{ATLAS:2015lsn} in the lepton+jets channel, as well as the ATLAS normalized $y_{t\bar{t}}$ distribution in the dilepton channel~\cite{ATLAS:2016pal}.
From CMS, it includes the normalized $y_{t\bar{t}}$ distribution in the lepton+jets channel~\cite{CMS:2015rld} as well as the normalized $1/\sigma d^2\sigma/dy_{t}dm_{t\bar t}$ distribution in the dilepton channel~\cite{CMS:2017iqf}.
Two measurements of $t\bar t$ total cross section from ATLAS~\cite{ATLAS:2014nxi} and CMS~\cite{CMS:2015lbj} are also considered. In addition to these measurements at $\sqrt{S}$= 8 TeV, the NNPDF4.0 analysis includes two CMS rapidity $y_t$ distributions at $\sqrt{S}$ = 13 TeV in the lepton+jets~\cite{CMS:2018htd} and dilepton channel~\cite{CMS:2018adi}, respectively, and other total cross section measurements at different center-of-mass energy: $\sigma_{t\bar t}$ at CMS~\cite{CMS:2017zpm} with $\sqrt{S}$ = 5.02 TeV, at ATLAS~\cite{ATLAS:2014nxi} and CMS~\cite{CMS:2015lbj} with $\sqrt{S}$ = 7 TeV, and  
at ATLAS~\cite{ATLAS:2020aln} and CMS~\cite{CMS:2015yky} with $\sqrt{S}$ = 13 TeV. $t$-channel single top   
total and differential production cross section measurements are also considered in the NNPDF4.0 global fit. The resulting gluon is more suppressed at $x \gtrsim 0.1$ as compared to CT18 and MSHT20.
total and differential production cross section measurements are also considered in the NNPDF4.0 global fit. The resulting gluon is more suppressed at $x \gtrsim 0.1$ as compared to CT18 and MSHT20.
However, top-quark pair production data have overall a
moderate impact on the NNPDF4.0 global fit. In a fit with no top-quark data, the gluon is slightly enhanced at $x \gtrsim 0.1$, but well within the NNPDF4.0 uncertainty. 

The ABMP16 analysis~\cite{Alekhin:2017kpj} considers selected measurements of top-quark pair production total cross section only at the LHC and Tevatron with different collision energies: one data point at CMS with $\sqrt S = 5.02$~\cite{CMS:2016pqu}, six data points at ATLAS~\cite{ATLAS:2014ixi,ATLAS:2014nxi,ATLAS:2012gpa,ATLAS:2015zgd,ATLAS:2012qtn,ATLAS:2012uma} and five at CMS~\cite{CMS:2016yys,CMS:2016csa,CMS:2012ahl,CMS:2013nie,CMS:2013yjt} at $\sqrt S = 7$ TeV, two data points at ATLAS~\cite{ATLAS:2014nxi,ATLAS:2015xtk} and four at CMS~\cite{CMS:2014btv,CMS:2015auz,CMS:2016yys,CMS:2016csa} at $\sqrt S = 8$ TeV, one data point at ATLAS~\cite{ATLAS:2016zet}, and four at CMS~\cite{CMS:2015yky,CMS:2015toa,CMS:2016hbk,CMS:2016rtp} 
at $\sqrt{S} = 13$ TeV. The ABMP16 PDF parameters are fitted simultaneously with $\alpha_s$ and $m_t$.
These total cross section measurements lead to an increase in the gluon central value at large $x$ in the $0.05\leq x \leq 0.35$ range, with $10-20\%$ increase at $x \gtrsim 0.1$ in both the $n_f=4$ and $n_f=5$ calculations, depending on the factorization scale choice. These variations are well within the ABMP PDF uncertainty. However, the impact on the gluon PDF uncertainty is found to be small.

\begin{table}
\centering
\footnotesize
\begin{tabular}{cc|c|c|c|c|c}
\hline
Data &Lumi~[\ifb] & Ref. & CT18 & MSHT20 & ABMP16 & NNPDF4.0  \\
\hline
ATLAS 8 TeV lep+jets (norm.) & 20.3  & \cite{ATLAS:2015lsn} & $p_{T,t}, m_{t\bar{t}}$ & --  & -- &  $ y_t, y_{t\bar t}$ \\ 
ATLAS 8 TeV lep+jets (abs.) & 20.3  & \cite{ATLAS:2015lsn} & -- & $m_{t\bar{t}}, y_{t\bar{t}}, p_{T,t}, y_t$  & -- &  -- \\ 
ATLAS 8 TeV dilep (norm.) & 20.2  & \cite{ATLAS:2016pal} & -- & $y_{t\bar t}$ & -- & $y_{t\bar t}$ \\
CMS 8 TeV lep+j and dilep (norm.)& 19.7 & \cite{CMS:2015rld} &-- & $p_{T,t}+y_t$  & -- &  $y_{t\bar t}$  \\
CMS 8 TeV dilep 2D (norm.) & 19.7 & \cite{CMS:2017iqf} & $(p_{T,t},y_t)$  & $(p_{T,t}, y_t)$ & --  &  $(y_t,m_{t\bar t})$  \\
ATLAS 13 TeV lep+jets      & 36  & \cite{ATLAS:2019hxz} & This work        & -- & -- &-- \\
ATLAS 13 TeV all hadronic &  36.1  & \cite{ATLAS:2020ccu} & This work & -- & -- & -- \\
CMS 13 TeV dilep &  35.9  & \cite{CMS:2018adi} & This work & --& -- & $y_t$ \\
CMS 13 TeV lep+jets & 35.8 & \cite{CMS:2018htd} & -- & --&  -- &  $y_t$ \\
CMS 13 TeV lep+jets & 137 & \cite{CMS:2021vhb} & This work & -- & -- & -- \\
\hline
\multicolumn{7}{c}{Total cross sections}\\
\hline
CMS 5.02 TeV  lep+jets   & 0.0274 & \cite{CMS:2017zpm} & -- & --& --& $\sigma_{t\bar t}$  \\
CMS 5.02 TeV lep+jets & 0.026 & \cite{CMS:2016pqu}& -- & --& $\sigma_{t\bar t}$ & -- \\
ATLAS 7 TeV  & 4.66/4.6/1.67/4.7 & \cite{ATLAS:2014ixi,ATLAS:2012gpa,ATLAS:2015zgd,ATLAS:2012qtn,ATLAS:2012uma}& --  &  -- & $\sigma_{t\bar t}$ &  --\\
ATLAS 7/8 TeV & 4.6/20.3  & \cite{ATLAS:2014nxi} & --  &  --
& $\sigma_{t\bar t}$ &  $\sigma_{t\bar t}$\\
ATLAS 8 TeV lep+jets & 20.3  & \cite{ATLAS:2015xtk} & --  &  $\sigma_{t\bar t}$
& $\sigma_{t\bar t}$ &  --\\
CMS 7 TeV & 2/2.2/3.9/3.54& \cite{CMS:2016yys,CMS:2016csa,CMS:2012ahl,CMS:2013nie,CMS:2013yjt} & --  & -- & $\sigma_{t\bar t}$ & -- \\
CMS 7/8 TeV & 19.6/19.7& \cite{CMS:2016yys,CMS:2016csa} & --  & -- & $\sigma_{t\bar t}$ & -- \\
CMS 8 TeV & 19.6/18.4& \cite{CMS:2014btv,CMS:2015auz} & --  & $\sigma_{t\bar t}$ & $\sigma_{t\bar t}$ & -- \\
CMS 8 TeV dilep & 5.3& \cite{CMS:2013hon} & --  & $\sigma_{t\bar t}$ & -- & -- \\
ATLAS 13 TeV dilep & 3.2 & \cite{ATLAS:2016zet} &  -- & --& $\sigma_{t\bar t}$ & -- \\
ATLAS 13 TeV lep+jets & 139 & \cite{ATLAS:2020aln} & -- & --&  -- & $\sigma_{t\bar t}$  \\
CMS 13 TeV lep+jets & 19.7 & \cite{CMS:2015yky} & -- & -- & $\sigma_{t\bar t}$ & $\sigma_{t\bar t}$ \\
CMS 13 TeV  & 0.042/2.2/2.53 & \cite{CMS:2015toa,CMS:2016hbk,CMS:2016rtp} & -- & -- & $\sigma_{t\bar t}$ & -- \\
\hline
\end{tabular}
\caption{$t\bar{t}$ measurements at the LHC used in the ABMP16~\cite{Alekhin:2017kpj}, CT18~\cite{Hou:2019efy}, MSHT20~\cite{Bailey:2020ooq}, and NNPDF4.0~\cite{NNPDF:2021njg} global analyses.}
\label{tab:LHCttbar-pdf-fits}
\end{table}

\subsection{The 13 TeV top-quark data in the post-CT18 era}
\label{sec:13TeVdata}

In this section we describe the top-quark pair production differential cross section measurements at ATLAS and CMS that are considered in this study. 

In 2018, the CMS collaboration published differential cross section measurements at 13 TeV in the dilepton~\cite{CMS:2018adi} and lepton+jets~\cite{CMS:2018htd} channels. The lepton+jets channel measurements of ref.~\cite{CMS:2018htd} have recently been superseded by new measurements with a higher IL of 137 fb$^{-1}$~\cite{CMS:2021vhb} which we use in this analysis.
In parallel, the ATLAS collaboration published two 13 TeV measurements based on the lepton+jets~\cite{ATLAS:2019hxz} and the all-hadronic~\cite{ATLAS:2020ccu} channels respectively.
These measurements are described below.

{\bf ATLAS lepton+jets channel (ATL13lj).} 
We explored the impact of absolute top-quark pair production single differential distributions based on lepton+jet events measured at ATLAS 13 TeV with 36 fb$^{-1}$ of IL~\cite{ATLAS:2019hxz}, which we label ``ATL13lj''. We use full phase-space results at parton-level, and consider reconstructed measurements at parton-level in the resolved topology which are expected to provide direct constraints on PDFs. 
We do not consider the transverse momentum $p_{T,t\bar{t}}$ distribution of the $t\bar t$ pair because final-state interactions (FSI) between hard partons and beam remnants from the initial state may lead to substantial corrections to pair invariant mass (PIM) kinematic distributions when recoiling radiation is suppressed (see for instance the discussion in ref.~\cite{Mitov:2012gt}).
These contributions may manifest as higher-order perturbative
corrections to the factorized cross section, and as nonperturbative corrections that are suppressed by powers of perturbative scales.

The correlated systematic uncertainties associated to these measurements~\cite{ATLAS:2019hxz} are presented in terms of both the covariance matrix and nuisance parameter representations. We adopt the nuisance parameter representation because this is the default treatment for correlated systematic uncertainties in all CTEQ PDF analyses.

We study the impact of bin-by-bin statistical correlations between these distributions released for various combinations. The results are shown in Sec.~\ref{sec:1dvs2d} where we find that the statistical correlations for these measurements have some impact on the $\chi^2$ description, but their overall impact on the PDFs and their errors is negligible.

In addition, we study the impact of combinations of double-differential distributions, such as $\dd\sigma/\dd m_{t\bar{t}}\dd y_{t\bar{t}}$, and confront the results with those obtained by including the corresponding statistically combined single differential $\dd\sigma/\dd m_{t\bar{t}}$, $\dd\sigma/\dd y_{t\bar{t}}$ distributions. 

To facilitate comparisons with measurements at CMS, the ATL13lj measurements are released by using two different bin resolutions: 1) the original ATLAS bin resolution, and 2) the CMS bin resolution which shares the size of its bins and the number of points with the CMS 13 TeV $t\bar t$ measurements in the dilepton channel~\cite{CMS:2018adi}.    
These two bin resolutions differ in number of bins and bin size. Moreover, bin-by-bin statistical correlations are made available only for bin choice 1). As we shall see in the next sections, this has a non-negligible impact on the data-vs-theory description for single distributions using these two bin resolutions. 

{\bf ATLAS all-hadronic channel (ATL13had).} For the ATLAS 13 TeV all-hadronic channel with an integrated luminosity of $36.1~\fb^{-1}$~\cite{ATLAS:2020ccu}, we consider absolute single differential cross sections for the reconstructed top quark in terms of $p_{T,t_1(t_2)}$, $|y_{t\bar{t}}|$, $m_{t\bar{t}}$, and $H_T^{t\bar{t}}=p_{T,t}+p_{T,\bar{t}}$, where $p_{T,t_1(t_2)}$ is the transverse momentum of the leading (trailing) top quark.
We label this data set as ``ATL13had''. Combinations of double-differential cross sections for this channel are not available. For these measurements, correlated systematic uncertainties are made available in terms of nuisance parameters which are included in our global analysis. 

{\bf CMS dilepton channel (CMS13ll).} We study the impact 
of differential cross sections of top-quark pair production in the dilepton channel measured at CMS 13 TeV with $35.9~\fb^{-1}$of IL~\cite{CMS:2018adi}. 
These are labeled as ``CMS13ll''. They are published in terms of absolute and normalized single-differential distributions for the reconstructed top quarks. 
We consider only absolute cross section measurements in the full phase space at the top-quark level. This allows us to simplify the calculation of the theoretical predictions. Measurements relative to the decayed particle in the fiducial phase space will be analyzed in a future work as they require an additional effort to obtain complete predictions at the NNLO accuracy. 
In this study, we consider single-differential cross sections in terms of the $ p_{T,t}(p_{T,\bar{t}})$, $y_t(y_{\bar{t}})$, $y_{t\bar{t}}$, and $m_{t\bar{t}}$ distributions.

Correlated systematic uncertainties are presented in terms of the covariance matrix. In accordance with the default treatment of systematic correlations in the CTEQ framework, we convert the covariance matrix into the nuisance parameter representation by using a version of the iterative $\Sigma+K$ decomposition method, adopted in the CT18 analysis~\cite{Hou:2019efy}. A slightly extended discussion is in Appendix~\ref{exp-unc-treatment}. 

Bin-by-bin statistical correlations between measurements are not available to date, to the best of our knowledge. Therefore, we cannot exploit combinations of single differential distributions for these measurements in the global fit.    

{\bf CMS lepton + jets channel (CMS13lj).} 
In 2018, the CMS collaboration published 13 TeV measurements with 35.8 fb$^{-1}$ of IL in the lepton+jets channel~\cite{CMS:2018htd}. These measurements are now superseded by new measurements with higher precision with 137 fb$^{-1}$ of IL~\cite{CMS:2021vhb}, which we use in this analysis and are labeled as ``CMS13lj''. 
We examine the impact from the $m_{t\bar{t}}$ and $y_{t\bar{t}}$ single differential distributions in the full phase space. Bin-by-bin statistical correlations are not provided also for these measurements. Statistical and correlated systematic uncertainties are given in terms of the covariance matrix which we convert to nuisance parameter representation as discussed above.
We have also examined the $p^t_T$ and $y_t$ distributions. These are for hadronically reconstructed tops. They are relative to a top quark decaying into a $b$ quark and a $W$ boson with a subsequent hadronic decay of the $W$ boson. The theory predictions at NNLO in QCD for these distributions through \texttt{MATRIX} are therefore challenging to be obtained as it produces distributions for a stable top. However, in our preliminary investigation we found that these distributions were either poorly described by the \texttt{MATRIX} theory (e.g., $p^t_T$), or their impact was negligible. Therefore, we do not consider these distributions.

\subsection{Theoretical framework}
\label{sec:theory}
Details of the theoretical framework used in this analysis are given below. Additional details and comparisons are given in Appendix~\ref{app:theory-comp}. 

Global PDF analyses necessitate fast, precise, and accurate theory predictions that are compared to experimental data in the $\chi^2$-minimization procedure. To reduce the CPU turn-around time, fast theory predictions are obtained as interpolating tables through the \texttt{FatNLO}~\cite{Kluge:2006xs,Wobisch:2011ij,Britzger:2012bs,Britzger:2015ksm} and~\texttt{APPLGrid}~\cite{Carli:2010rw} frameworks. 

\textbf{NNLO QCD and NLO EW corrections.} 
The theory predictions at NNLO in QCD used in this work are based on two independent calculations. One is the numerical calculation described in refs.~\cite{Czakon:2015owf,Czakon:2016ckf}, based on the \texttt{STRIPPER} subtraction method~\cite{Czakon:2010td,Czakon:2011ve,Czakon:2014oma} and implemented in \texttt{fastNLO} tables~\cite{Czakon:2017dip,repo}; the other one is described in refs.~\cite{Catani:2019iny,Catani:2019hip,Catani:2020tko}, based on the $q_T$-subtraction method~\cite{Catani:2007vq} and implemented in the computer program \texttt{MATRIX}~\cite{Grazzini:2017mhc,Matrixrepo}, that is publicly available. 

The theory predictions utilized for the CMS13ll~\cite{CMS:2018adi} differential distributions which share the same bin resolution of some of the ATL13lj~\cite{ATLAS:2019hxz} distributions, are generated at NNLO with \texttt{fastNLO}~\cite{repo}. 
The theory predictions for the ATL13lj distributions resolved in terms of the ATLAS bins, are instead obtained with \texttt{MATRIX}. 
To optimize the calculation and minimize the CPU's turn-around time in the global fit, the \texttt{MATRIX} NNLO theory is constructed through bin-by-bin NNLO/NLO $K$-factors defined as $K=(\hat{\sigma}^{\rm(NNLO)}\otimes {\cal L}^{\rm (NNLO)})/(\hat{\sigma}^{\rm(NLO)}\otimes {\cal L}^{\rm(NNLO)})$, 
where ${\cal L}^{\rm(NNLO)}$ represents the corresponding NNLO PDF luminosity.  
The NLO theory calculation is obtained from fast lookup \texttt{APPLGrid} tables~\cite{Carli:2010rw} generated with \texttt{MCFM}~\cite{Campbell:2015qma,Campbell:2012uf}. A detailed comparison between the NNLO theory obtained with \texttt{fastNLO} tables~\cite{Czakon:2017dip} and that obtained with \texttt{MATRIX}, as well as a consistency check, are in Appendix~\ref{app:QCDNNLO}.

The theory predictions relative to the ATL13had~\cite{ATLAS:2020ccu} and CMS13lj~\cite{CMS:2018htd,CMS:2021vhb} measurements in the global fit are obtained with \texttt{MATRIX} in similar manner. 

The EW corrections and their implementation are discussed in Appendix~\ref{app:EWcorr}. Overall, the impact of the EW corrections on PDF determination is found to be negligible given the current size of the experimental errors. 

\begin{figure}
\centering
\includegraphics[width=0.49\textwidth]{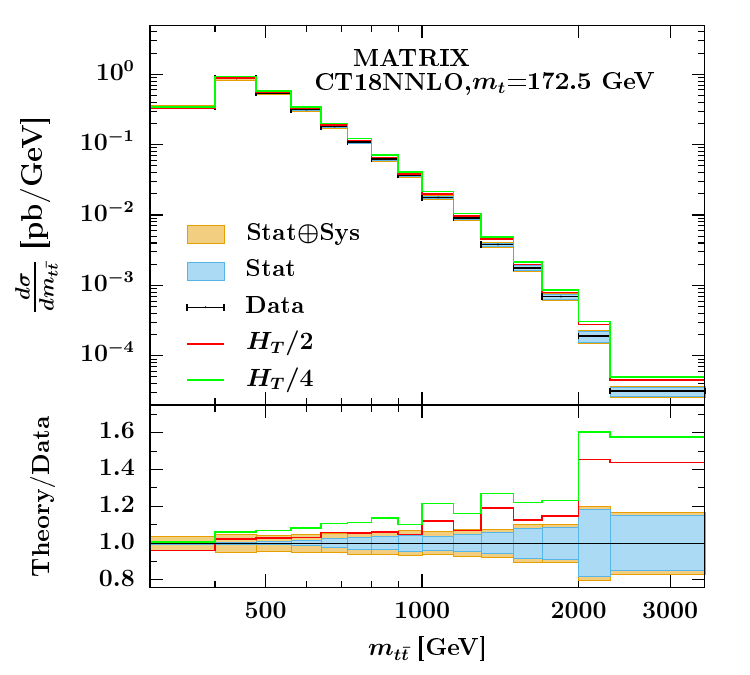}
\includegraphics[width=0.49\textwidth]{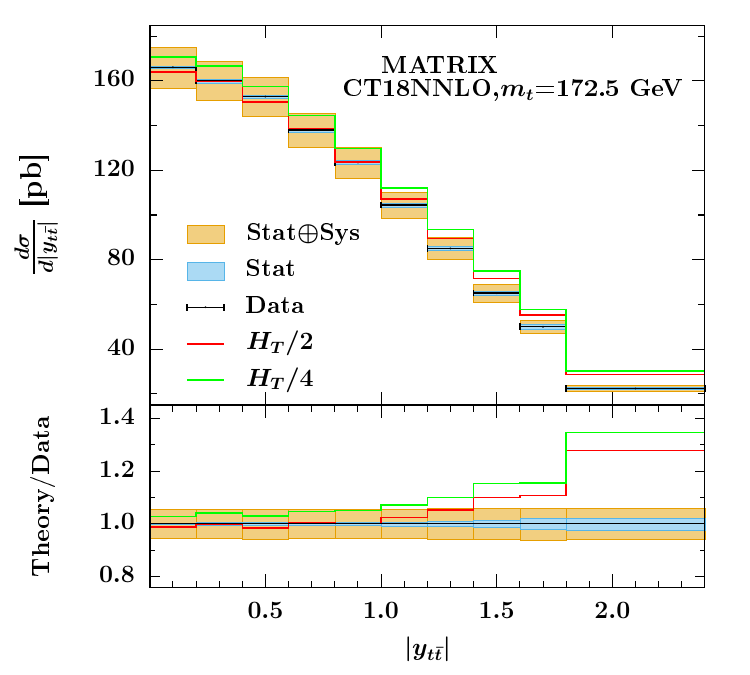}
\caption{Central scale choices for the $m_{t\bar{t}}$(left) and $|y_{t\bar t}|$(right) theory predictions compared to the CMS13lj measurements. The two central scales are represented by solid lines of different colors: $\mu_R=\mu_F=H_T/2$ is red, while $H_T/4$ is green.}
\label{fig:mtt-ytt-HTO24-CMS137}
\end{figure}

{\bf Impact of different central scales in the theory prediction.}
It is interesting to explore the impact of different choices for the central scale in the theory prediction for the $t\bar t$ differential cross sections at 13 TeV and make a comparison with the experimental uncertainties. To this purpose, we consider the CMS13lj measurements~\cite{CMS:2021vhb} with 137 fb$^{-1}$ of IL as they are the most precise data in this analysis. In Fig.~\ref{fig:mtt-ytt-HTO24-CMS137}, we confront to the experimental data with theory predictions for the $m_{t\bar{t}}$ (left) and $|y_{t\bar t}|$ (right) distributions calculated with the CT18NNLO PDFs and with different scale choices. 
The theory predictions are computed with central scales $\mu_F=\mu_R=H_T/2$ and $H_T/4$ respectively, represented by solid lines of different color in Fig.~\ref{fig:mtt-ytt-HTO24-CMS137}. 
For both distributions, the agreement with data deteriorates at large $m_{t\bar t}$ ($m_{t\bar t} \gtrsim 1.5$ TeV) and large $|y_{t\bar t}|$ ($|y_{t\bar t}| \gtrsim 1.5$),
although it is better in correspondence of $H_T/2$.   
These two CMS13lj measurements are very sensitive to the gluon and play a major role in the global fit which is discussed later.   

In this work, independent fits with central-scale choices $H_T/4$, $H_T/2$, and $H_T$ are examined. Differences in the predictions obtained with these scale choices can be used to quantify part of the theoretical uncertainty in the $t\bar t$ theory predictions in the global analysis.

\textbf{Hessian profiling with the \texttt{ePump} framework.}
A preliminary assessment of the impact of new measurements on existing PDFs sets before performing the global fit can be done by using different tools relying on a variety of statistical procedures based on the Monte Carlo or the Hessian method. Examples of these are: \texttt{ePump} (error PDF Updating Method Package)~\cite{Schmidt:2018hvu,Hou:2019gfw}, Hessian profiling~\cite{Paukkunen:2014zia}, Bayesian reweighting techniques~\cite{Giele:1998gw,Ball:2011gg,Ball:2010gb}, and the \texttt{PDFSense} method~\cite{Wang:2018heo}. 

We perform a preliminary investigation to assess the impact of the 13 TeV $t\bar t$ on PDFs using the \texttt{ePump} package. \texttt{ePump} allows one to obtain the updated best-fit PDF and relative PDF errors using the Hessian-profiling method. Previous studies using \texttt{ePump} are presented in refs.~\cite{Willis:2018yln,Hou:2019gfw,Yalkun:2019gah,Czakon:2019yrx,Ablat:2020wzd,Czakon:2019yrx,Kadir:2020yml}. \texttt{ePump} calculates the minimum of the $\chi^2$ function within the Hessian method using some approximations (e.g.,  quadratic dependence on the parameters for the original data) which may restrict its applicability in certain cases, especially in presence of tensions among data (see related discussion in Sec. II E of ref.~\cite{Schmidt:2018hvu}). \texttt{ePump} is used in this work as a fast and powerful tool to assess the impact of new data on PDFs, but it cannot replace a full PDF global analysis of experimental data and deviations from the results obtained in a global fit are expected, in particular, when subsets of data in the baseline are incompatible (see Appendix~\ref{CMS13lj-ytt-gluon-postfit} for more details).

\begin{figure}
\centering
\includegraphics[width=0.45\textwidth]{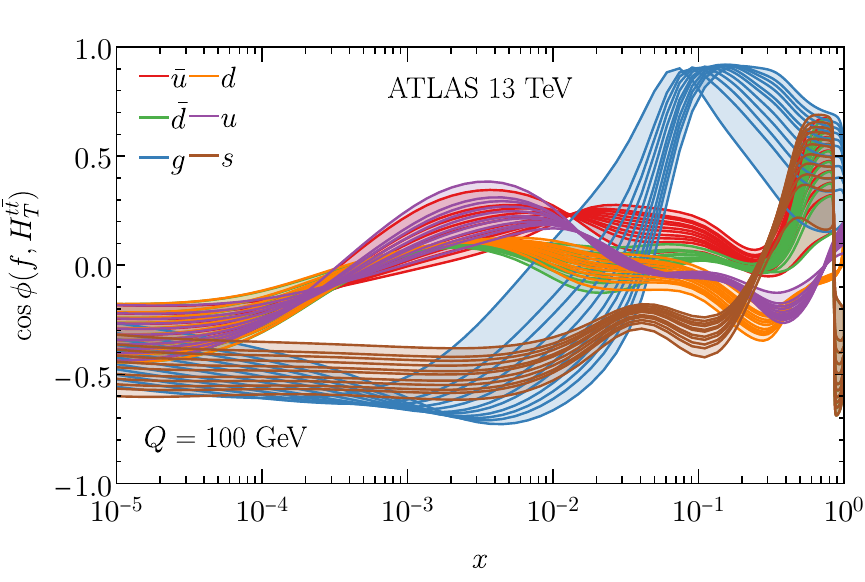}
\includegraphics[width=0.45\textwidth]{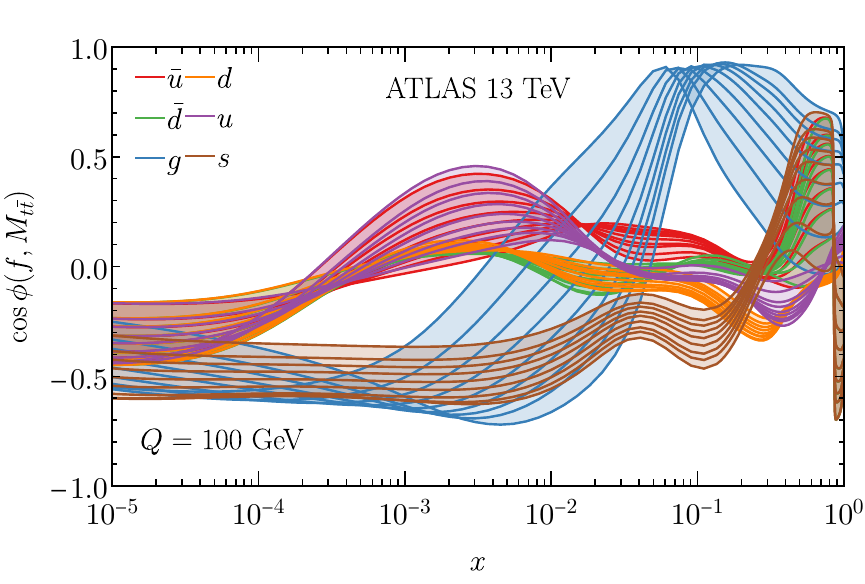}
\includegraphics[width=0.45\textwidth]{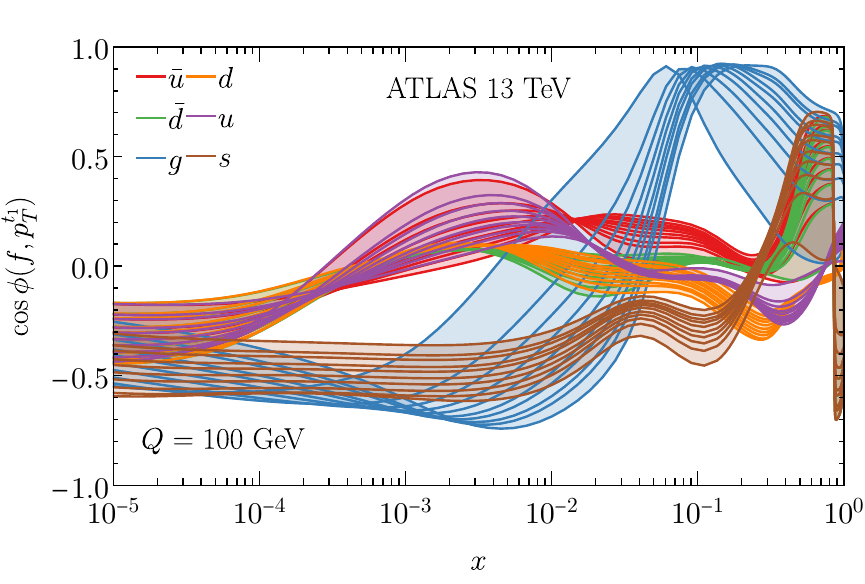}
\includegraphics[width=0.45\textwidth]{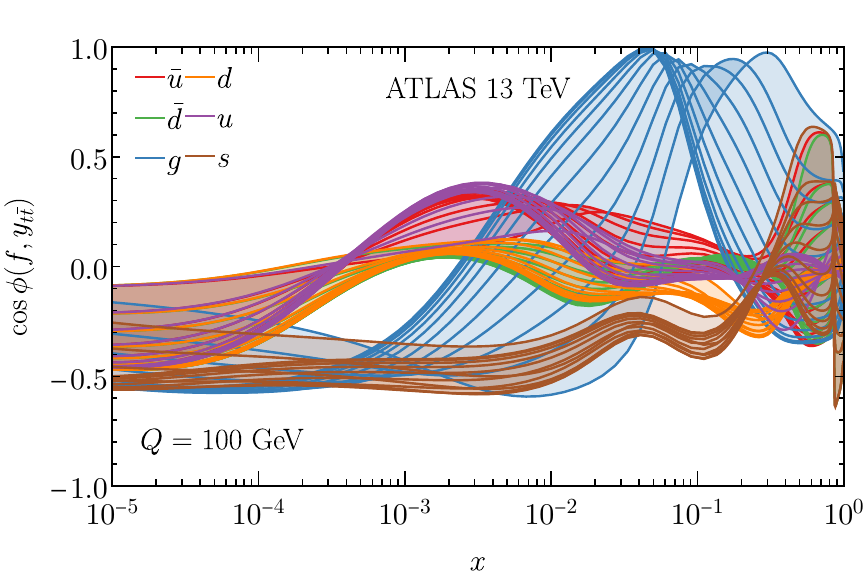}  
\caption{PDF correlation-cosine plots~\cite{Nadolsky:2008zw} as a function of the momentum fraction $x$ for the ATL13had~\cite{ATLAS:2020ccu} measurements added on top of the CT18NNLO baseline at $Q=100$ GeV.}
\label{fig:CosPhi-ATL-all-had}
\end{figure}

\begin{figure}
\centering
\includegraphics[width=0.45\textwidth]{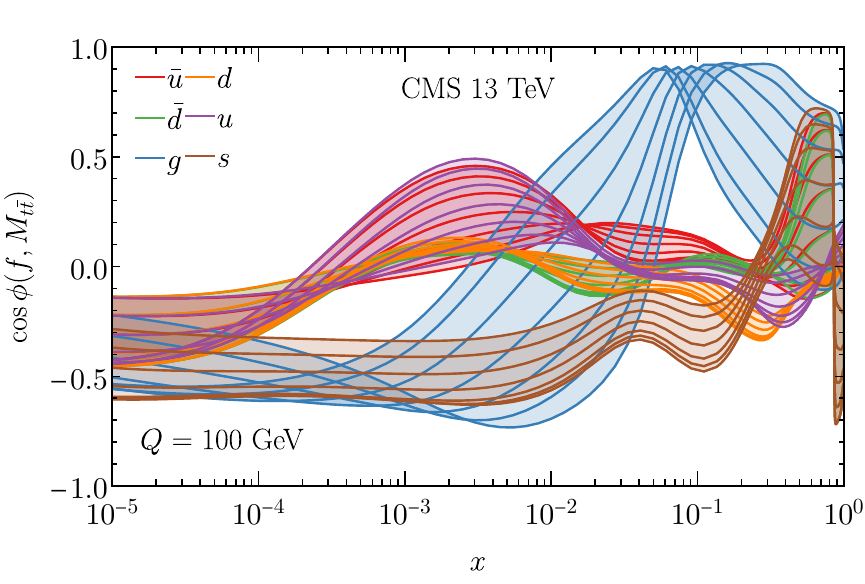}
\includegraphics[width=0.45\textwidth]{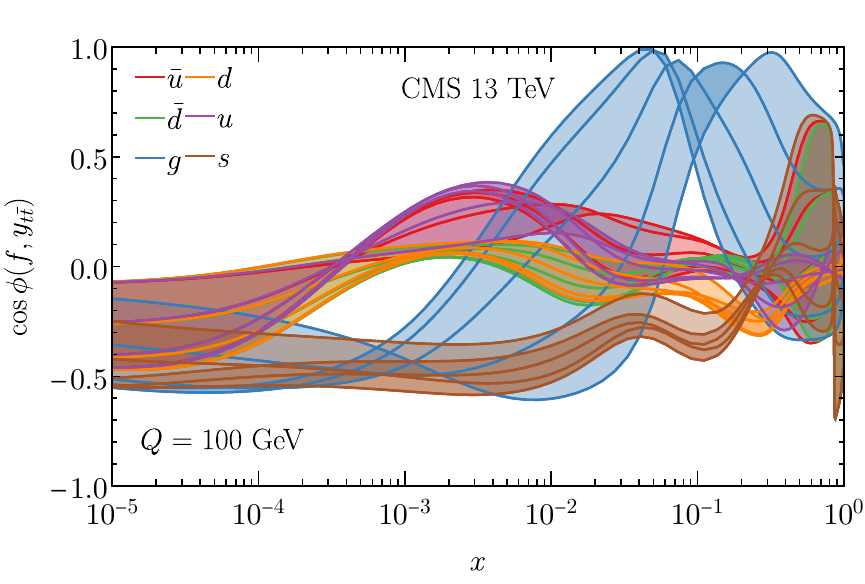}
\includegraphics[width=0.45\textwidth]{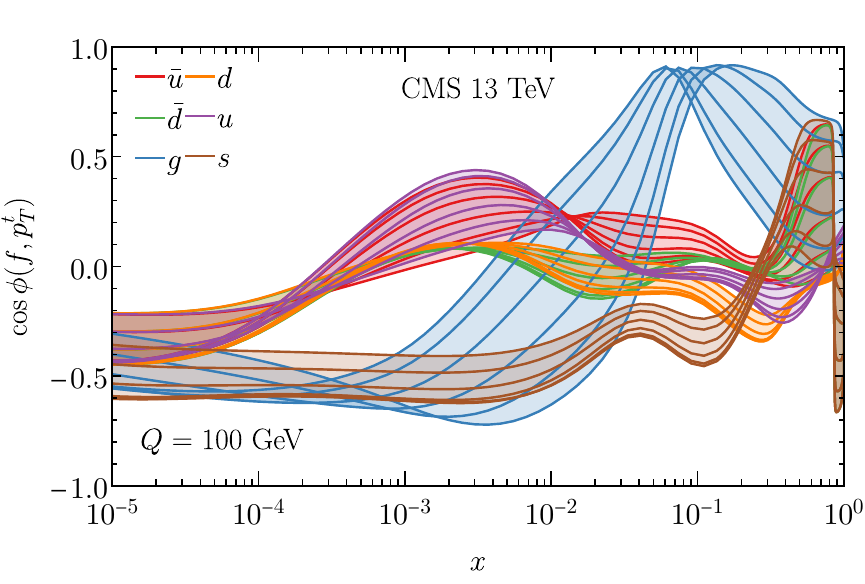}
\includegraphics[width=0.45\textwidth]{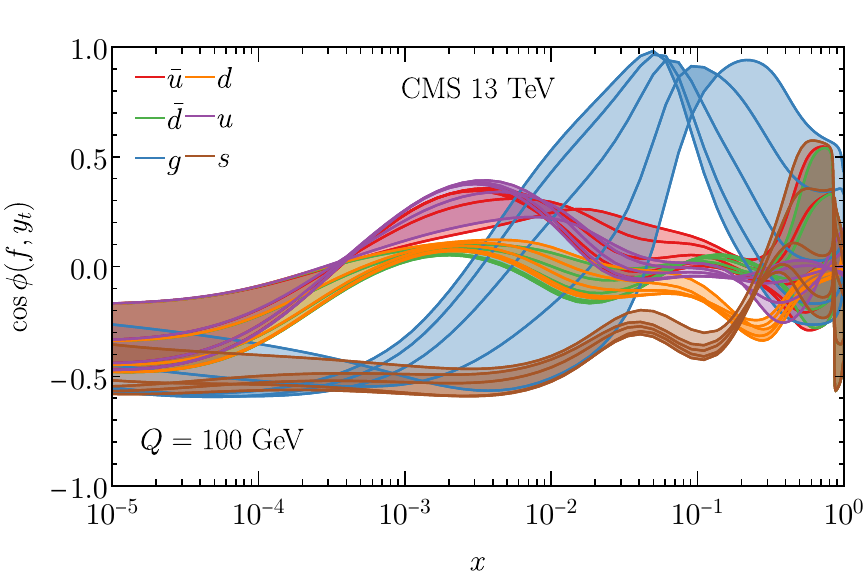}  
\caption{Same as in Fig.~\ref{fig:CosPhi-ATL-all-had}, but for the  CMS13ll measurements~\cite{CMS:2018adi}.}
\label{fig:CosPhiCMS-2lep}
\end{figure}

In Fig.~\ref{fig:CosPhi-ATL-all-had} and~\ref{fig:CosPhiCMS-2lep}, we present plots for the correlation-cosine between experimental data and PDFs for various differential distributions as a function of the parton momentum fraction $x$ for the ATL13had~\cite{ATLAS:2020ccu} and CMS13ll~\cite{CMS:2018adi} measurements, added on top of the CT18 baseline. PDFs in the various insets are represented by different colors and each solid line corresponds to different bins of the distribution under scrutiny. 
In both experiments, the gluon PDF appears to be strongly correlated at large $x$ in the interval $0.05\lesssim x\lesssim 0.4$ for all distributions. 
The correlation plots relative to the other experimental data considered in this work are very similar.

The results from \texttt{ePump} for the preliminary assessment of impact of individual experiments are reported in Tab.~\ref{global-fit-res}, where they are shown for different values of the central scale in the theory predictions, and can be compared to those from the global fit which will be discussed later in Sec.~\ref{sec:global-fit}. We study the quality-of-fit in terms of $\chi^2/N_{pt}$ and use it as the criterion for an initial investigation of the data.  

Looking at the $\chi^2/N_{pt}$ values from \texttt{ePump} in Tab.~\ref{global-fit-res}, we note that the $y_{t\bar t}$ and $p_{T,t_{1,2}}$ distributions produce an acceptable fit quality (with $\mu_F=\mu_R=H_T/2$ or $H_T/4$) among the ATL13had measurements, while for the $H_T^{t\bar t}$ and $m_{t \bar t}$ distributions $\chi^2/N_{pt}>1.5$ regardless of the scale choice. 

The $y_{t\bar t}$ and $y_{t}$ distributions are the only two that can be well described (for $\mu_F=\mu_R=H_T/2$ or $H_T/4$) among the CMS13ll measurements, and the $m_{t\bar t}$ distribution is the only acceptable candidate in the CMS13lj measurements as $y_{t\bar t}$ cannot produce a good fit, regardless of the scale choice.   

The ATL13lj case deserves more efforts because of the two bin resolutions and the bin-by-bin statistical uncertainties.
In the ATL13lj measurements with the CMS bin resolution, we observe that the $y_{t\bar t}$, $m_{t\bar t}$ and $y_{t}$ distributions produce acceptable $\chi^2/N_{pt}$ values in Tab.~\ref{global-fit-res}, while the $p_{T,t}$ distribution produces $\chi^2/N_{pt}>2$, independently of the scale choice. 

We discuss the ATL13lj measurements with the ATLAS binning resolution separately in Sec.~\ref{sec:ATL-stat-corr}, where we analyze the results from \texttt{ePump} with and without inclusion of bin-by-bin statistical correlations.

\subsection{Statistical correlations in the ATLAS lepton+jets data}
\label{sec:ATL-stat-corr}

The impact of bin-by-bin statistical correlations in the ATL13lj measurements~\cite{ATLAS:2019hxz} available on the HEPData~\cite{ATLleppjetrepo} repository is first analyzed with \texttt{ePump}.
The theoretical predictions are obtained with the \texttt{MATRIX} computer program as described in Sec.~\ref{sec:theory}, with top-quark mass set to $m_t=172.5$ GeV in the pole mass approximation. The impact of central-scale dependence is analyzed for the three different scales $\mu_R=\mu_F=\{H_T,H_T/2,H_T/4\}$. 

In Tab.~\ref{stat-corr-tab} we report the $\chi^2/N_{pt}$ values from \texttt{ePump} obtained with and without inclusion of statistical correlations. The implementation using bin-by-bin statistical correlations is labeled as ``WSC'' while the one without statistical correlations is ``NSC''. In addition, we use the notation adopted in the ATLAS publication where $y^B_{t\bar t}$ denotes the boosted topology for the rapidity of the $t\bar t$ system, while $H_{T}^{t\bar t}$ denotes the scalar sum of the transverse momenta of the hadronic and leptonic top quarks, 
$H_{T}^{t\bar t}=p^{had}_{T,t} +p^{lep}_{T,t}$. 

To consistently account for the statistical correlations, the WSC implementation adopts the \texttt{ePump} ``error type 2'' option. This option corresponds to a prescription of  $\chi^2$  calculation in  \texttt{ePump},   in which nuisance parameters, statistical and systematical correlation matrices, can be read from file as inputs. In this case, \texttt{ePump} treats the correlated systematic uncertainties in terms of their absolute (rather than percent) contribution to each data point, and therefore as additive errors. 
The NSC implementation adopts the \texttt{ePump} ``error type 4'' option, where the nuisance parameters representation is used and correlated systematic uncertainties are normalized to data\footnote{This choice can in principle introduce the D'Agostini bias~\cite{DAgostini:1993arp,DAgostini:1999gfj}, where data fluctuations can result in a bias in favor of experimental data points with lower central values. We do not discuss this here because the overall impact of the bin-by-bin statistical correlations of the ATL13lj measurements is negligible, and the same analysis is performed in the global fit where the systematic error for each multiplicative term is determined by multiplying the fractional uncertainty times the theoretical prediction for that bin.  (See Appendix C2 of the CT18 study~\cite{Hou:2019efy}.)}.

The ``error type 2'' and ``error type 4'' choices in \texttt{ePump} are used to facilitate the treatment and inclusion of statistical correlations in both \texttt{ePump} and the global fitting code (as we shall see later) which uses the nuisance parameter's $\chi^2$ definition. Since we studied the impact of two different scale choices (i.e., two theories), ``error type 2'' and ``error type 4'' are easy to handle as there is no need to normalize to two theories. The results from the global fits presented in the next sections are obtained using the multiplicative definition for the correlated systematic uncertainties, normalized to theory predictions, in the $\chi^2$ definition (which corresponds to ``error type 1'' in \texttt{ePump})\footnote{The multiplicative approach for the correlated systematic uncertainties in this choice suppresses the D'Agostini bias.}.
For more details about the \texttt{ePump} implementation, we refer the reader to the online manual~\cite{epumprepo}.

On the one hand, from the values in Tab.~\ref{stat-corr-tab} we note that regardless of the inclusion of statistical correlations and scale choice, there is some tension between various distributions, {\it e.g.},  $m_{t\bar t}$ and $H_T^{t\bar t}$, $m_{t\bar t}$ and $y_{t\bar t}$, and $m_{t\bar t}$ and $y^B_{t\bar t}$. Taken individually, all of these single differential distributions produce acceptable $\chi^2/N_{pt}$. In the \texttt{ePump} environment, tensions are amplified when statistical correlations are included. In addition, we note that 
$y_B + H^{t\bar t}_T$ produce a net effect that is similar to that of $y_{t\bar t} + m_{t\bar t} + y_B + H^{t\bar t}_T$
when they are statistically combined with the scale choice $H_T/4$.

\begin{figure}
\centering
\includegraphics[width=0.49\textwidth]{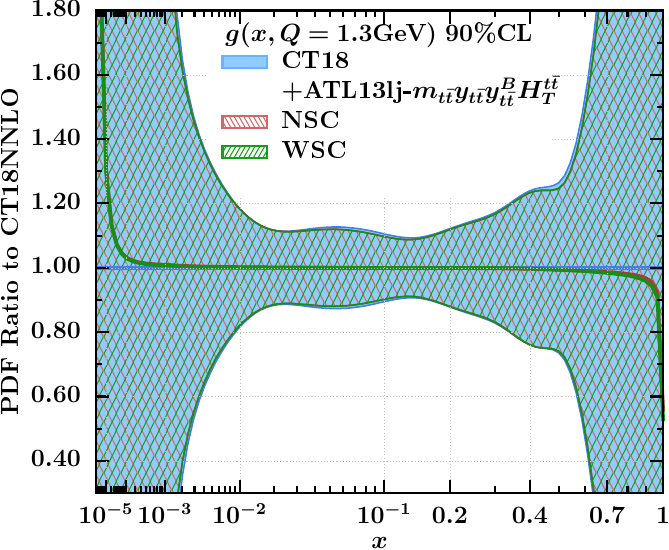}
\includegraphics[width=0.49\textwidth]{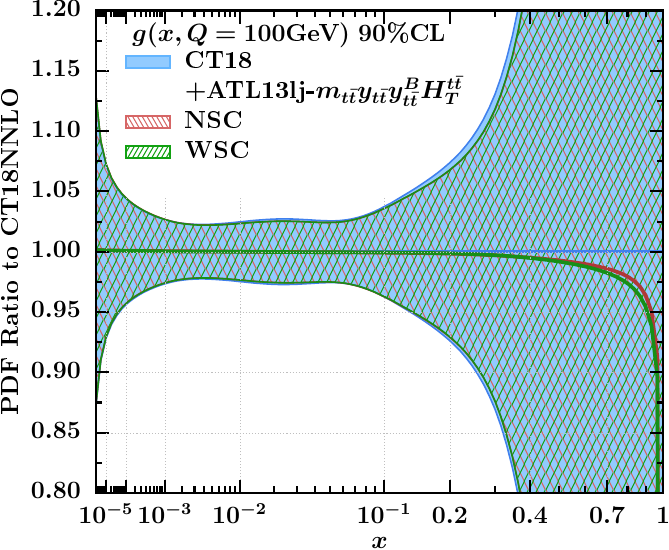}
\caption{Impact of bin-by-bin statistical correlations in the ATLAS lepton+jet channel measurements~\cite{ATLAS:2019hxz} on the CT18NNLO PDFs with \texttt{ePump}. PDF uncertainties are obtained at the 90\% C.L. }
\label{stat-corr-ATL-PDFs}
\end{figure}

On the other hand, as illustrated in Fig.~\ref{stat-corr-ATL-PDFs}, the impact of the bin-by-bin statistical correlations on PDFs is negligible. In addition, even if the $m_{t\bar t}$, $y_{t\bar t}$, $y^B_{t\bar t}$, and $H_T^{t\bar t}$ distributions are simultaneously included in \texttt{ePump}, the impact of these distributions on the gluon is very small. As we shall see in Sec.~\ref{sec:global-fit}, a similar behavior is observed in the true global QCD analysis where the overall impact from statistical correlations in the ATL13lj data is very small.

A general discussion on the importance of making statistical models and the associated data publicly available can be found in ref.~\cite{Cranmer:2021urp}. 

\begin{table}
\begin{tabular}{c|c||c|c||c|c||c|c}
\hline
\multicolumn{2}{c||}{Scale}  & \multicolumn{2}{c||}{$H_T/4$} & \multicolumn{2}{c||}{$H_T/2$} & \multicolumn{2}{c}{$H_T$} \\
\hline
Observable & $N_{\rm pt}$ &WSC &NSC &WSC &NSC &WSC &NSC\\
\hline
$m_{t\bar t}$          &   9  &1.150/1.146 &1.105/1.101 &1.309/1.300 &1.260/1.251 &1.249/1.231 &1.234/1.214\\
$y_{t\bar t}$          &   7  &1.027/1.022 &1.002/0.997 &0.713/0.706 &0.718/0.710 &0.671/0.622 &0.687/0.677\\
$y^{B}_{t\bar t}$           &   9  &1.153/1.147 &1.087/1.081 &0.861/0.851 &0.855/0.845 &0.853/0.838 &0.862/0.847\\
$H_{T}^{t\bar t}$         &   9  &0.923/0.920 &0.864/0.862 &0.832/0.831 &0.781/0.780 &1.185/1.183 &1.106/1.104 \\
$m_{t\bar t}$+$y_{t\bar t}$        &  16  &1.312/1.304 &1.204/1.193 &1.198/1.193 &1.139/1.132 &1.175/1.166 &1.151/1.141\\
$m_{t\bar t}$+$y^{B}_{t\bar t}$       &  18  &1.331/1.318 &1.246/1.230 &1.205/1.198 &1.143/1.135 &1.186/1.178 &1.144/1.135\\
$m_{t\bar t}$+$H_{T}^{t\bar t}$      &  18  &1.610/1.594 &1.462/1.449 &1.696/1.640 &1.206/1.187 &2.364/2.257 &1.511/1.472\\
$y_{t\bar t}$+$y^{B}_{t\bar t}$        &  16  &0.994/0.985 &0.906/0.900 &0.703/0.692 &0.633/0.625 &0.663/0.651 &0.581/0.571\\
$y_{t\bar t}$+$H_{T}^{t\bar t}$      &  16  &0.950/0.942 &0.892/0.886 &0.648/0.645 &0.625/0.622 &0.761/0.758 &0.725/0.721\\
$y^{B}_{t\bar t}$+$H_{T}^{t\bar t}$       &  18  &1.531/1.521 &1.255/1.247 &1.152/1.147 &0.967/0.962 &1.130/1.127 &0.961/0.956\\
$m_{t\bar t}$+$y_{t\bar t}$+$y^{B}_{t\bar t}$     &  25  &1.098/1.085 &1.002/0.991 &0.925/0.919 &0.864/0.858 &0.906/0.900 &0.855/0.848\\
$m_{t\bar t}$+$y_{t\bar t}$+$H_{T}^{t\bar t}$   &  25  &1.657/1.648 &1.431/1.425 &1.680/1.658 &1.087/1.082 &2.402/2.356 &1.377/1.362\\
$m_{t\bar t}$+$y^B_{t\bar t}$+$H_{T}^{t\bar t}$    &  27  &1.694/1.687 &1.424/1.417 &1.780/1.766 &1.162/1.157 &2.480/2.447 &1.475/1.464\\
$y_{t\bar t}$+$y^B_{t\bar t}$+$H_{T}^{t\bar t}$    &  25  &1.257/1.248 &1.001/0.994 &0.949/0.944 &0.729/0.724 &0.966/0.962 &0.716/0.712\\
$m_{t\bar t}$+$y_{t\bar t}$+$y^B_{t\bar t}$+$H_{T}^{t\bar t}$ &  34  &1.571/1.564 &1.233/1.227 &1.672/1.654 &0.954/0.949 &2.466/2.429 &1.220/1.210\\
\hline
\end{tabular}
\caption{$\chi^{2}/N_{\rm pt}$ values for the ATLAS 13 TeV $t\bar{t}$ production in the lepton+jets channel~\cite{ATLAS:2019hxz} before/after the ePump updating of the CT18 PDFs, with (WSC) and without (NSC) statistical correlation, and with central scales set to $H_T/4$, $H_T/2$, and $H_T$, respectively.}
\label{stat-corr-tab}
\end{table}

\subsection{Single vs double distributions at the LHC 13 TeV}
\label{sec:1dvs2d}

The ATLAS and CMS collaborations published several measurements of top-quark pair production double differential distributions at 13 TeV at both particle and parton level. A question then arises about the optimal strategy to be used to best exploit these new high-precision measurements. That is, whether to use statistically-combined single differential distributions, or double differential distributions directly. While a more exhaustive analysis specifically related to this point is beyond the scope of this paper and will be addressed elsewhere, here we explore the impact on CT18 PDFs in both cases by using \texttt{ePump}. As a case study, we consider the ATL13lj~\cite{ATLAS:2019hxz} measurements. In particular, we investigate the impact on the CT18 gluon PDF from adding two single-differential (1d) distributions (e.g. $\dd\sigma/\dd m_{t\bar t}$ and $\dd\sigma/\dd y_{t\bar t}$) on top of the CT18 baseline as well as the impact from the double differential (2d) distribution $\dd^2\sigma/\dd m_{t\bar t}\dd y_{t\bar t}$. 

\begin{figure}
\includegraphics[width=0.49\textwidth]{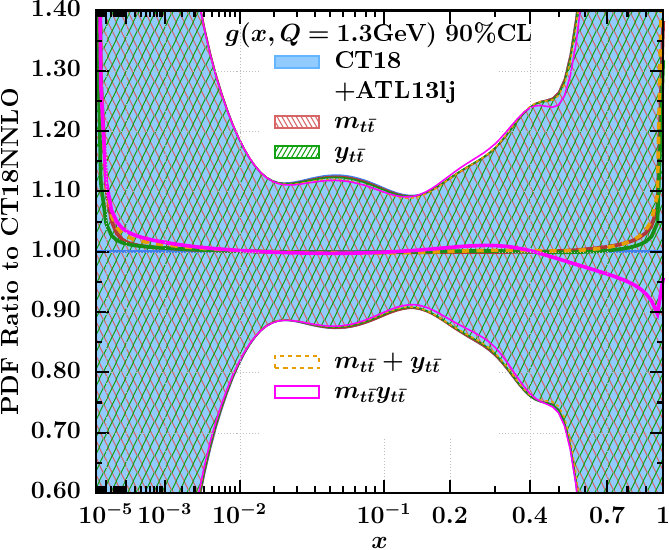}
\includegraphics[width=0.49\textwidth]{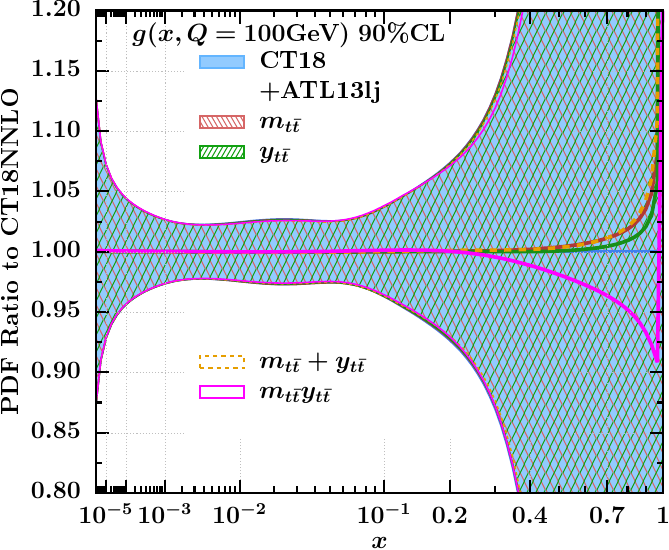}
\caption{Impact on the CT18 gluon PDF from 1d and 2d distributions at ATLAS~\cite{ATLAS:2019hxz} at 13 TeV with \texttt{ePump}. Here $m_{t\bar{t}}+y_{t\bar t}$ indicates two 1d distributions added on top of the CT18 baseline, while $m_{t\bar{t}} ~ y_{t\bar t}$ refers to the inclusion of a 2d distribution. The blue band represents the CT18NNLO PDF uncertainty at 90\% CL.}
\label{1dvs2d}
\end{figure}

In Fig.~\ref{1dvs2d} we illustrate the \texttt{ePump} results for the CT18 gluon in four different cases: 1) when the $m_{t\bar{t}}+y_{t\bar t}$ 1d distributions are added together on top of the CT18 baseline (orange dashed curve), 2) the $y_{t\bar t}$ distribution only is added (green curve), 3) the $m_{t\bar{t}}$ distribution only is added (red curve), 4) the $\dd^2\sigma/\dd m_{t\bar t}\dd y_{t\bar t}$ 2d distribution is added (purple curve). The blue band represents the CT18NNLO PDF uncertainty at 90\% C.L.
In this example, when the $m_{t\bar{t}}$ and $y_{t\bar t}$ 1d distributions are individually added, or added together on top of the CT18 baseline, they generate opposite pulls on the large-$x$ gluon as compared to the 2d distribution. At $Q=1.3$ GeV, the 2d distribution has a preference for a softer gluon in the $x\gtrsim 0.4$ region, while at $Q=100$ GeV this preference is in the $x\gtrsim 0.3$ region.
In both cases, the 1d distributions have opposite trends at large $x$ with a much milder effect.

Because of these opposite trends at large $x$, and because the gluon PDF uncertainty is large in this region, where there is essentially no data, we conclude that there is no obvious preference between the 1d $m_{t\bar{t}}+y_{t\bar t}$ combination, and the 2d $\dd^2\sigma/\dd m_{t\bar t}\dd y_{t\bar t}$ distribution. Ultimately, the impact from 2d and other multiple differential distributions must be assessed in the more general environment of a global PDF fit and explored against that of 1d distributions and their combinations.
This will be analyzed in a future work.

\section{Global QCD analysis: Impact from individual single differential distributions at the LHC 13 TeV}
\label{sec:global-fit}

In this section, we extend the analysis of the 13 TeV $t\bar t$ data to the global fit and extract post-CT18 PDFs at NNLO.  
In particular, we illustrate the results of multiple global QCD analyses at NNLO using the 1d absolute differential cross sections from the ATL13had, ATL13lj, CMS13ll, and CMS13lj measurements individually added on top of the CT18 baseline. 
We analyze the impact of these high-precision measurements on the gluon PDF and other relevant PDF combinations, and compare the results to the CT18 fit.  

The setup of the global fits is the default one used in the CT18 study, where we use the same tolerance criterion to estimate the post-CT18 PDF uncertainties, and the same PDF parametrizations. 
The inclusion of these new 13 TeV $t \bar t$ data that place constraints on the gluon at large $x$ would require a further analysis of PDF parametrizations and in particular the study of a more flexible parametrization for the gluon PDF. This requires substantial effort and an extended analysis using more flexible PDF parametrizations will be presented in a forthcoming study. 

The value of the top-quark mass is set to $m^{(\textrm{pole})}_t=172.5$ GeV, and the strong coupling constant at the $Z$-boson mass is set to $\alpha_s(M_Z)=0.118$. 

To find optimal combinations of measurements that allow us to include as much information as possible and, at the same time, minimize tension between observables, we first perform analyses including the 13 TeV $t\bar t$ data sets from every single channel one at a time, in individual global fits. Distributions are included in a combined manner only for the ATL13lj measurements where bin-by-bin statistical correlations are studied in advance in \texttt{ePump} preliminary assessments. 

In addition, we consider different choices for the central scale in the 13 TeV $t\bar t$ theory predictions and investigate their impact. The quality-of-fit in terms of $\chi^2/N_{pt}$ (as well as the same values previously obtained with \texttt{ePump}), are reported in Tab.~\ref{global-fit-res}. 

The results of individual global fits using the 13 TeV ATLAS and CMS $t\bar t $ measurements in the various channels are discussed below. 

\begin{table}
\begin{tabular}{c|cc|ccc|cc}
\hline
\hline
\multirow{2}{*}{Exp}&  \multirow{2}{*}{Obs} & \multirow{2}{*}{$N_{\rm pt}$} & \multicolumn{3}{c}{ ePump}  &\multicolumn{2}{c}{Global fit} \\ 
\cline{4-8}
 & &  & $H_T$ & $H_T/2$ & $H_T/4$  & $H_T/2$  & $H_T/4$\\ \hline
& $m_{t\bar{t}}$ & 9 & 1.75 &  1.57 & 1.60  & 1.53 &  1.47  \\
& $H_T^{t\bar{t}} $ & 11  & 1.98  & 1.77 &  1.59  &  1.50 & 1.74\\
& $y_{t\bar{t}}$ & 12  & 1.28 & 1.15 &  0.94  &  1.05  & 1.07 \\
& $p_{T,t_1}$  & 10  & 1.30 & 1.19&  1.12  &  1.20      & 1.33 \\
\multirow{-5}{*}{ATL13had}& $p_{T,t_2}$ & 8   & 1.13 &  0.84 & 1.05  &  0.84 & 1.59 \\ \hline
& $m_{t\bar{t}}$ & 7 & 3.46 &  3.07 & 3.14  &  3.12& 3.23\\
& $y_{t\bar{t}}$ & 10  & 1.66 & 0.97 &  0.68  & 0.94 &  0.67\\
& $p_{T,t}$ & 6 &  3.60 & 3.70 & 3.68  & 3.56 &  3.05 \\
\multirow{-4}{*}{CMS13ll} & $y_t$ & 10  & 1.33 & 0.94 &  0.87  & 1.00 &  0.69\\ 
\hline
& $m_{t\bar{t}}$ & 15  & 1.49 &  1.38 & 1.81  &  1.20 & 1.67\\
\multirow{-2}{*}{CMS13lj} & $y_{t\bar{t}}$ & 10  & 6.47 &  6.24 & 6.42  & 6.01 &  5.88 \\ \hline
& \multicolumn{7}{c}{\cellcolor[HTML]{E8F2A1} CMS bins}\\ 
& $m_{t\bar{t}}$& 7 & 2.40 & 1.17 &  0.68  & 0.83 &  0.66 \\
& $y_{t\bar{t}}$& 10  & 0.91 & 0.69 &  0.62  &  0.74 & 0.75\\ 
&$p_{T,t}$ &  6 & 2.34 & 2.01 & 2.47 & 1.35 & 1.43\\
&$y_{t}$ & 10 & 1.30 & 1.07 & 1.10 & 1.16 & 0.68\\
& \multicolumn{7}{c}{\cellcolor[HTML]{E8F2A1} ATLAS bins without statistical correlation (NSC)}\\ 
& $m_{t\bar{t}}$& 9 & 1.55& 1.12 &  0.94  & 1.27&  0.92\\
& $y_{t\bar{t}}$& 7   & 0.91 &  0.74 & 0.80  &  0.76 & 0.90\\
& $y^{B}_{t\bar{t}}$ & 9   & 1.40 &  1.27 & 1.53  &  0.85 & 0.93 \\ 
& $H_{T}^{t\bar{t}}$ & 9   & 1.35 &  0.91 & 0.93  & 0.81 &  0.80\\ & $m_{t\bar{t}}+y_{t\bar{t}}+y^{B}_{t\bar{t}}+H_{T}^{t\bar{t}}$ & 34  & 1.87 &  1.28 & 1.46  &  0.93 & 1.06 \\ 
& \multicolumn{7}{c}{\cellcolor[HTML]{E8F2A1} ATLAS bins with statistical correlations (WSC)} \\  
& $m_{t\bar{t}}$ & 9   &  1.68 & 1.35 &  0.98  & 1.29 &  0.96\\ 
& $y_{t\bar{t}}$ & 7   &  0.88 &  0.75 & 0.92  &  0.75 & 0.92\\ 
& $y^{B}_{t\bar{t}}$ & 9   &  1.06 & 0.87 & 1.01  & 0.86    & 0.99 \\ 
& $H_{T}^{t\bar{t}}$ & 9   &  1.40 & 0.85 & 0.85  & 0.86 & 0.86 \\
\multirow{-15}{*}{ATL13lj} & $m_{t\bar{t}}+y_{t\bar{t}}+y^{B}_{t\bar{t}}+H_{T}^{t\bar{t}}$ & 34  &  3.10 & 1.61 & 1.32  & 1.59 & 1.32 \\ 
\hline
\hline
\end{tabular}
\caption{Results of the NNLO global QCD analysis for all the 13 TeV measurements from ATLAS and CMS included one at a time, and in a combined manner with statistical correlations when these are available. Included are also the $\chi^2/N_{pt}$ values from \texttt{ePump}. The analysis is performed by adopting different central-scale choices for the $t\bar t$ 13 TeV measurements.}
\label{global-fit-res}
\end{table}

\subsection{Impact from the ATLAS lepton+jets channel} 

We start by analyzing the impact from the ATL13lj measurements~\cite{ATLAS:2019hxz} using the two bin resolutions, {\it i.e.}, the ATLAS and CMS binning, published by the ATLAS collaboration and available on the HEPData repository~\cite{ATLleppjetrepo}.   

{\bf ATL13lj resolved with CMS bins.} 
The ATLAS measurements resolved in terms of CMS bins share the bin size with the CMS13ll data that have equal IL~\cite{CMS:2018adi}. Bin-by-bin statistical correlations are not available for this specific binning resolution.
Looking at the individual $\chi^2/N_{pt}$ values in Tab.~\ref{global-fit-res}, the 1d distributions that are better described by the theory are $m_{t\bar t}$ and $y_{t\bar t}$, while the description of the $p_{T,t}$ and $y_t$ distributions is more challenging due, in part, to the fact that they refer to reconstructed hadronically-decayed top quarks.  
We note that both the $m_{t\bar t}$ and $y_{t\bar t}$ distributions have $\chi^2/N_{pt}\lesssim 1$ independently of the choice of the central scale in their theory predictions. The $y_{t\bar t}$ spectrum appears to be more stable when the central scale is varied, with $\chi^2/N_{pt}\approx 0.75$ in both cases. The description of the $y_t$ distribution is largely affected by the central-scale choice. 

The $m_{t\bar t}$ and $y_{t\bar t}$ distributions provide us with most of the information and introduce the least tension when they are independently added on top of the CT18 baseline.   

\begin{figure}
\includegraphics[width=0.49\textwidth]{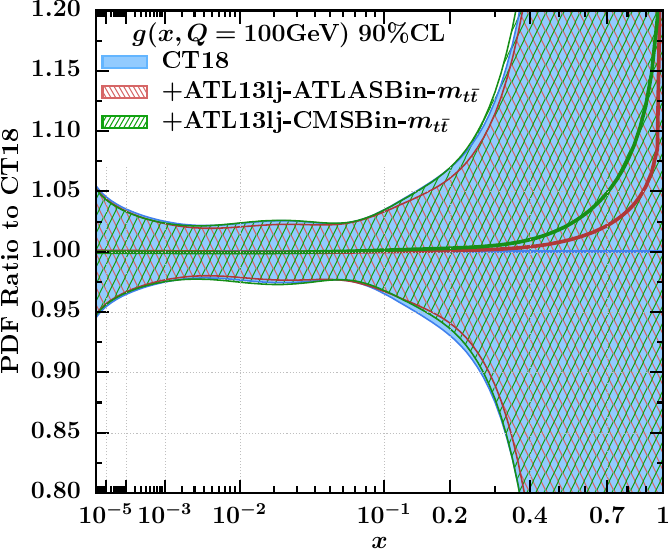}
\includegraphics[width=0.49\textwidth]{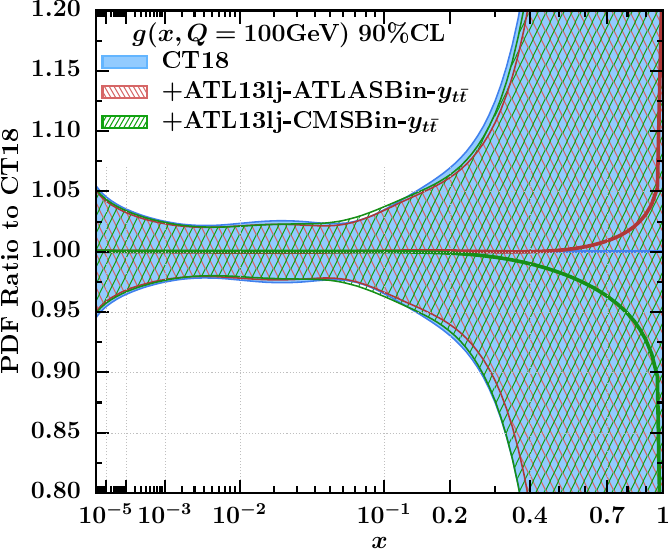}
\caption{{\bf ATLAS lepton+jet channel.} Impact of the $m_{t\bar t}$ and $y_{t\bar t}$ distributions on the gluon PDF in a NNLO global fit. Distributions are shown for two different binning resolutions labeled by ``ATLASBin'' and ``CMSBin'' respectively. Error bands with different hatching represent PDF uncertainties at $90\%$ CL. The central scale in their theory predictions is set to $H_T/4$.}
\label{ATL-lep+jet-cms-atl-bin}
\end{figure}

In Fig.~\ref{ATL-lep+jet-cms-atl-bin} we compare the individual impact on the large-$x$ gluon obtained using the two bin resolutions in a global fit at NNLO. Error bands with different hatching represent PDF uncertainties at $90\%$ CL, and the central-scale choice in the 13 TeV $t\bar t$ theory predictions is set to $H_T/4$ (results are very similar for $H_T/2$).
We observe that the constraints placed by the $m_{t\bar t}$ and $y_{t\bar t}$ distributions resolved with the CMS13ll bins have pulls in opposite directions above $x\gtrsim 0.3$. In particular, the $m_{t\bar t}$ distribution prefers a harder gluon while the $y_{t\bar t}$ one prefers a softer gluon as compared to CT18. Overall, the impact on the gluon PDF error is small. As reflected in Tab.~\ref{global-fit-res}, both distributions exhibit $\chi^2/N_{pt}\lesssim 1$ with small variations when the central-scale choice in their theory prediction is varied.

{\bf ATL13lj resolved with ATLAS bins.}
The impact from both the individual and cumulative (in terms of bin-by-bin statistically combined) 1d ATL13lj distributions, resolved with ATLAS bins, is analyzed together with the impact of statistical correlations. The most relevant information is obtained from the $m_{t\bar t}$, $y_{t\bar t}$, $H_{T}^{t\bar t}$, and $y^{B}_{t\bar t}$ distributions for which bin-by-bin statistical correlations are available. Their $\chi^2/N_{pt}$ values from the NNLO global fit in Tab.~\ref{global-fit-res} is of order 1 or less for all distributions, regardless of the central-scale choice in their theory predictions, with some deterioration when the four spectra are statistically combined and fitted together. This is expected as statistical correlations impose further constraints in the fit. Correlated systematic uncertainties are given in terms of the nuisance parameter representation. 

By looking at the $m_{t\bar t}$ and $y_{t\bar t}$ distributions in Fig.~\ref{ATL-lep+jet-cms-atl-bin}, we observe that the pulls on the gluon at large $x$ are in the same direction, {\it i.e.}, both $m_{t\bar t}$ and $y_{t\bar t}$ prefer a harder gluon at $x\gtrsim 0.4$ and distortions are more pronounced for $m_{t\bar t}$. 
However, these distortions are milder as compared to those obtained from the same distributions resolved with the CMS13ll bins. The behavior of the gluon at large-$x$ constrained by the inclusion of the $y^{B}_{t\bar t}$ and $H_{T}^{t\bar t}$ distributions in the fit, is very similar to that of the $m_{t\bar t}$ and $y_{t\bar t}$ distributions, respectively, and it is not shown here. The individual distribution impact on the PDF uncertainty is negligible also in this case. 

\begin{figure}
\includegraphics[width=0.49\textwidth]{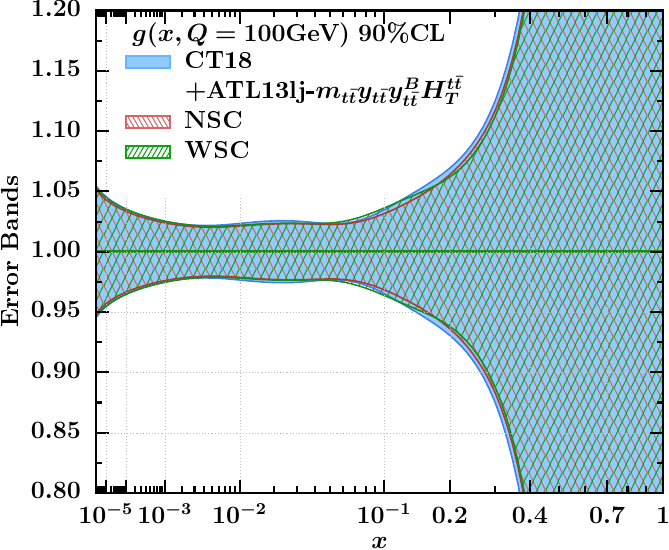}
\includegraphics[width=0.49\textwidth]{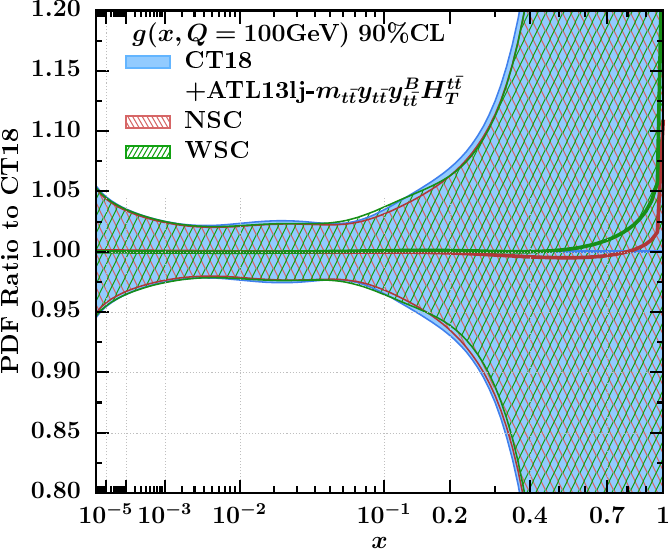}
\caption{{\bf PDF Impact of statistical correlations in the ATL13lj data.} Impact of statistical correlations on the gluon PDF in a NNLO global QCD analysis. 
Left: Impact on the error bands. Right: PDF ratios to CT18 PDFs. 
The $ y_{t\bar t}, y^B_{t\bar t}, m_{t\bar t}$, and $H_T^{t\bar t}$ distributions are added with (green) and without (red) bin-by-bin statistical correlations. The error bands with different hatching represent PDF uncertainties at $90\%$ CL. The central-scale choice in the 13 TeV $t\bar t$ theory predictions is set to $H_T/4$. }
\label{ATL-lep+jet-stat}
\end{figure}

{\bf Impact of bin-by-bin statistical correlations in the global fit.}
In Fig.~\ref{ATL-lep+jet-stat} we illustrate the impact on the gluon PDF from the $m_{t\bar t}$, $y_{t\bar t}$, $H_{T}^{t\bar t}$, and $y^{B}_{t\bar t}$ distributions of the ATL13lj data added together on top of the CT18 baseline, and added together including statistical correlations. 
The gluon PDF is shown at a scale of $Q=100$ GeV to emphasize the impact, while error bands with different hatching represent PDF uncertainties at $90\%$ CL. Also here, the central-scale choice in the 13 TeV $t \bar t$ theory predictions is set to $H_T/4$. Considerations are similar for the scale choice $H_T/2$.
The left panel in Fig.~\ref{ATL-lep+jet-stat} shows that statistical correlations have a negligible impact on the gluon PDF errors. In the right panel, we show distortions in the gluon central value at large $x$ in fits with and without statistical correlations. The impact of statistical correlations is very small and mostly localized at $x\gtrsim 0.6$ where there is weak or essentially no constraint from data. In addition, we observe that when the four distributions are added together, their effect on the gluon central value is diluted in the fit, and the quality-of-fit improves ($\chi^2/N_{pt} \approx 1.3$ in Tab.~\ref{global-fit-res}) when the central scale is set to $H_T/4$ as compared to $H_T/2$, in presence of statistical correlations. Without statistical correlations, a good description of the combined spectra is obtained by using either the $H_T/2$ ($\chi^2/N_{pt} \approx 0.93$), or the $H_T/4$ ($\chi^2/N_{pt} \approx 1.06$) scale choice, with a mild improvement in the $H_T/2$ case.  

As discussed in Sec.~\ref{sec:ATL-stat-corr}, the small impact of statistical correlations from the ATL13lj data is confirmed by \texttt{ePump}. However, the observed trend of pulls on the gluon from \texttt{ePump} in Fig.~\ref{stat-corr-ATL-PDFs} is in the opposite direction as compared to the global fit in Fig.~\ref{ATL-lep+jet-stat}. This is plausible, because the statistical procedure to update PDFs and their errors in \texttt{ePump} poses more restrictions as compared to a global PDF fit. 
In addition, $\chi^2/N_{pt}$ may differ depending on the treatment of the errors ({\it i.e.}, the ``error type'' option) in \texttt{ePump} (see for instance the $\chi^2/N_{pt}$ values in Tab.~\ref{stat-corr-tab} and Tab.~\ref{global-fit-res}). In Tab.~\ref{global-fit-res}, the \texttt{ePump} $\chi^2/N_{pt}$ values are obtained by using the ``error type 1'' option, where uncorrelated statistical and systematic errors are given, and correlated systematic errors (with percent correlation to each data point) are also given. The correlated systematic errors are treated multiplicatively, {\it i.e.}, by multiplying the percentages by the original best-fit theory predictions for each data point. 

The interplay between the two bin resolutions in the ATL13lj measurements is further discussed in Sec.~\ref{optimal-comb} where optimal data combinations are selected.  
The two bin resolutions differ in number of bins and bin size. We have cross checked the theory predictions in both cases, with independent calculations from \texttt{MATRIX}~\cite{Catani:2019hip,Catani:2020tko} and the NNLO version of \texttt{fastNLO} tables~\cite{Czakon:2017dip}, and we find consistency. Overall, we find that the gluon impact of the ATL13lj measurements in the global analysis is negligible.  

\subsection{Impact from the ATLAS all-hadronic channel} 

\begin{figure} 
\includegraphics[width=0.49\textwidth]{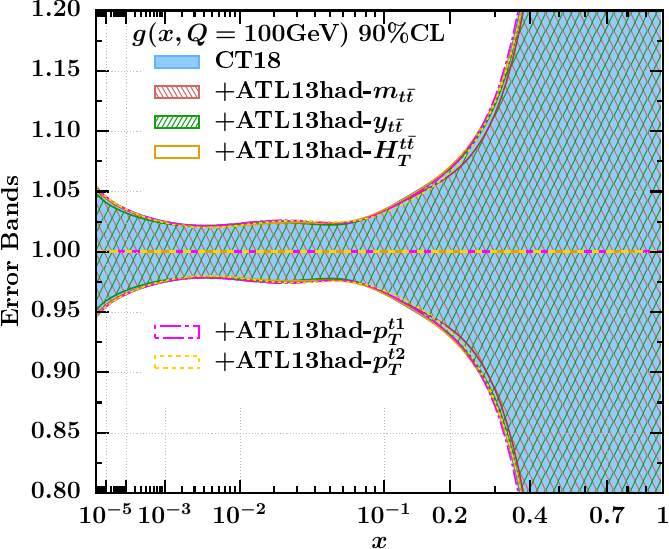}
\includegraphics[width=0.49\textwidth]{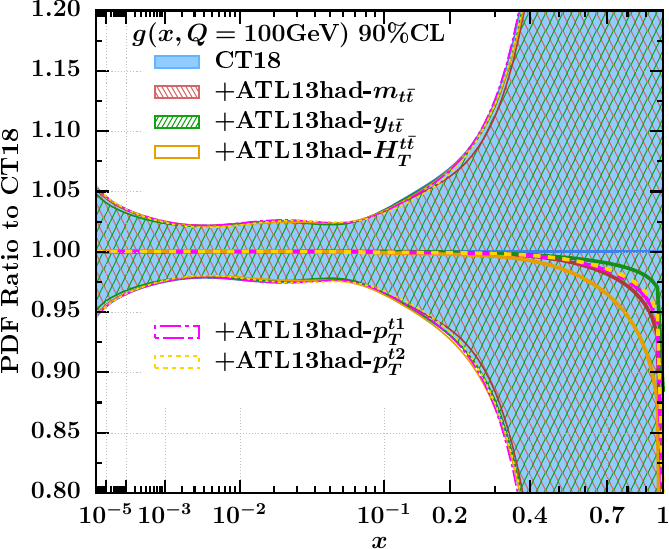}
\caption{{\bf ATLAS all-hadronic channel}. Impact of the $y_{t\bar t}, m_{t\bar t}, H_T^{t\bar t}$ , $p_{T,t_1}$ and $ p_{T,t_2}$  distributions, added one at a time in the NNLO global QCD analysis. Error bands with different hatching represent PDF uncertainties at $90\%$ CL. The central-scale choice in the theory predictions is set to $H_T/4$. }
\label{ATL-hadronic}
\end{figure}

The impact on the gluon when the ATL13had measurements~\cite{ATLAS:2020ccu} are added on top of the CT18 baseline is illustrated in Fig.~\ref{ATL-hadronic}.
The most relevant information in the global fit is obtained from the $y_{t\bar t}$, $m_{t\bar t}$, $H_T^{t\bar t}$, $p_{T,t_1}$, and $p_{T,t_2}$ distributions which we study here. 
The $y_{t\bar t}$, $m_{t\bar t}$, and $H_T^{t\bar t}$ distributions produce visible impact on the large-$x$ gluon at $x\gtrsim 0.5$. Pulls are all in the same direction and there is a preference for a softer gluon at large $x$, with a more pronounced effect from $H_T^{t\bar t}$ at $x\gtrsim 0.5$. 

A milder impact is observed for $y_{t\bar t}$ and $m_{t\bar t}$ at $x\gtrsim 0.6$. The $p_{T,t_1}$ and $p_{T,t_2}$ distributions produce an almost identical behavior, with most of the impact located at $x\gtrsim 0.5$. 
The constraining power of these data in this kinematic region is limited by large experimental uncertainties that affect these distributions.

For all distributions, the impact on PDF uncertainties is negligible. The $\chi^2/N_{pt}$ in Tab.~\ref{global-fit-res} is of order 1 for the $y_{t \bar t}$ distribution regardless of the central scale choice in the theory predictions. 
Moreover, the $m_{t\bar t}$ and $H_T^{t\bar t}$ distributions are in general not well described, while the $\chi^2/N_{pt}$ for individual fits using $p_{T,t_1}$ and $p_{T,t_2}$ depends on the central-scale choice. As before, correlated systematic uncertainties are given in terms of nuisance parameters. 
As for the ATL13lj data, the overall individual impact of the ATL13had measurements is negligible.

\subsection{Impact from the CMS dilepton channel} 

\begin{figure}
\includegraphics[width=0.49\textwidth]{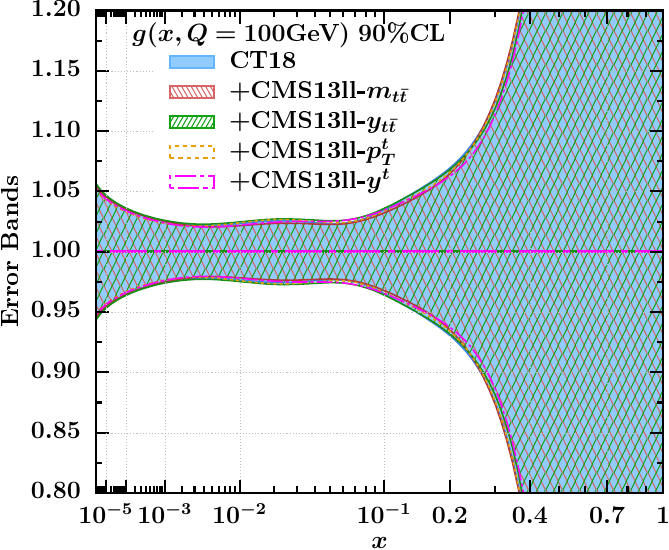}
\includegraphics[width=0.49\textwidth]{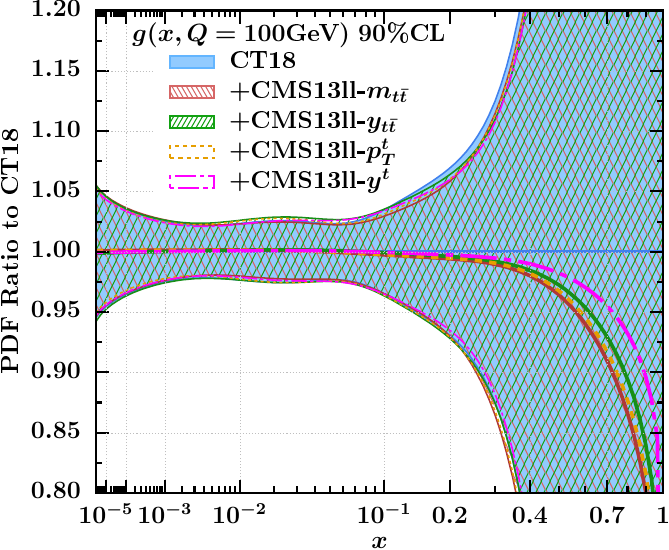}
\caption{{\bf CMS dilepton channel.} Impact of the $y_{t\bar t}$, $p_{T,t}$, $y_{t}$, and $m_{t\bar t}$ distributions added one at a time on top of the CT18 baseline in the global fit. 
Error bands with different hatching represent PDF uncertainties at $90\%$ CL. The central-scale choice in the 13 TeV $t\bar t$ theory predictions is $H_T/4$.}
\label{CMS-2lep}
\end{figure}

In Fig.~\ref{CMS-2lep} we illustrate the individual PDF impact of the  CMS13ll measurements~\cite{CMS:2018adi}.
The subset from which we gather the most relevant information to constrain the gluon is obtained by considering the $y_{t \bar t}$, $m_{t\bar t}$, $y_{t}$, and $p_{T,t}$ distributions. 
As discussed in Sec.~\ref{sec:13TeVdata}, the correlated systematic uncertainties for these measurements are published in terms of the covariance matrix representation. Therefore, we perform conversion to nuisance parameters as discussed in Appendix~\ref{exp-unc-treatment}. 

From the $\chi^2/N_{pt}$ values in Tab.~\ref{global-fit-res}, the $m_{t\bar t}$ and $p_{T,t}$ distributions do not produce a good fit, regardless of the central-scale choice in their theory predictions. This is also confirmed by the results from \texttt{ePump}. 
The $y_{t\bar t}$ and $y_t$ spectra are the distributions for which we obtain the best description with $\chi^2/N_{pt}\approx 1$ using the central-scale choice $H_T/2$, and with $\chi^2/N_{pt}\approx 0.7$ with central-scale set to $H_T/4$.  
We note in particular that the fit quality for the $y_t$ distribution is similar to that observed in the NNPDF4.0 global fit ($\chi^2/N_{pt} =0.52$).

The gluon central value with central-scale choice $H_T/4$ in Fig.~\ref{CMS-2lep} is affected by the CMS13ll data in the $x\gtrsim 0.25$ region with a strong preference for a much softer large-$x$ gluon as compared to the ATL13lj and ATL13had data. However, the impact on the gluon PDF uncertainty is very modest and comparable to that of the ATL13lj data.

In general, the $y_{t}$ and $y_{t\bar t}$ distributions are expected to show less sensitivity to radiative corrections and $m_t$ variations~\cite{Czakon:2016dgf}. The $\chi^2/N_{pt}$ description of these measurements in the global fit from Tab.~\ref{global-fit-res} is overall good for ATL13had, CMS13ll, and ATL13lj, with a very mild dependence on the scale choice for ATL13had and ATL13lj resolved with CMS bins. However, the constraining power of these data sets is modest due to larger errors as compared to the CMS13lj~\cite{CMS:2021vhb} data which are obtained with higher IL.      

\subsection{Impact from the CMS lepton+jets channel}
\label{subsec: Impact from CMS13lj}

The CMS collaboration released two sets of measurements in the lepton+jets channel, obtained with 35.8 fb$^{-1}$ of IL~\cite{CMS:2018htd} and with 137 fb$^{-1}$ IL~\cite{CMS:2021vhb}, respectively. We focus on the 137 fb$^{-1}$ measurements as they extend the previous ones~\cite{CMS:2018htd} and their precision is significantly improved. In addition, we observe that they place stronger constraints on the gluon in the global analysis, due to their improved systematic uncertainties which results in higher precision. 
Bin-by-bin statistical correlations are not available for these measurements. 

\begin{figure} 
\includegraphics[width=0.49\textwidth]{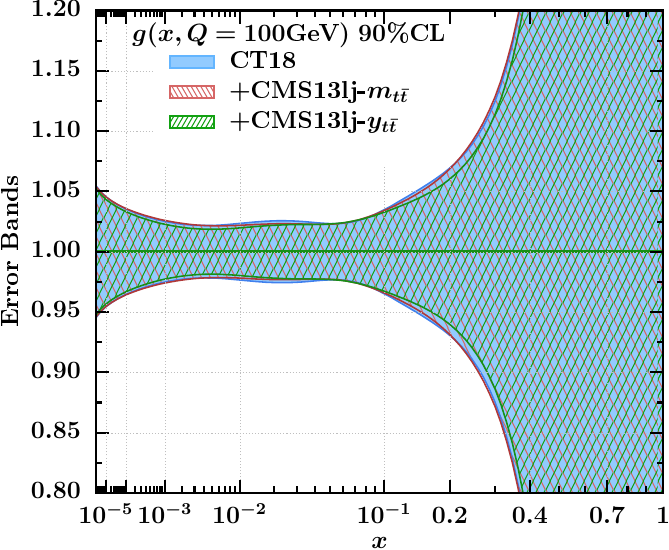}
\includegraphics[width=0.49\textwidth]{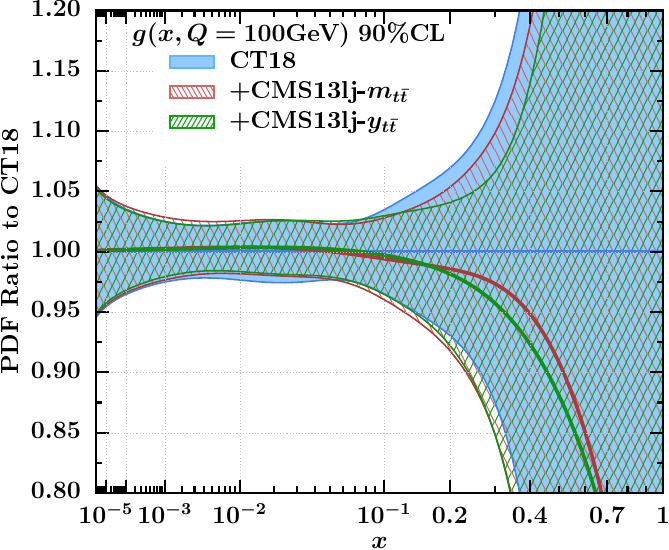}
\caption{{\bf CMS lepton + jets channel.} Impact of the $y_{t\bar t}$, and $m_{t\bar t}$ distributions added one at a time in the NNLO global fit. 
Error bands with different hatching represent PDF uncertainties at $90\%$ CL. }
\label{CMS-lep+jet}
\end{figure}

We study the individual impact of the $m_{t\bar{t}}$ and $y_{t\bar{t}}$ distributions, which provide us with the most relevant information among the CMS13lj measurements in the global fit. In Fig.~\ref{CMS-lep+jet}, we illustrate the impact from these two measurements on the gluon PDF. 

From the $\chi^2/N_{pt}$ values in Tab.~\ref{global-fit-res}, we observe that the $y_{t\bar t}$ distribution cannot be well described in either the global QCD analysis or with \texttt{ePump}, with  $\chi^2/N_{pt}\approx 6$ or more, regardless of the  scale choice. 
This can be in part ascribed to tensions with the jet production data and the $t\bar t$ pair production data at 8 TeV previously included in the CT18 analysis. However, another cause of disagreement may be related to the correlated systematic uncertainties associated to these measurements. 
We observe that even by varying the  scale in the theory prediction, the $y_{t\bar t}$ distribution seems to have a preference for a too soft gluon as compared to jet data.
A thorough study of the fit quality of the $y_{t\bar t}$ distribution would also require an investigation of amended versions of the CT18 gluon parametrizations. This will be presented in a forthcoming work.  
The $m_{t\bar t}$ distribution instead produces a better fit with a preference for a slightly harder gluon in the $0.2\lesssim x\lesssim 0.65$ range as compared to the $y_{t\bar t}$ spectrum.  
Moreover, we note that the description of $m_{t\bar t}$ improves when the scale in the theory prediction is set to $H_T/2$. 

We note that this dynamical scale choice is different from the default $H_T/4$ discussed in ref.~\cite{Czakon:2016dgf}. 
Throughout this work, the criterion according to which the optimal central scale for the theory prediction is chosen, is based on improvements in the quality-of-fit.

In Fig.~\ref{CMS-lep+jet}, we observe that these measurements have stronger impact on the gluon uncertainty as compared to those previously analyzed. This is mainly ascribed to a better control of experimental uncertainties which in turn, enhances the constraining power of these data. It would be useful to examine the impact of bin-by-bin statistical correlations, should they become available. 

\section{Optimal combinations of 13 TeV top-quark pair production measurements in the global analysis} 
\label{optimal-comb}

The analysis of impact from individual 13 TeV $t\bar t$ differential cross section measurements in various channels at ATLAS and CMS, previously discussed in Sec.~\ref{sec:global-fit}, allows us to select optimal combinations of measurements from which we can extract maximum information to constrain the gluon PDF, and minimize tensions among data in the extended baseline.
The guiding principle for this selection is essentially based on the quality of impact and quality of description in the global fit, in terms of $\chi^2$. That is, we select those measurements that place effective constraints and produce a good quality fit, and that do not deteriorate the description of data in the pre-existing baseline.

Looking at the $\chi^2/N_{pt}$ values and central-scale choice for the 13 TeV $t\bar t$ theory predictions in Tab.~\ref{global-fit-res}, we observe that the $y_{t\bar t}$ distribution is well described (i.e., $\chi^2/N_{pt}\approx 1$) in both the ATL13had and CMS13ll cases. We therefore select the $y_{t\bar t}$ distribution from these measurements.

We select the $m_{t\bar t}$ distribution from the CMS13lj measurements, because the $y_{t\bar t}$ one does not produce a good fit ($\chi^2/N_{pt} \geq 5$ regardless of the central-scale choice). 

Because of the differences observed in the two bin resolutions of the ATL13lj measurements, we consider two separate cases: in one case we include the $y_{t \bar t}$ distribution with the CMS bin resolution, whose $\chi^2$ appears to be more stable against central-scale changes in the theory prediction as compared to $m_{t\bar t}$ (see Tab.~\ref{global-fit-res}); in the other, we include the $m_{t\bar t}$, $y_{t\bar t}$, $y^B_{t\bar t}$, and $H^T_{t\bar t}$ distributions added together without statistical correlations as the latter produce negligible impact on PDF uncertainties. In addition, this greatly simplifies the global analysis. 

Therefore, we identify two optimal combinations which we refer to as CT18+nTT1 and CT18+nTT2, respectively. The CT18+nTT1 combination includes the ATL13had-$y_{t\bar{t}}$, CMS13ll-$y_{t\bar{t}}$, CMS13lj-$m_{t\bar{t}}$, and ATL13lj-$y_{t\bar{t}}$ distributions (resolved in terms of the CMS bins), while the CT18+nTT2 combination includes the same distributions from the ATL13had, CMS13ll, CMS13lj measurements, and the $y_{t\bar{t}}+y_{t\bar{t}}^B+m_{t\bar{t}}+H_T^{t\bar{t}}$ combination without statistical correlations from ATL13lj, resolved with ATLAS bins.    

\subsection{Main results from the post-CT18 global analysis with extended baseline}
\begin{figure}
\includegraphics[width=0.49\textwidth]{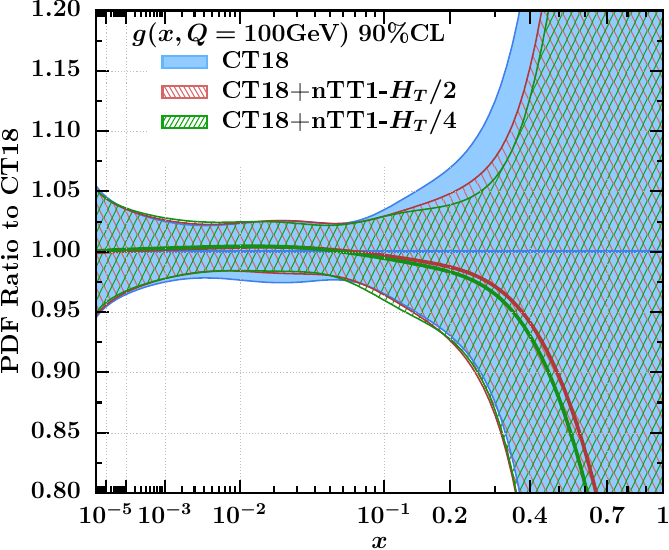}
\includegraphics[width=0.49\textwidth]{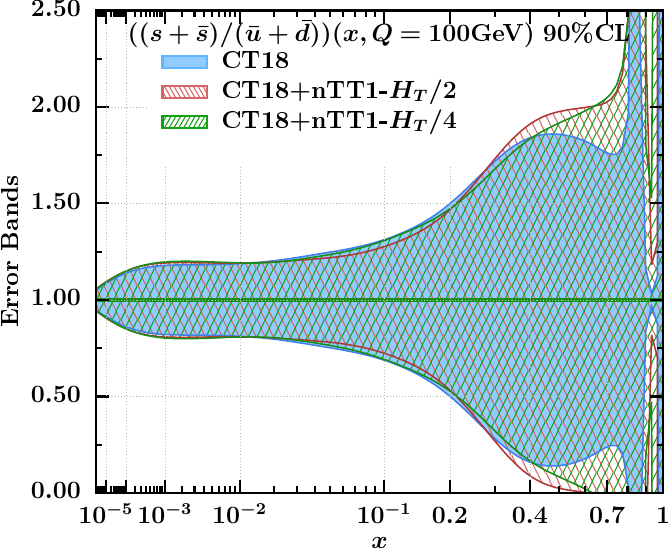}
\caption{Impact of CT18+nTT1 on the global QCD analysis at NNLO. {\bf Left}: gluon PDF ratio to CT18 NNLO. {\bf Right}: $R_s(x,Q)=(s+\bar s)/({\bar u}+{\bar d})$.
Error bands with different hatching represent different choices for the central scale in the 13 TeV $t \bar t$ theory predictions: $H_T/4$ (green), and $H_T/2$ (red). PDF uncertainties are evaluated at the $90\%$ CL.}
\label{Optimal-comb-1}
\end{figure}

\begin{figure} 
\includegraphics[width=0.49\textwidth]{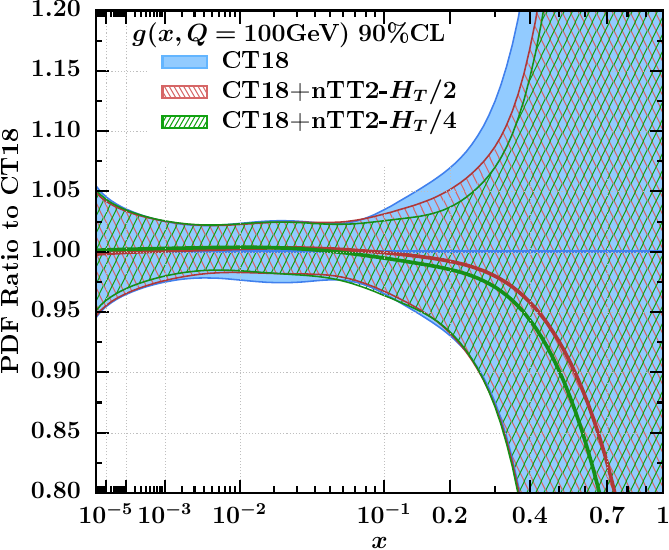}
\includegraphics[width=0.49\textwidth]{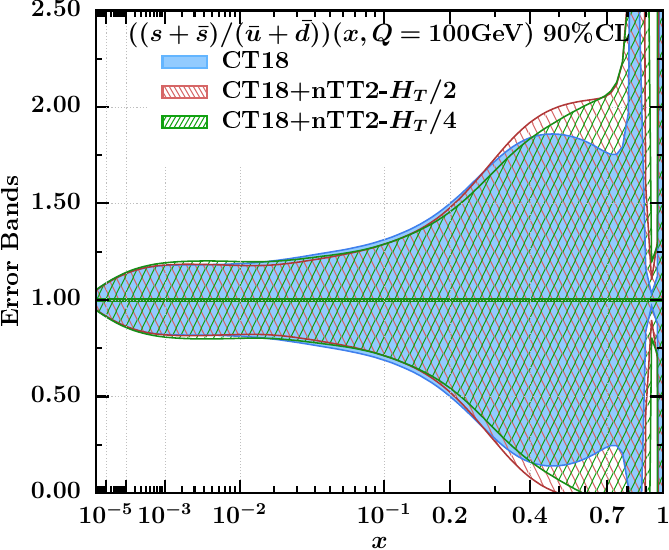}
\caption{Same as in Fig.~\ref{Optimal-comb-1} but for CT18+nTT2.}
\label{Optimal-comb-2}
\end{figure}

The results of the NNLO global analysis obtained with the nTT1 and nTT2 combinations are summarized in Tab.~\ref{global-fit-opt}, and illustrated in Figs.~\ref{Optimal-comb-1} and~\ref{Optimal-comb-2} where we show the gluon PDF ratio to CT18, and the $R_s=(s+\bar{s})/(\bar{u}+\bar{d})$ ratio. $g(x,Q)$ and $R_s(x,Q)$ are selected as representative cases as they show the most visible impact from the inclusion of the new data in the global fit. 

In Tab.~\ref{global-fit-opt} we compare the $\chi^2/N_{pt}$ values obtained from the CT18+nTT1 and CT18+nTT2 to CT18. 
There, we only report the CT18 data sets that exhibit a noticeable change in $\chi^2/N_{pt}$. The remaining CT18 data sets have the same $\chi^2/N_{pt}$ as in Tabs.~I and II in ref.~\cite{Hou:2019efy}.

The CT18 NNLO fit~\cite{Hou:2019efy} has $N_{pt}=3681$ and $\chi^2=4293$ with $\chi^2/N_{dof}=1.16$. In the CT18+nTT1 fit, the total number of points is $N_{pt}=3728$ and the resulting total $\chi^2$ from the global fit is $\chi^2=4341$ when the central scale of the 13 TeV $t\bar t$ theory is set to $H_T/2$, while we obtain $\chi^2=4346$ when the same scale is set to $H_T/4$. 
In the CT18+nTT2 fit, the total number of points is $N_{pt}=3752$,  $\chi^2=4366$ for the central-scale choice $H_T/2$, and $\chi^2=4376$ for $H_T/4$.

Comparing the $\chi^2$ values of the individual experiments to those of CT18 in Tab.~\ref{global-fit-opt}, we observe that the most noticeable quality-of-fit deterioration happens for both nTT1 and nTT2, regardless of the scale choice, in the case of the CMS 8 TeV single inclusive jet cross section~\cite{CMS:2016lna} measurements, and for the CMS13lj $m_{t\bar t}$ distribution when the  scale in the 13 TeV $t\bar t$ theory is set to $H_T/4$. 

In addition, a different PDF suppression preference over a wide range of $x$ for both nTT1 and nTT2 is observed between the LHCb 8 TeV $Z \rightarrow e^- e^+$ forward rapidity cross section~\cite{LHCb:2015kwa} and the CMS13lj $m_{t\bar t}$ distribution. 
The LHCb 8 TeV measurements have an impact on the strange PDF, anti-quarks (e.g., $\bar{u}, \bar{d}$), and their errors at both small and large $x$. 
In particular, the $\bar{u}$ and $\bar{d}$ PDFs are impacted in the $10^{-4} \lesssim x \lesssim 10^{-2}$ range.
This interplay as well as that between the top-quark and the jet data can be further understood in terms of the $L_2$ sensitivity~\cite{Jing:2023isu}, that is a statistical tool to explore the pulls from individual measurements on the best-fit PDFs, and to identify tensions between competing data sets.
It is important to point out that the use of the $L_2$ sensitivity in this study follows a different approach because it is used to study PDF suppression preferences using PDFs that are obtained after  a global fit. 

\begin{figure}
\includegraphics[width=1.00\textwidth]{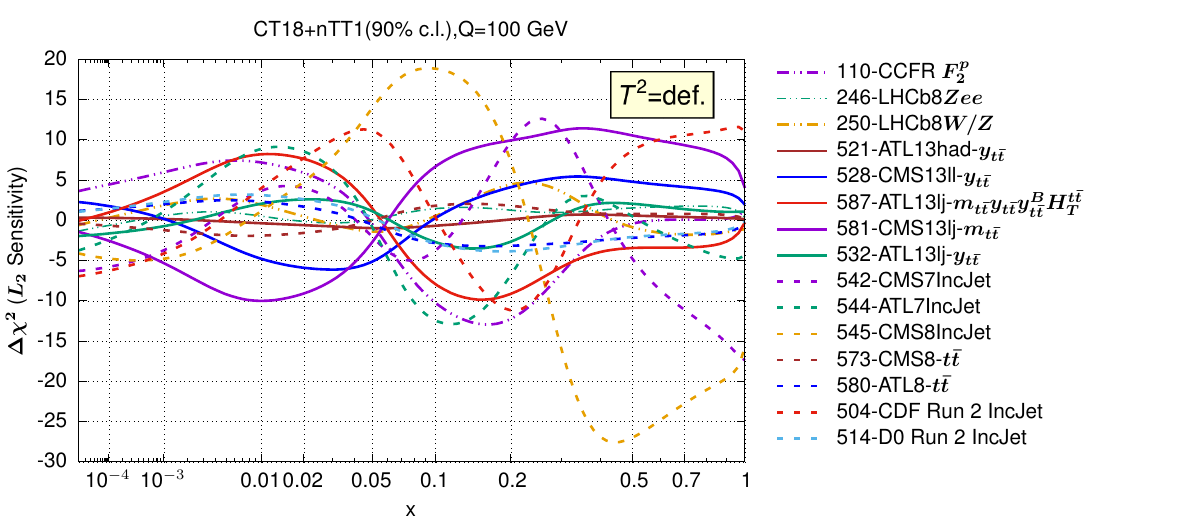}
\caption{$L_2$ sensitivity study for the gluon PDF: ${t\bar t}$ and jet production in the CT18+nTT1 fit.}
\label{L2-TopJet-data}
\end{figure}

To illustrate the different suppression-pattern preference between the new $t\bar t$ and jet production data, in Fig.~\ref{L2-TopJet-data} we present the $L_2$ sensitivity plot for the gluon PDF in the CT18+nTT1 fit. The plot relative to the CT18+nTT2 gluon is very similar and we omit it here. The $T^2=\textrm{def}.$ in the inset label refers to the default  tolerance criteria adopted in the CT18 analysis. 
From the gluon suppression patterns in Fig.~\ref{L2-TopJet-data}, we observe for example that the CMS13lj-$m_{t\bar{t}}$ and the CMS13ll-$y_{t\bar{t}}$ data have opposite trend with respect to the CMS 8 TeV inclusive jet production data in the 
 $x > 0.3$ range, where CMS13lj-$m_{t\bar{t}}$ and CMS13ll-$y_{t\bar{t}}$ prefer a softer gluon. We note that a positive value of $\Delta\chi^2$ in the $L_2$ sensitivity plot, cf. Fig.~\ref{L2-TopJet-data}, indicates the preference of decreasing the gluon PDF. Namely, a positive sensitivity indicates a negative pull on the corresponding PDF (see further discussion in ref.~\cite{Jing:2023isu}).
Another different suppression-pattern trend is observed for the ATLAS 7 TeV inclusive jet data, which prefer a softer gluon in the $10^{-3} \lesssim x \lesssim 0.05$ range as compared to CMS13lj-$m_{t\bar{t}}$ and CMS13ll-$y_{t\bar{t}}$.

\begin{figure} 
\includegraphics[width=0.49\textwidth]{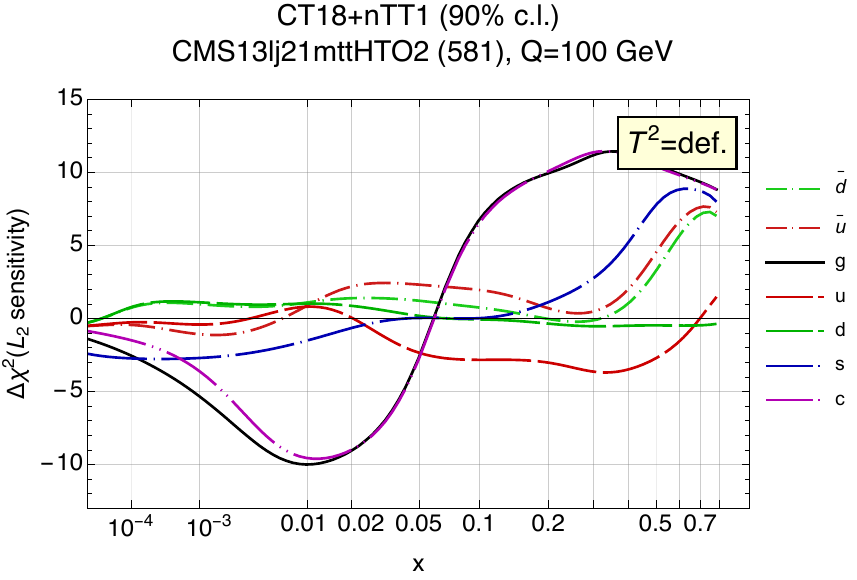}
\includegraphics[width=0.49\textwidth]{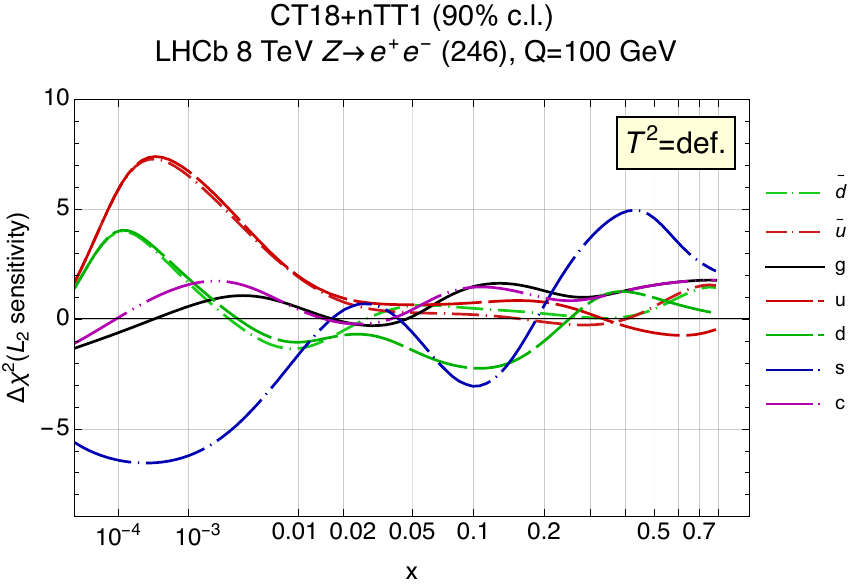}
\caption{$L_2$ sensitivity study to explore the interplay between the CMS13lj (left) and LHCb 8 TeV 2.0 fb$^{-1}$ $Z\rightarrow e^{-} e^{+}$ forward rapidity cross section measurements (right).}
\label{L2-nTT1}
\end{figure}

\begin{figure} 
\includegraphics[width=0.7\textwidth]{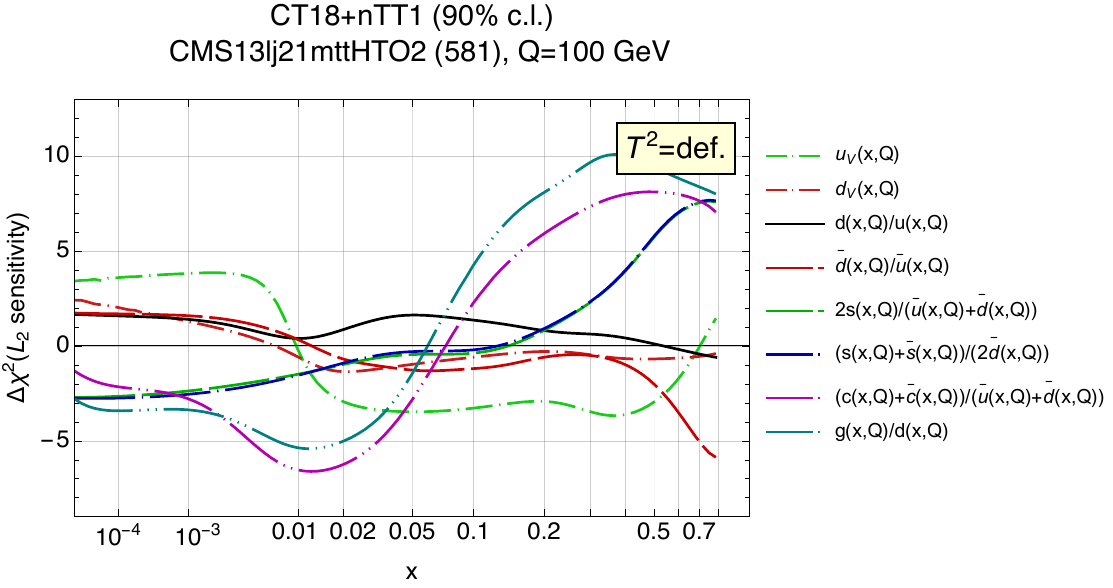}
\includegraphics[width=0.7\textwidth]{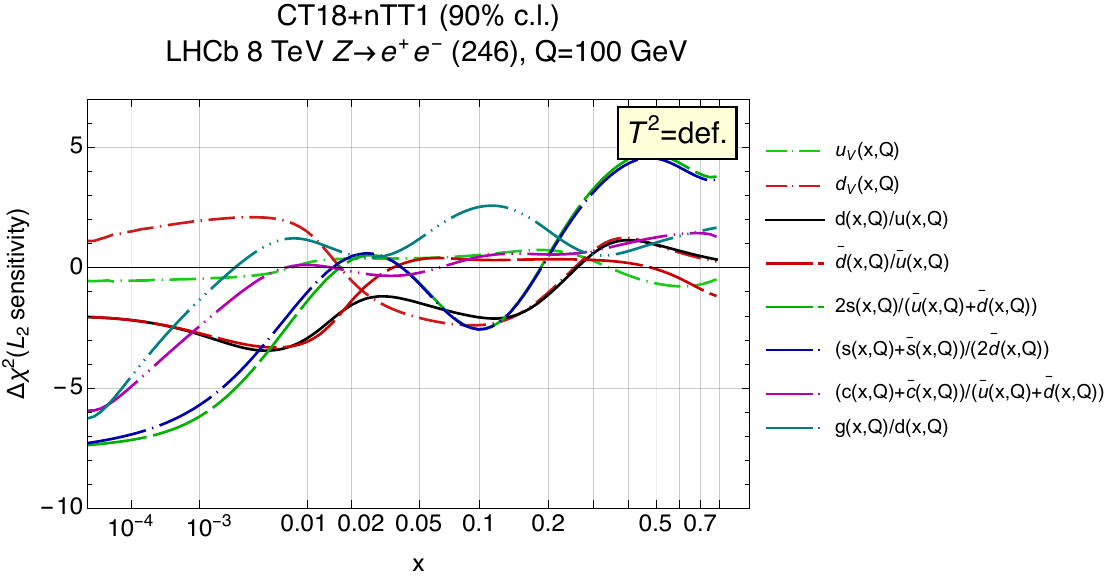}
\caption{Same as in Fig.~\ref{L2-nTT1}, but for different PDF ratios.}
\label{L2-Rs-nTT1}
\end{figure}

In Figs.~\ref{L2-nTT1} and~\ref{L2-Rs-nTT1}, we illustrate the $L_2$ sensitivity for the CMS13lj-$m_{t\bar t}$ distribution and the LHCb 8 TeV 2.0 fb$^{-1}$ $Z\rightarrow e^{-} e^{+}$ forward rapidity cross section measurements in the CT18+nTT1 global fit. 
As before, the same figures for the CT18+nTT2 global fit are very similar and we do not show them here.  

In Fig.~\ref{L2-nTT1}, we observe that the $m_{t\bar t}$ distribution in CMS13lj has a strong preference for a softer gluon in the $0.1 \lesssim x\lesssim 0.7$ range, as compared to the LHCb 8 TeV $Z\rightarrow e^{-} e^{+}$ large rapidity data, which prefer a harder gluon in the same range. In addition, we observe a different suppression preference for the $u$, $d$, $\bar{u}$, $\bar{d}$, and $s$ quarks across the entire $x$ range, where in particular we note  a higher preference for softer 
$u$- and $\bar u$-quark
PDFs in the small-$x$ ($10^{-4} \lesssim x \lesssim 10^{-2}$) range by the LHCb 8 TeV $Z$ rapidity measurements.

These different suppression preferences have an impact on quark ratios, in particular on $R_s$.     
The $L_2$ sensitivity for the $R_s$ ratio is illustrated in Fig.~\ref{L2-Rs-nTT1} where $u_V$, $d_V$ and other PDF ratios are also shown\footnote{In the CT18 analysis $s=\bar{s}$, therefore $R_s=2 s/(\bar u + \bar d)$.}.

The increase in the $\chi^2$ values in Tab.~\ref{global-fit-opt} and the different suppression preferences discussed above are in part reflected by an increase in the uncertainty of the $R_s$ ratio, as compared to CT18. This is illustrated in Figs.~\ref{Optimal-comb-1} and~\ref{Optimal-comb-2}. A more flexible choice for the strange-quark parametrization allowing $s\neq\bar{s}$ in the global fit, such as CT18As or CT18As\_Lat NNLO analysis with lattice QCD constraints \cite{Hou:2022onq}, may in general be beneficial to the description of the data. However, the different suppression preferences for the $u$, $d$, $\bar{u}$, and $\bar d$ quarks across the entire range of $x$ also play an important role in the determination of the uncertainty associated to $R_s$.

The $R_s$ uncertainty is shown in the right plot of Figs.~\ref{Optimal-comb-1} and~\ref{Optimal-comb-2} for CT18+nTT1 and CT18+nTT2, respectively. We note that the uncertainty increase in the $0.3 \lesssim x \lesssim 0.8$ region has a different shape in both the CT18+nTT1 and CT18+nTT2 fits for the two different scale choices $H_T/2$ and $H_T/4$.    
The left plots in Figs.~\ref{Optimal-comb-1} and~\ref{Optimal-comb-2} illustrate the resulting changes in the gluon PDF and its uncertainty, as well as (small) variations induced by selecting the two central scales $H_T/2$, and $H_T/4$, in the $t \bar t$ theory predictions at 13 TeV. For both scale choices we obtain a reduction in the gluon uncertainty in the $0.2 \lesssim  x \lesssim 0.5$ range
and above in the extrapolation region, and a much less pronounced reduction in the $2 \times 10^{-3} \lesssim x \lesssim 5\times 10^{-2}$ range.

We observe a preference for a softer gluon as compared to CT18 in the $10^{-1} \lesssim x \lesssim 0.6$ region. This constraint is mainly driven by the $m_{t\bar t}$ distribution from the CMS13lj data, with an improved description when the central scale is set to $H_T/2$.  
A data vs theory comparison for the CMS13lj $m_{t\bar t}$ and $y_{t\bar t}$ distributions, obtained with CT18nTT1 and CT18+nTT2 PDFs, respectively, is shown in Fig.~\ref{CMS13lj-distr-postfit} in Appendix \ref{CMS13lj-postfit}.

While the quality-of-fit of the two combinations CT18+nTT1 and CT18+nTT2 is essentially the same as that of CT18 with $\chi^2/N_{dof}\approx 1.16$, small differences between the CT18+nTT1 and CT18+nTT2 global fits are noticeable in Figs.~\ref{Optimal-comb-1} and~\ref{Optimal-comb-2} as compared to CT18. These are discussed in more detail in Sec.~\ref{nTT1-nTT2-diff}.

\begin{widetext}
\begingroup
\squeezetable
\begin{table}
\begin{tabular}{|l|lr|c|c|c|c|c|c|}
\hline
 &  &  &   &  & $H_T/2$ & $H_T/2$ & $H_T/4$ & $H_T/4$ \tabularnewline
\hline
\textbf{ID\# }  & \textbf{Experimental data set} &  & $N_{pt}$  & CT18 & nTT1 & nTT2 & nTT1 & nTT2 \tabularnewline
\hline
\hline
 110 & CCFR $F_{2}^{p}$                                                                    & \cite{CCFRNuTeV:2000qwc}          &   69  &  1.1& 1.1 & 1.1  & 1.2 & 1.2 \tabularnewline\hline
 545 & CMS 8 TeV 19.7 fb$^{-1}$, single incl. jet cross sec., $R=0.7$, (extended in y)     & \cite{CMS:2016lna}  & 185   & 1.1 & 1.2 & 1.2  & 1.2 & 1.2 \tabularnewline\hline
 573 & CMS 8 TeV 19.7 fb$^{-1}$, $t\bar{t}$ norm. double-diff. top $p_T$ and $y$ cross sec.& \cite{CMS:2017iqf}     &  16   & 1.2 & 1.2 & 1.2  & 1.1 & 1.1 \tabularnewline\hline
 580 & ATLAS 8 TeV 20.3 fb$^{-1}$, $t\bar{t}$ $p_{T,t}$ and $m_{t\bar{t}}$ abs. spectrum & \cite{ATLAS:2015lsn}          &  15   & 0.6 & 0.7 & 0.7  & 0.7 & 0.7 \tabularnewline\hline
 521 & ATLAS 13 TeV 36.1 fb$^{-1}$, $t\bar{t}$ abs. $y_{t\bar t}$ cross sec. all-hadronic  & \cite{ATLAS:2020ccu}        &  12   &  -   & 1.0 & 1.0  & 1.1 & 1.1 \tabularnewline\hline
 528 & CMS 13 TeV 35.9 fb$^{-1}$, $t\bar{t}$ abs. $y_{t\bar t}$ cross sec. dilepton ch.    & \cite{CMS:2018adi}          &  10   &  -   & 0.8 & 0.8  & 0.5 & 0.7 \tabularnewline\hline
 532 & ATLAS 13 TeV 36 fb$^{-1}$, $t\bar{t}$ abs. $y_{t\bar t}$ cross sec. l+j ch. cms-bin & \cite{ATLAS:2019hxz}        &  10   &   -  & 0.7 &  -    & 0.8 &  - \tabularnewline\hline
 587 & ATLAS 13 TeV 36 fb$^{-1}$, $t\bar{t}$ abs. $y_{t\bar t}, m_{t\bar t},y_{t\bar t}^B, H^{t\bar t}_{T}$ cross secs. l+j ch.                    & \cite{ATLAS:2019hxz}        &  34   &  -  &  -  &  1.0  & - & 1.1 \tabularnewline\hline
 581 & CMS 13 TeV 137 fb$^{-1}$, $t\bar{t}$ abs. $m_{t\bar t}$ cross sec. l+j ch.          & \cite{CMS:2021vhb}          &  15   &  -   & 1.1 &  1.1  & 1.6 & 1.7 \tabularnewline
\hline
\end{tabular}
\caption{Data sets of the extended NNLO global QCD analysis including the optimal combinations CT18+nTT1 and CT18+nTT2. Here we directly compare the quality-of-fit found for CT18+nTT1 and CT18+nTT2 vs. CT18 NNLO on the basis of $\chi^2/N_{pt}$ and scale choices 
$\{H_T/2,H_T/4\}$ for the central theory predictions of the $t\bar t$ data at 13 TeV. We only report the CT18 data sets for which the $\chi^2/N_{pt}$ exhibits a noticeable change.}
\label{global-fit-opt}
\end{table}
\endgroup
\end{widetext}

\subsection{Interplay between top-quark and jet production data} 
\label{interplay}

Top-quark pair production and inclusive jet production at the LHC place complementary constraints on the gluon and other PDFs as these two processes overlap in the $Q-x$ kinematic plane. It is therefore interesting to further examine the interplay between top-quark pair and jet production data in NNLO global analyses with the CT18+nTT1 and CT18+nTT2 combinations. In particular, we analyze the individual constrains placed on the gluon from these processes separately. 

To this purpose, we perform three main fits in each of which we study the impact of different central scales in the 13 TeV $t\bar t$ theory predictions. These global fits are performed by using the CT18 baseline as the starting point, where we remove either the jet or the 8 TeV $t\bar t$ data, or both, and add the nTT1 combination with scale choice $H_T/2$ or $H_T/4$. The same fits performed with the nTT2 combination produce similar results.

The complementarity and tensions of jet and top data has also been analyzed before for 8 TeV top-quark pair production measurements in the context of the PDF4LHC21 studies~\cite{Cridge:2021qjj,PDF4LHCWorkingGroup:2022cjn,Jing:2023isu} as well as in the recent PDF analyses in Refs.~\cite{AbdulKhalek:2020jut,NNPDF:2021njg,Bailey:2020ooq,Hou:2019efy}.    

{\bf Global fits without inclusive jet data.}
In a first global fit, which we label CT18mJet, PDFs are obtained from the CT18 fit by removing the inclusive jet production data. To explore the impact of the new 13 TeV $t \bar t$ combinations we consider two variants of this fit including the nTT1 data sets and where the theory predictions for the latter are included with different central scales. These fit variants are: $i$) the CT18 baseline without QCD jets and with nTT1 data sets, with central scale $H_T/2$, labeled as CT18mJet+nTT1-$H_T/2$; $ii$) the CT18 baseline without QCD jets and with nTT1 data sets, with central scale $H_T/4$, labeled as CT18mJet+nTT1-$H_T/4$. 

\begin{figure}
\includegraphics[width=0.50\textwidth]{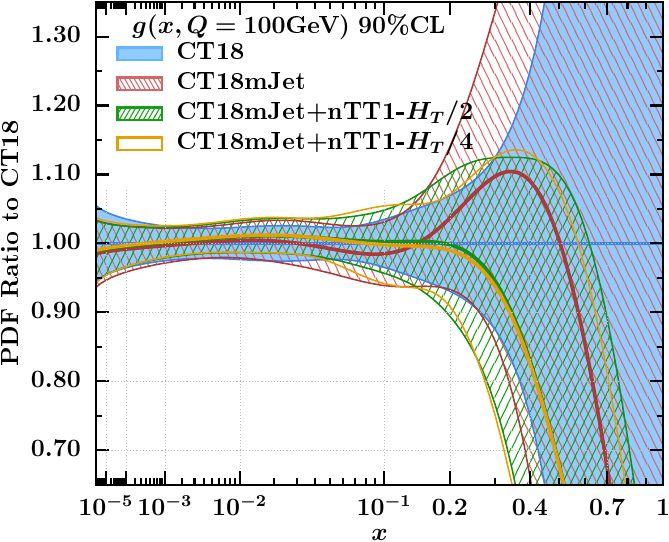}
\includegraphics[width=0.475\textwidth]{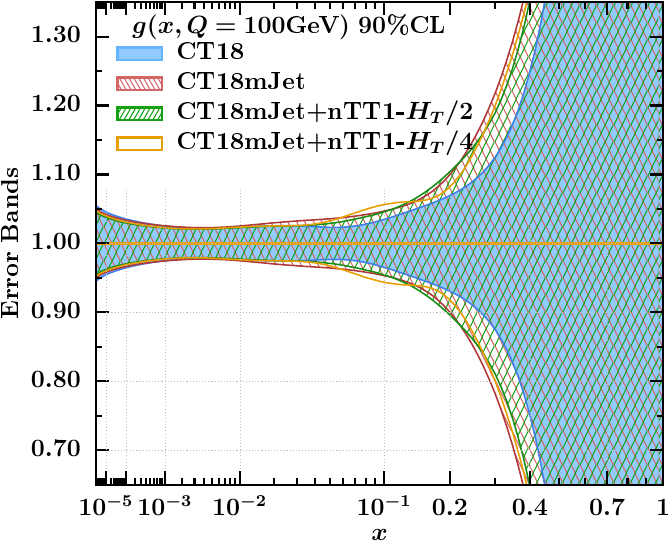}
\caption{Global fit without jet data vs CT18NNLO. Left: PDF ratio to CT18. Right: Error bands comparison.}
\label{CT18mQCDJet}
\end{figure}

We focus our attention on the gluon which is the PDF receiving most of the impact. The results for the gluon PDF and its uncertainties resulting from these global analyses are illustrated in Fig.~\ref{CT18mQCDJet} where they are compared to the CT18 gluon at $Q=100$ GeV.  

Looking at the PDF ratio plot in Fig.~\ref{CT18mQCDJet} (left), we observe that removing the QCD jet data from the CT18 baseline (CT18mJet) results in an increase of the PDF uncertainty at large $x$ in the $0.1 \lesssim x \lesssim 0.5$ range where there is a preference for a harder gluon with a bump in the $0.2 \lesssim x \lesssim 0.6$ region. Upon inclusion of the nTT1 combination in the CT18mJet+nTT1-$H_T/2$ and CT18mJet+nTT1-$H_T/4$ global fits, we observe a softer gluon in the $x\gtrsim 0.1$ range, regardless of the central-scale choice. 

Comparing the gluon error bands in Fig.~\ref{CT18mQCDJet} (right), we note that uncertainties are only marginally reduced when nTT1 is included, with a small increase in the $0.06 \lesssim x \lesssim 0.15$ region for scale choice $H_T/4$. This is due to small tensions with several data sets in the baseline, in particular the D0 run II 9.7 fb$^{-1}$ electron charge asymmetry $A_{ch}$, with $p_{Tl}>25$ GeV measurements~\cite{D0:2014kma}, and the structure function measurements $F^p_2$~\cite{BCDMS:1989qop} and $F^d_2$~\cite{BCDMS:1989ggw} from the BCDMS collaboration.
From an $L_2$ sensitivity study, one can see that the latter have a preference for a harder gluon PDF at large $x$ as compared to the nTT1 data, while the D0 run II electron charge asymmetry $A_{ch}$ data are sensitive to the $u$- and ${\bar u}$-quark PDFs with opposite trend as compared to BCDMS $F^d_2$.

\begin{figure}
\includegraphics[width=0.50\textwidth]{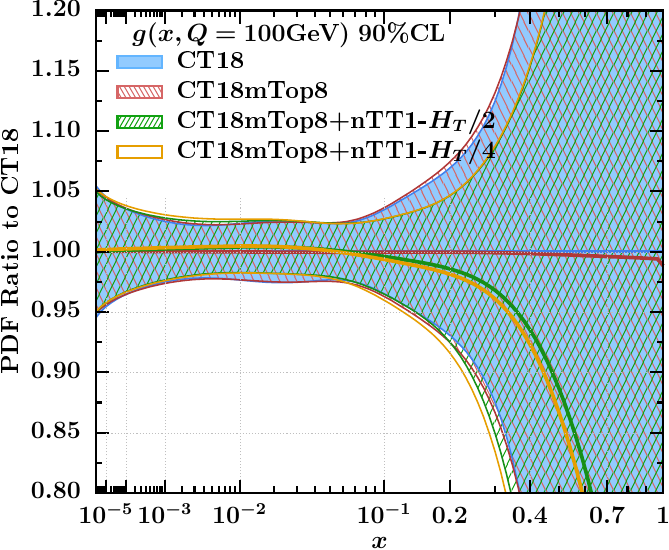}
\includegraphics[width=0.475\textwidth]{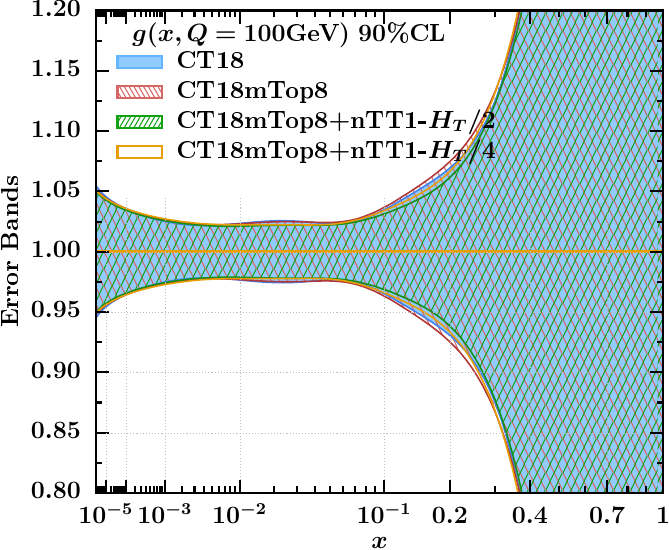}
\caption{Global fit without $t\bar t$ data vs CT18NNLO. Left: PDF ratio to CT18. Right: Error bands comparison.}
\label{CT18mttb}
\end{figure}

{\bf Global fits without $t\bar t$ data.}
In a second global fit, labeled as CT18mTop8, PDFs are obtained from the CT18 fit by removing top-quark pair production data (at $\sqrt{S}=$8 TeV only in CT18).  
As before, we consider two fit variants to assess the impact of the 13 TeV $t\bar t$ data, where the nTT1 combination is included with scale choices $H_T/2$ and $H_T/4$ in the theoretical predictions for the nTT1 data subset. These fit variants are labeled as CT18mTop8+nTT1-$H_T/2$ and CT18mTop8+nTT1-$H_T/4$, respectively. The results for the gluon PDF are shown in Fig.~\ref{CT18mttb} where we illustrate the impact. This is complementary to that in Fig.~\ref{CT18mQCDJet}. 

We note that when the 8 TeV $t\bar t$ measurements are removed from the CT18 baseline, the impact is negligible on both the central value and uncertainty of the gluon. When the nTT1 combination is included, there is a preference for a softer gluon at large $x$ in the $0.15 \lesssim x \lesssim 0.9$ region. However, the impact on the gluon uncertainty is negligible, as shown in Fig.~\ref{CT18mttb} (right), regardless of the scale choice. Moreover, we observe that the central value of the gluon in the nTT1 fits is harder as compared to the  CT18mJet+nTT1-$H_T/2$ and CT18mJet+nTT1-$H_T/4$ fits. This is due to the impact of inclusive jet production data which prefer a softer gluon PDF in the large-$x$ region. 

{\bf Global fits without QCD jets and $t\bar t$ production.}
Finally, in a third fit labeled as CT18mJet\&Top8, PDFs are obtained by removing both the inclusive jet and $t\bar t$ data from the CT18 baseline.  
The results of this fit are then compared to the fit variants obtained by including the nTT1 data subset with central scale choice $H_T/2$ for the nTT1 theory predictions labeled as CT18mJet\&Top8+nTT1-$H_T/2$, and to that obtained with $H_T/4$, labeled as CT18mJet\&Top8+nTT1-$H_T/4$. 

\begin{figure} 
\includegraphics[width=0.49\textwidth]{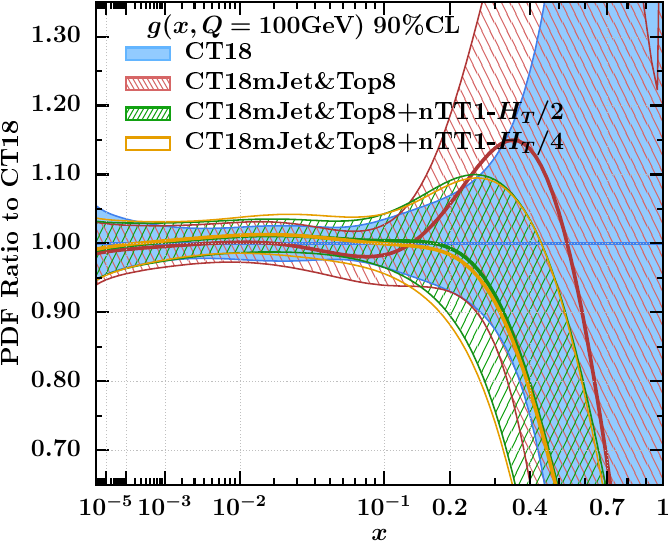}
\includegraphics[width=0.49\textwidth]{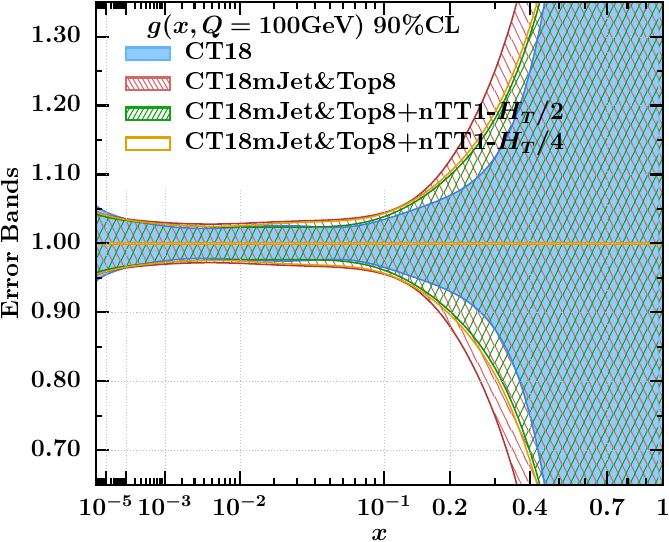}
\caption{Global fit without $t\bar t$ and jet data vs CT18NNLO. Left: PDF ratio to CT18. Right: Error bands comparison.
}
\label{CT18mttbandJet}
\end{figure}

The results are shown in Fig.~\ref{CT18mttbandJet} where we observe that the general trend is similar to that in Fig.~\ref{CT18mQCDJet}. However, we note that the 8 TeV $t\bar t$ data lead to an approximately 8\% reduction in the gluon central value in the $0.2 \lesssim x \lesssim 0.6$ region (compare the red solid line in Fig.~\ref{CT18mQCDJet}(left) to that in Fig.~\ref{CT18mttbandJet}(left)).

Looking at the PDF uncertainties in Fig.~\ref{CT18mttbandJet}(right), we note only a small increase in the $0.15 \lesssim x \lesssim 0.4$ range as compared to the CT18mJet fit, which reflects the small impact of the 8 TeV $t\bar t $ data in the CT18 fit. When the nTT1 combination is included, the gluon central value becomes softer in the $0.2 \lesssim x \lesssim 0.6$ range similar to that of the CT18mJet+nTT1-$H_T/2$ and CT18mJet+nTT1-$H_T/4$ fits with a reduction of the uncertainty in the $0.2 \lesssim x \lesssim 0.5$ range as compared to the CT18mJet\&Top8 fit (see Fig.~\ref{CT18mttbandJet}(right)). 

The conclusions we draw by comparing the results of the three main fits discussed above is that inclusive jet data place stronger constraints on the gluon as compared to top-quark pair production. This is mainly ascribed to the much larger number of data points in the inclusive jet measurements which tend to dilute the impact from $t\bar t$ production. 

However, the impact from the 13 TeV $t\bar t$ production data results in a softer gluon at large $x$, similar to that of the LHC jet data, but with a very different degree of suppression. In fact, the phase-space suppression in the hard scattering contributions for these two processes is different and generates different shapes and suppression in Fig.~\ref{CT18mQCDJet}(left), and Fig.~\ref{CT18mttb}(left). Moreover, we note that most of the gluon suppression at large $x$ is driven by the CMS13lj data with 137 fb$^{-1}$ of IL.  

 In Tab.~\ref{global-fit-opt2} we summarize the most noticeable changes in the $\chi^2/N_{pt}$ values, as described in Tab.~\ref{global-fit-opt}, but for the CT18mTop, CT18mJet, and CT18mJet$\&$Top8 global fits. Overall, the description of the new 13 TeV $t\bar{t}$ data improves after removing the 8 TeV $t\bar{t}$ and/or jet data from the CT18 baseline. In particular, one can consider the CT18+nTT1-$H_T/2$ global fit where the $\chi^2/N_{pt}=1.1$ for the CMS13lj-$m_{t\bar{t}}$ data. In the CT18mJet fit the $\chi^2/N_{pt}$ decreases to 0.9 and in CT18mJet$\&$Top8+nTT1-$HT/2$ it further reduces to 0.8. Similarly, for the CMS13ll-$y_{t\bar{t}}$ data, one finds $\chi^2/N_{pt}=0.8$ in the CT18+ntt1-$HT/2$ fit and this reduces to $\chi^2/N_{pt}=0.6$ in the CT18mJet$\&$Top8+ntt1-$HT/2$ fit.
In addition, an $L_2$ sensitivity study is shown in Sec.~\ref{optimal-comb},  where  Fig.~\ref{L2-TopJet-data} displays the impact on the gluon PDF from both $t \bar t$ and jet production data  as well as the different suppression preferences among the experiments.

Finally, we observe that all of the changes in shape and magnitude in the gluon central value from the impact of the new measurements are within the 90\% uncertainty band obtained from the CT18 NNLO fit. 
\begin{widetext}
\begingroup
\squeezetable
\begin{table}
\begin{tabular}{|l|lr|c|c|c|c|c|c|}
\hline
 &  &  &   &  & $H_T/2$ & $H_T/2$ & $H_T/4$ & $H_T/4$ \tabularnewline
\hline
\textbf{ID\# }  & \textbf{Experimental data set} &  & $N_{pt}$  & CT18mTop & nTT1 & nTT2 & nTT1 & nTT2 \tabularnewline
\hline

\hline
 110 & CCFR $F_{2}^{p}$                                                                                                                  & \cite{CCFRNuTeV:2000qwc}    &  69   &  1.1  &     1.1   &      1.1&            1.2&          1.2   \\
 545 & CMS 8 TeV 19.7 fb$^{-1}$, single incl. jet cross sec., $R=0.7$, (extended in y)                                                   & \cite{CMS:2016lna}          & 185   &  1.1  &     1.2   &      1.2&            1.2&          1.2   \\
   521 & ATLAS 13 TeV 36.1 fb$^{-1}$, $t\bar{t}$ abs. $y_{t\bar t}$ cross sec. all-hadronic                                               & \cite{ATLAS:2020ccu}       &  12   &  - &    1.0   &   1.0   &    1.1   &  1.1  \\  
  528 & CMS 13 TeV 35.9 fb$^{-1}$, $t\bar{t}$ abs. $y_{t\bar t}$ cross sec. dilepton ch.                                                 & \cite{CMS:2018adi}          &  10   &  - &    0.8   &   0.9   &    0.5   &  0.6  \\  
  532 & ATLAS 13 TeV 36 fb$^{-1}$, $t\bar{t}$ abs. $y_{t\bar t}$ cross sec. l+j ch. cms-bin                                              & \cite{ATLAS:2019hxz}        &  10   &  - &    0.7   & -       &    0.7   &  -    \\
  587 & ATLAS 13 TeV 36 fb$^{-1}$, $t\bar{t}$ abs. $y_{t\bar t}, m_{t\bar t},y_{t\bar t}^B, H^{t\bar t}_{T}$ cross secs. l+j ch.         & \cite{ATLAS:2019hxz}        &  34   &  - &    -     &   1.0   &     -    &  1.1  \\
  581 & CMS 13 TeV 137 fb$^{-1}$, $t\bar{t}$ abs. $m_{t\bar t}$ cross sec. l+j ch.                                                       & \cite{CMS:2021vhb}          &  15   &  - &    1.1   &   1.2   &    1.6   &  1.6  \\  
\hline
\hline
     &                                                        CT18mJet                                                                   &                             &            &  CT18mJet     &     &      &     &       \\
\hline
 110 & CCFR $F_{2}^{p}$                                                                                                                  & \cite{CCFRNuTeV:2000qwc}    &  69   &   1.2  &     1.2&                 1.2&            1.2&          1.2   \\
 573 & CMS 8 TeV 19.7 fb$^{-1}$, $t\bar{t}$ norm. double-diff. top $p_T$ and $y$ cross sec.                                              & \cite{CMS:2017iqf}          &  16   &   1.2  &     1.2&                 1.2&            1.2&          1.1   \\
 580 & ATLAS 8 TeV 20.3 fb$^{-1}$, $t\bar{t}$ $p_{T,t}$ and $m_{t\bar{t}}$ abs. spectrum                                                 & \cite{ATLAS:2015lsn}        &  15   &   0.7  &     0.7&                 0.7&            0.7&          0.7   \\
 521 & ATLAS 13 TeV 36.1 fb$^{-1}$, $t\bar{t}$ abs. $y_{t\bar t}$ cross sec. all-hadronic                                               & \cite{ATLAS:2020ccu}        &  12   &  -          &      1.0   &   1.0   &    1.0   &  1.0       \\  
 528 & CMS 13 TeV 35.9 fb$^{-1}$, $t\bar{t}$ abs. $y_{t\bar t}$ cross sec. dilepton ch.                                                 & \cite{CMS:2018adi}          &  10   &  -          &      0.6   &   0.7   &    0.4   &  0.5       \\     
 532 & ATLAS 13 TeV 36 fb$^{-1}$, $t\bar{t}$ abs. $y_{t\bar t}$ cross sec. l+j ch. cms-bin                                              & \cite{ATLAS:2019hxz}        &  10   &  -          &      0.5   &   -     &    0.6   &  -         \\
 587 & ATLAS 13 TeV 36 fb$^{-1}$, $t\bar{t}$ abs. $y_{t\bar t}, m_{t\bar t},y_{t\bar t}^B, H^{t\bar t}_{T}$ cross secs. l+j ch.         & \cite{ATLAS:2019hxz}        &  34   &  -          &      -     &   0.9   &    -     &  1.1       \\
 581 & CMS 13 TeV 137 fb$^{-1}$, $t\bar{t}$ abs. $m_{t\bar t}$ cross sec. l+j ch.                                                       & \cite{CMS:2021vhb}          &  15   &  -          &      0.9   &   1.0   &    1.3   &  1.5       \\
\hline
\hline
     &                                                        CT18mJet$\&$Top                                                            &                             &      &   CT18mJet$\&$Top           &           &      &     &         \\
\hline 
 110 & CCFR $F_{2}^{p}$                                                                                                                  & \cite{CCFRNuTeV:2000qwc}    &  69   &  1.2  &     1.2&                 1.2&            1.2&          1.2    \\
 521 & ATLAS 13 TeV 36.1 fb$^{-1}$, $t\bar{t}$ abs. $y_{t\bar t}$ cross sec. all-hadronic                                               & \cite{ATLAS:2020ccu}        &  12   &  -          &      1.0   &   1.0   &    1.0   &  1.0         \\
 528 & CMS 13 TeV 35.9 fb$^{-1}$, $t\bar{t}$ abs. $y_{t\bar t}$ cross sec. dilepton ch.                                                 & \cite{CMS:2018adi}          &  10   &  -          &      0.5   &   0.7   &    0.3   &  0.5         \\
 532 & ATLAS 13 TeV 36 fb$^{-1}$, $t\bar{t}$ abs. $y_{t\bar t}$ cross sec. l+j ch. cms-bin                                              & \cite{ATLAS:2019hxz}        &  10   &  -          &      0.4   &    -    &    0.5   &  -           \\
 587 & ATLAS 13 TeV 36 fb$^{-1}$, $t\bar{t}$ abs. $y_{t\bar t}, m_{t\bar t},y_{t\bar t}^B, H^{t\bar t}_{T}$ cross secs. l+j ch.         & \cite{ATLAS:2019hxz}        &  34   &  -          &       -    &   1.0   &    -     &  1.1         \\
 581 & CMS 13 TeV 137 fb$^{-1}$, $t\bar{t}$ abs. $m_{t\bar t}$ cross sec. l+j ch.                                                       & \cite{CMS:2021vhb}          &  15   &  -          &      0.8   &   1.0   &    1.2   &  1.4         \\
\hline

\end{tabular}
\caption{Same as in Tab.~\ref{global-fit-opt}, but for the CT18mTop, CT18mJet, and CT18mJet$\&$Top8 global fits.}
\label{global-fit-opt2}
\end{table}
\endgroup
\end{widetext}

\begin{figure} 
\includegraphics[width=0.49\textwidth]{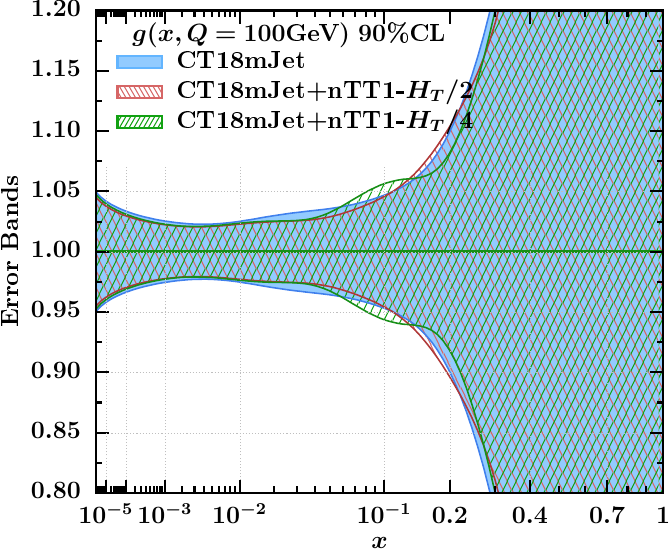}
\includegraphics[width=0.49\textwidth]{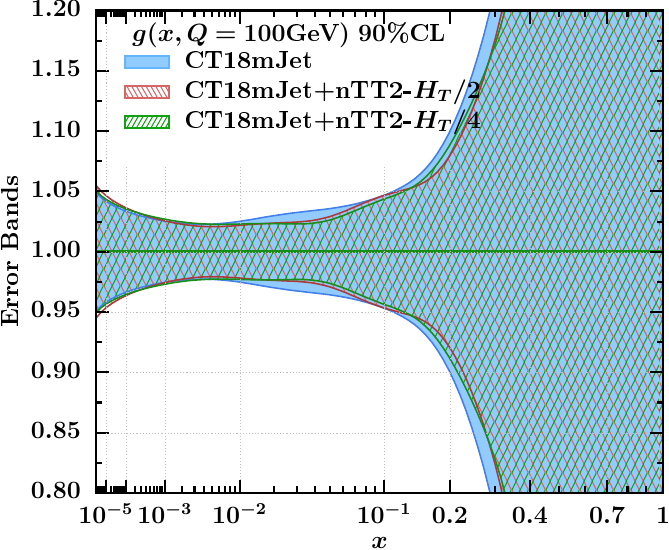}
\caption{Interplay between 13 TeV and 8 TeV top-quark data in a fit without QCD jets. Left: CT18 without QCD jets, and with nTT1. Right: CT18 without QCD jets, and with nTT2. }
\label{CT18mjets-vs-ntt1-ntt2}
\end{figure}

\textbf{Differences between the nTT1 and nTT2 data subsets.}
\label{nTT1-nTT2-diff}
To conclude this section, we wish to discuss differences between the nTT1 and nTT2 data subsets and their interplay with the 8 TeV top-quark data in the CT18 baseline. To facilitate this analysis and better emphasize the differences, we study the impact of the nTT1 and nTT2 data subsets on the gluon PDF uncertainty in the CT18mJet global fit, where inclusive jet data are removed. The results of this study are illustrated in Fig.~\ref{CT18mjets-vs-ntt1-ntt2}. 

In Fig.~\ref{CT18mjets-vs-ntt1-ntt2}(left) we observe an increase in the gluon uncertainty in the $0.06\lesssim x \lesssim 0.15$ region when the central scale in the nTT1 theory predictions is set to $H_T/4$. The same increase is not present for the nTT2 combination in Fig.~\ref{CT18mjets-vs-ntt1-ntt2}(right). This is one of the main differences which emerges in the nTT1 combination containing the $y_{t\bar t}$ distribution from the ATL13lj measurements resolved in terms of CMS bins that uses $H_T/4$ as the central scale in the nTT1 theory predictions. This tension disappears when the $H_T/2$ central scale choice is used. 

By looking at the $\chi^2/N_{pt}$ values in Tab.~\ref{global-fit-opt}, we argue that this mild tension is generated in part by the CMS 8 TeV 19.7 fb$^{-1}$ single inclusive jet cross section, whose $\chi^2/N_{pt}$ increases from 1.1 to 1.2 (regardless of scale choice); in part by the ATLAS 8 TeV 20.3 fb$^{-1}$ top-quark $p_{T,t}$ and $m_{t\bar t}$ absolute distributions, with a $\chi^2/N_{pt}$ increase from 0.6 to 0.7 (regardless of scale choice), and finally, by the ATL13lj data in nTT2 with a $\chi^2/N_{pt}$ increase from 0.7 to 1.1 when the central scale goes from $H_T/2$ to $H_T/4$, and by the CMS13lj data in nTT1 and nTT2 with a $\chi^2/N_{pt}$ increase from 1.1 to 1.6/1.7 when the central scale is reduced to $H_T/4$.  
As already pointed out in Sec.~\ref{subsec: Impact from CMS13lj},
this indicates a preference for the $H_T/2$ scale choice in the 13 TeV $t\bar t$ theory predictions in contrast to the suggested $H_T/4$ scale choice discussed in ref.~\cite{Czakon:2016dgf}. 
In our global fits, the optimal central scale for the 13 TeV $t\bar t$ theory predictions is chosen according to the improvements it produces in the quality-of-fit.

\begin{figure} 
\includegraphics[width=0.49\textwidth]{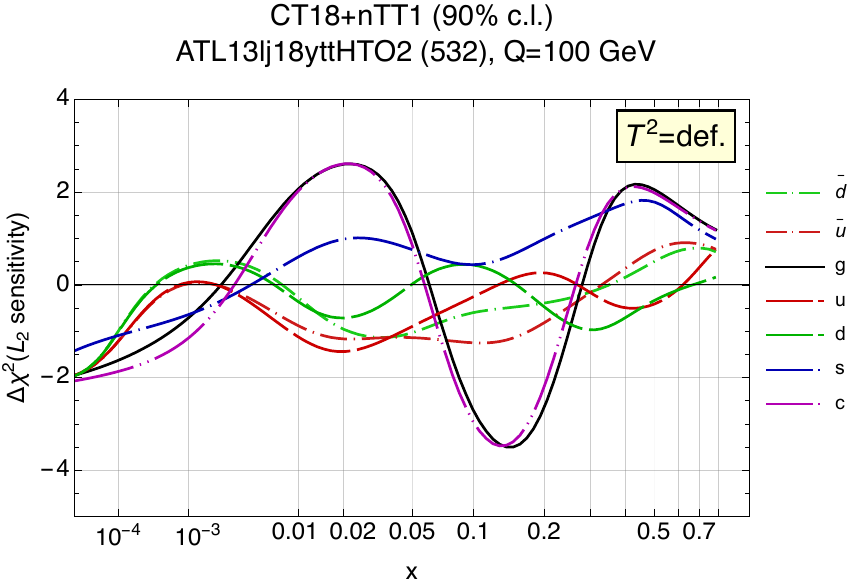}
\includegraphics[width=0.49\textwidth]{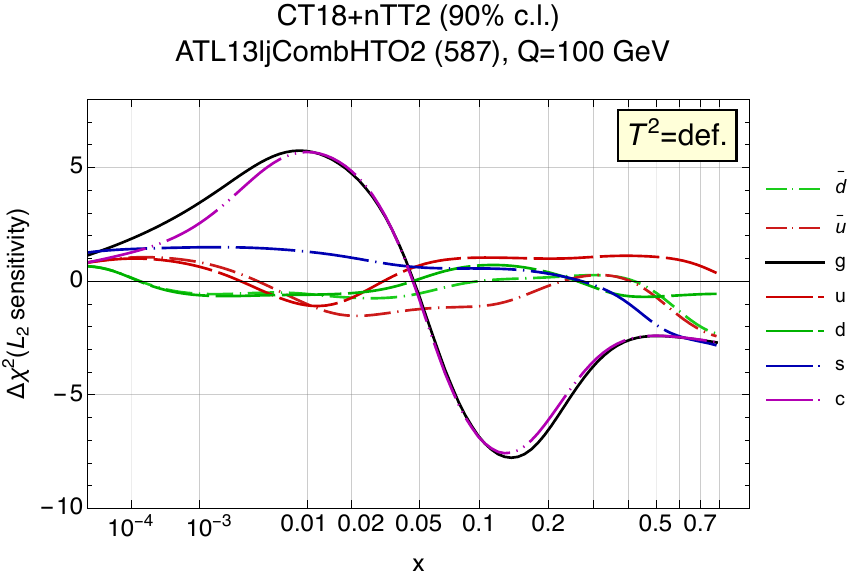}
\caption{$L_2$ sensitivity of the various PDFs for the ATL13lj measurements in the CT18+nTT1 (left) and CT18+nTT2 (right) global fits. Here, the central scale in the 13 TeV $t\bar t$ theory predictions is set to $H_T/2$.}
\label{L2-ntt1-vs-ntt2}
\end{figure}

It is also interesting to look at the $L_2$ sensitivity plots for the ATL13lj measurements in the full CT18+nTT1 and CT18+nTT2 gloabl fits which we illustrate in Fig.~\ref{L2-ntt1-vs-ntt2}. There, in the left panel we show the ALT13lj $y_{t\bar t}$ distribution resolved in terms of CMS bins. In the right panel, we show the $y_{t\bar t}$ resolved with ATLAS bins, combined with the $y^B_{t\bar t}$, $H^{t\bar t}_{T}$, and $m_{t\bar t}$ distributions without bin-by-bin statistical correlations. As already argued in previous discussions, the ATL13lj bin-by-bin statistical correlations have negligible impact and can essentially be ignored. 
We note that the two different treatments of the ATL13lj measurements lead to different preferences for the gluon and strange PDFs in the $0.2\lesssim x \lesssim 0.7$ range: the ATL13lj $y_{t\bar t}$ distribution resolved with CMS bins prefers softer gluon and softer strange PDFs, while the ATL13lj combination has opposite behavior.

{\bf \texttt{ePump} Optimization.} To further investigate differences in the CT18+nTT1 and CT18+nTT2 combinations, it is interesting to identify eigenvector (EV) directions that have maximal PDF sensitivity. The \texttt{ePump} framework allows us to identify a reduced set of error PDFs that contain the majority of the PDF dependence of the observables under consideration. 
The procedure is entirely based on the Hessian method and is documented in ref.~\cite{Schmidt:2018hvu}.
The new eigenvectors contain exactly the same information as the original eigenvectors, but are
optimized so that a smaller set of error PDFs can be chosen for use with the set of observables to any required PDF-sensitivity. In turn, this allows us to assess and validate the data that place the strongest constraints on PDF errors.

\begin{figure} 
\includegraphics[width=0.49\textwidth]{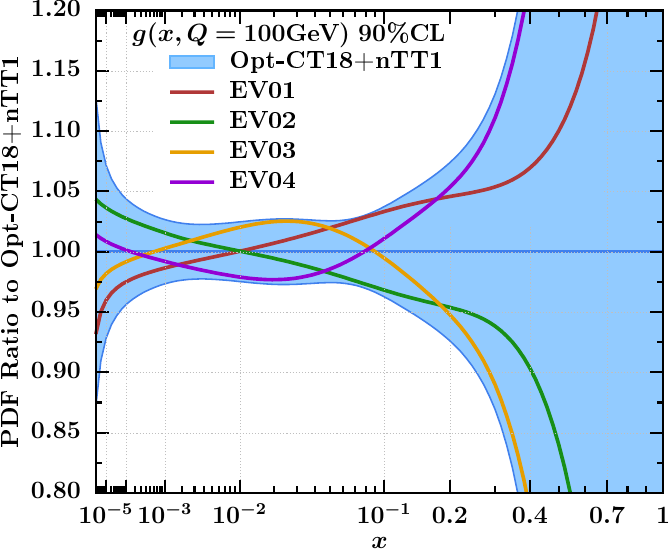}
\includegraphics[width=0.49\textwidth]{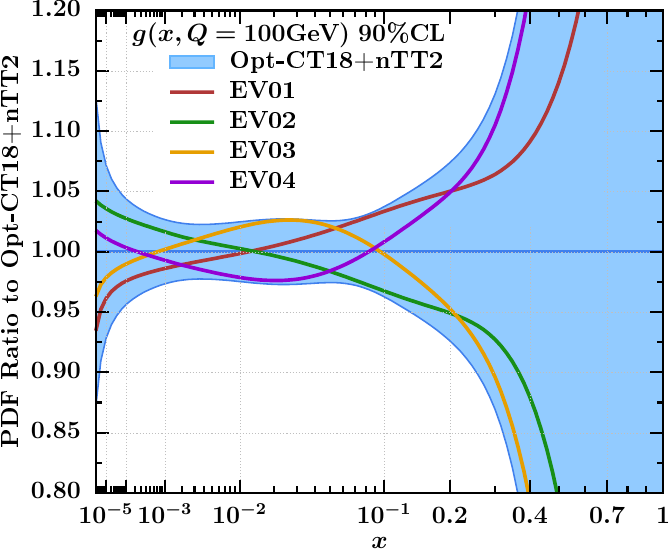}
\caption{Optimized eigenvector directions with \texttt{ePump} for the CT18+nTT1(left) and CT18+nTT2(right) combinations.}
\label{ePump-Optmization}
\end{figure}

In Fig.~\ref{ePump-Optmization}, we illustrate the behavior of the optimized eigenvector directions for the CT18+nTT1 and CT18+nTT2 combinations, while fractional contributions to the PDF error from the leading eigenvectors for each individual data set are reported in Tabs.~\ref{opt-epump-CT18-nTT1} and~\ref{opt-epump-CT18-nTT2}. 
With \texttt{ePump}, we find six optimized PDF error sets for both CT18+nTT1 and CT18+nTT2.  

In the case of the CT18+nTT1 optimized directions, we observe that the three leading eigenvectors approximately account for 99\% of the PDF error band. Among them, the largest contribution is from CMS13lj $m_{t\bar t}$, with fractional uncertainty of approximately 31\%. The second largest contribution is from ATL13had $y_{t\bar t}$, with roughly 25\%, and the smallest contributions are from CMS13ll $y_{t\bar t}$ and ATL13lj $y_{t\bar t}$ resolved in terms of CMS bins, both with fractional uncertainty of approximately 22\%.

In the case of CT18+nTT2, the data with the largest contribution is the ATL13lj combination of $m_{t\bar t}$ $y_{t\bar t}$, $y^B_{t\bar t}$, and $H_T^{t\bar t}$, which account for approximately 47\% of fractional uncertainty. This is due to the larger number of data points. 
The second largest contribution is from CMS13lj $m_{t\bar t}$ with approximately 21\%, and ATL13had $y_{t\bar t}$ with 16\%. The smallest contribution is from CMS13ll $y_{t\bar t}$, with fractional uncertainty of 13\%.

\begin{table}
\begin{tabular}{c|c|c|c|c|c}
\hline\hline
\multicolumn{6}{c}{CT18+nTT1  (nTT1-$N_{\rm pt}=47$)}\\
\hline
 Leading EV No. & Leading EV (\%) & ATL13had & CMS13ll & CMS13lj & ATL13lj   \tabularnewline\hline
  1 & 65.80& 16.96 & 13.42 & 22.00 &13.42\tabularnewline\hline
  2 & 31.64& 7.96  &  7.24 & 9.20 &7.24 \tabularnewline\hline
  3 & 1.69 & 0.49  &  0.47 & 0.27 &0.47\tabularnewline \hline 
  4 & 0.79 & 0.10  &  0.14 & 0.42 & 0.14 \tabularnewline\hline
  5  & 0.07 & 0.02  &  0.01 & 0.02 & 0.01  \tabularnewline 
\hline
\hline
\end{tabular}
\caption{Fractional contribution of leading eigenvectors (EV) from \texttt{ePump} optimization for each data set in the CT18+nTT1 combination. The second column shows the sum of fractional contributions from individual data sets.}
\label{opt-epump-CT18-nTT1}
\end{table}

\begin{table}
\begin{tabular}{c|c|c|c|c|c}
\hline\hline
\multicolumn{6}{c}{CT18+nTT2 (nTT2-$N_{\rm pt}=71$)} \\ 
 \hline
 Leading EV No. & Leading EV (\%) & ATL13had & CMS13ll & CMS13lj & ATL13lj   \tabularnewline\hline
  1 & 68.43& 10.53 &  8.64& 15.53 &33.73\tabularnewline\hline
  2 & 28.92&  5.91 &  4.99&  5.24 &12.77\tabularnewline\hline
  3 &  1.52&  0.36 &  0.33&  0.12 & 0.71\tabularnewline\hline 
  4 &  1.00&  0.08 &  0.12&  0.22 & 0.58\tabularnewline\hline
  5  & 0.11 & 0.02  & 0.01&  0.01 & 0.07\tabularnewline 
\hline
\hline
\end{tabular}
\caption{Same as in Tab.~\ref{opt-epump-CT18-nTT2}, but for the CT18+nTT2 combination.}
\label{opt-epump-CT18-nTT2}
\end{table}

These differences in the treatment of the ATL13lj measurements prompted us to identify the two optimal combinations CT18+nTT1 and CT18+nTT2, but overall they have mild impact in the global fits we have analyzed.  

\section{Conclusions}
\label{sec:conclusions}
We presented a comprehensive study of the impact of recent high-precision LHC top-quark pair production measurements at $\sqrt{S}=13$ TeV of collision energy on PDFs of the proton, in particular the gluon, in global analyses at NNLO in QCD.
This extensive analysis of post-CT18 PDFs is relevant for the next release of CTEQ-TEA PDFs which will be challenged by the inclusion of high-precision forthcoming measurements at the LHC for a multitude of standard candle processes like top-quark pair production.   

Besides the PDF impact of the 13 TeV $t\bar t$ differential cross section measurements from ATLAS and CMS, we studied their interplay with inclusive jet production measurements in global PDF fits. 

Due to differences in the binning resolution of the $t\bar t$ 13 TeV ATLAS lepton+jet data, we identified two optimal combinations of measurements that maximize the information to constrain the gluon, and minimize conflict with the other data sets in the extended baseline.

Overall, the impact of these measurements in reducing the uncertainty in the gluon PDF is found to be mild. However, their role is important as they constrain the behavior of the gluon PDF at large momentum fraction $x$ in a way that complements that of inclusive jet production data.
In fact, the $t\bar{t}$ and inclusive jet production processes overlap in the $Q$--$x$ plane, but their matrix elements and phase-space suppression are different, and constraints on the gluon and other PDFs are placed at different values of $x$.

We analyzed the impact of bin-by-bin statistical correlations whenever possible, as well as that of central-scale variations in the theory prediction for the 13 TeV top-quark data. The criterion according to which a particular scale is chosen, is based on the improvements produced in the quality-of-fit description ($\chi^2$ description) in the global analysis. We observed that the $\mu_F=\mu_R=H_T/2$ choice of the central scale improves the description of the 13 TeV lepton+jet data at CMS with 137 fb$^{-1}$ of integrated luminosity. These are the most precise data included in this work and place stronger constrains on PDFs as compared to the other measurements. Future analyses and extensions of this work will include measurements from ATLAS with similar or higher integrated luminosity as well as implications due to novel PDF parametrizations.    

The global analyses performed in this work have been challenged 
by the interpretation of the correlated systematic uncertainties published by the ATLAS and CMS collaborations. The default treatment of correlated systematic errors in the CTEQ-framework is in terms of nuisance parameters. The top-quark pair production measurements from the CMS collaborations have been recently published in terms of the covariance matrix representation. We performed conversion between the covariance matrix and nuisance parameter representation using a similar strategy to that used in the CT18 study. This allowed us to obtain identical $\chi^2$ values in both representations. 
However, this conversion is not unique.
Detailed information on both the covariance and nuisance parameter representations for experimental errors is critical to fully exploit constraints from the data in global QCD analyses for PDF determinations, and is critical to perform a simultaneous determination of $m_t$, $\alpha_s$, and the PDFs as well as their correlations in future analyses.

\acknowledgments
We are indebted to Pavel Nadolsky for his critical reading of the manuscript, numerous discussions, and useful suggestions. 
We thank Otto Hindrichs and Maria Aldaya from the CMS collaboration for correspondence and discussions about the CMS $t\bar t$ differential cross section measurements at 13 TeV with 137 fb$^{-1}$ of integrated luminosity.
We thank Mandy Cooper-Sarkar from the ATLAS collaboration for discussions about the ATLAS lepton+jets measurements and their bin-by-bin statistical correlations.
We thank Klaus Rabbertz for discussions and for suggestions on future analyses.   
We thank Joey Huston, and other CTEQ-TEA members for discussions. 
The work of AA, SD  and IS are supported by the National Natural Science Foundation of China under Grant No.11965020 and Grant No. 11847160, respectively.
The work of MG is supported by the National Science Foundation under Grant No. PHY2112025. MG would like to thank the Erwin Schr\"odinger International Institute for Mathematics and Physics (ESI) in Vienna, for hospitality, discussions, and partial support during his stay at the ESI-QFT 2023 workshop.
The work of KX was supported by the U.S. Department of Energy under grant No. DE-SC0007914, the U.S. National Science Foundation under Grants No. PHY-2112829 and also in part by the Pittsburgh Particle Physics Astrophysics and Cosmology Center (PITT PACC). 
The work of KX was also performed in part at the Aspen Center for Physics, which is supported by National Science Foundation grant PHY-2210452. 
The work of TH is supported by the Natural Science Foundation of Hunan province of China under Grant No.2023JJ30496. 
CPY was supported by the U.S. National Science Foundation under Grant
No.~PHY-2013791 and~PHY-2310291. CPY is also grateful for the support from the Wu-Ki Tung endowed chair in particle physics. 
This work was supported by high-performance computing (HPC) resources at KSU (KSU HPC), SMU (SMU M2/M3), MSU (MSU HPCC), and the Pittsburgh Center For Research Computing (PITT CRC).


\appendix

\section{Details of the theoretical calculations} 
\label{app:theory-comp}
In this section, we provide more details about the calculation of the theory predictions used in this work.  
We point out again that global QCD analyses to determine PDFs of the proton necessitate fast, precise, and accurate theory predictions to be confronted to the experimental data in the minimization procedure to reduce the CPU turn-around time. 
For this reason, frameworks like \texttt{fastNLO} and \texttt{APPLGrid} for the generation of fast tables are indispensable tools.

\subsection{NNLO QCD prediction benchmarks}
\label{app:QCDNNLO}
As discussed in Sec.~\ref{sec:theory}, two independent theory calculations for the top-quark pair production differential cross section at NNLO in QCD have been used in this work. One is obtained with fast tables~\cite{Czakon:2017dip,repo} based on the \texttt{STRIPPER}~\cite{Czakon:2011ve,Czakon:2010td} subtraction method, that are produced at NNLO with \texttt{fastNLO}. The other one is obtained by using NNLO/NLO $K$-factors, where fast tables for the NLO cross section in the denominator are generated with \texttt{APPLgrid}~\cite{Carli:2010rw} using the \texttt{MCFM}~\cite{Campbell:2015qma,Campbell:2012uf} program, while the NNLO cross section in the numerator is computed with the \texttt{MATRIX} program~\cite{Catani:2019hip,Catani:2019iny,Catani:2020tko}, based on the $q_T$-subtraction method~\cite{Catani:2007vq}.
Currently, there are no fast tables available for \texttt{MATRIX}.

\begin{figure}
\includegraphics[width=0.49\textwidth]{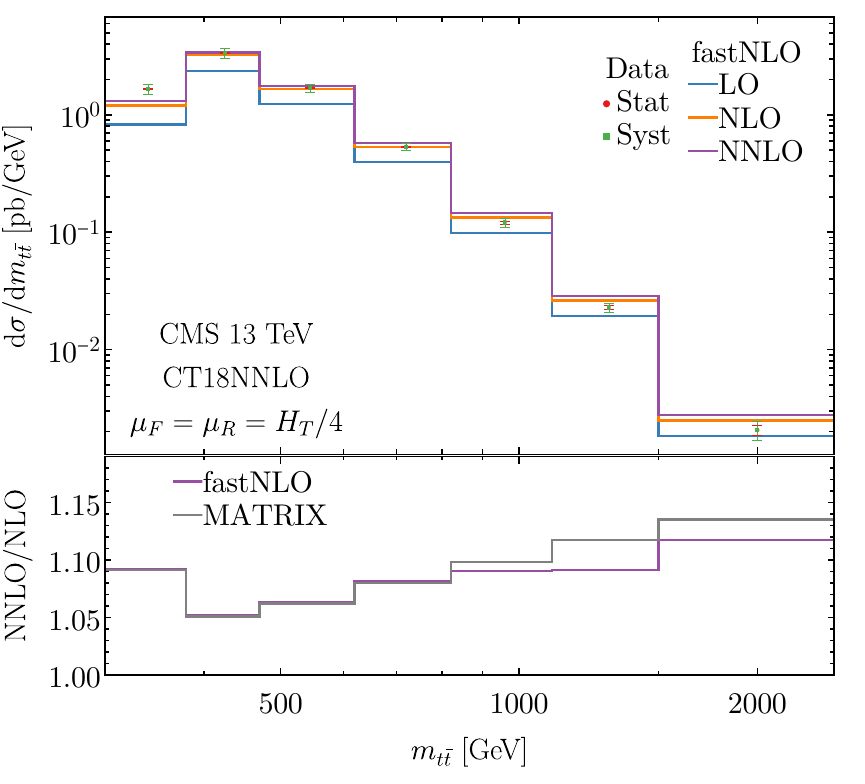}
\includegraphics[width=0.49\textwidth]{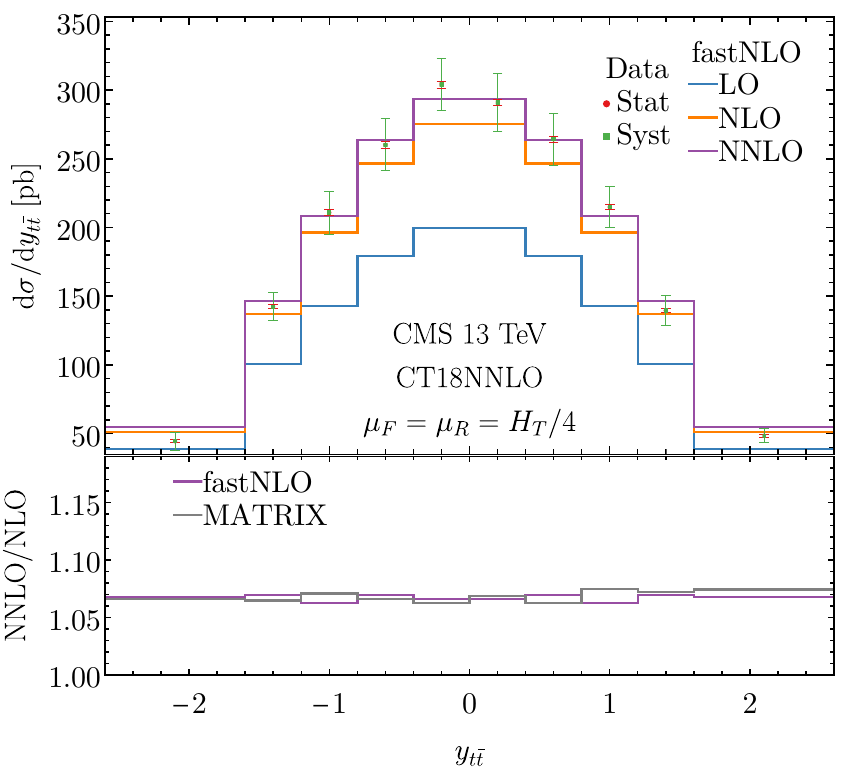}
\includegraphics[width=0.49\textwidth]{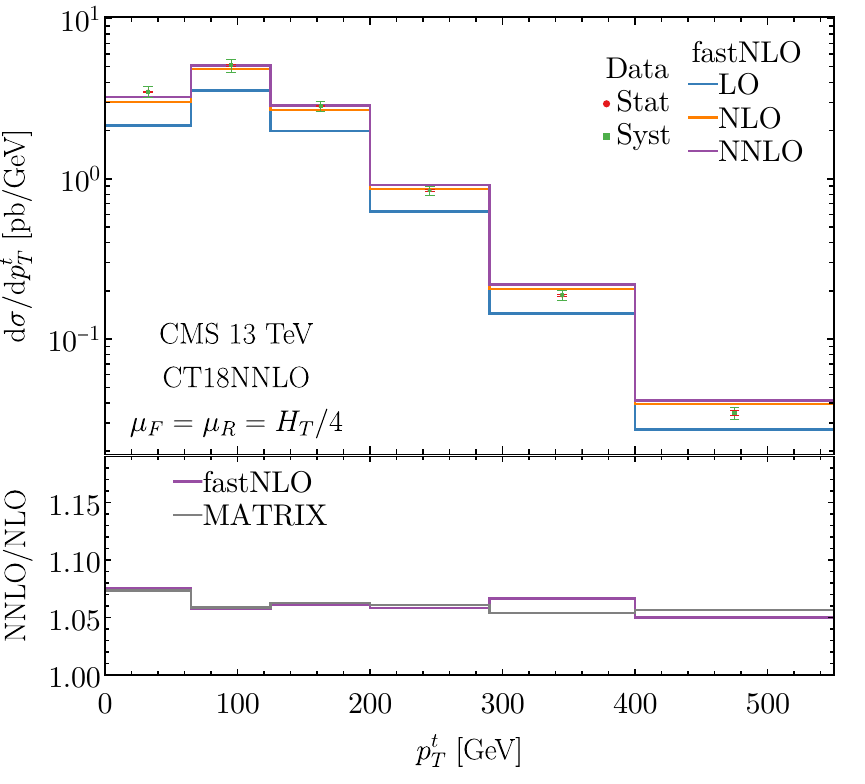}
\includegraphics[width=0.49\textwidth]{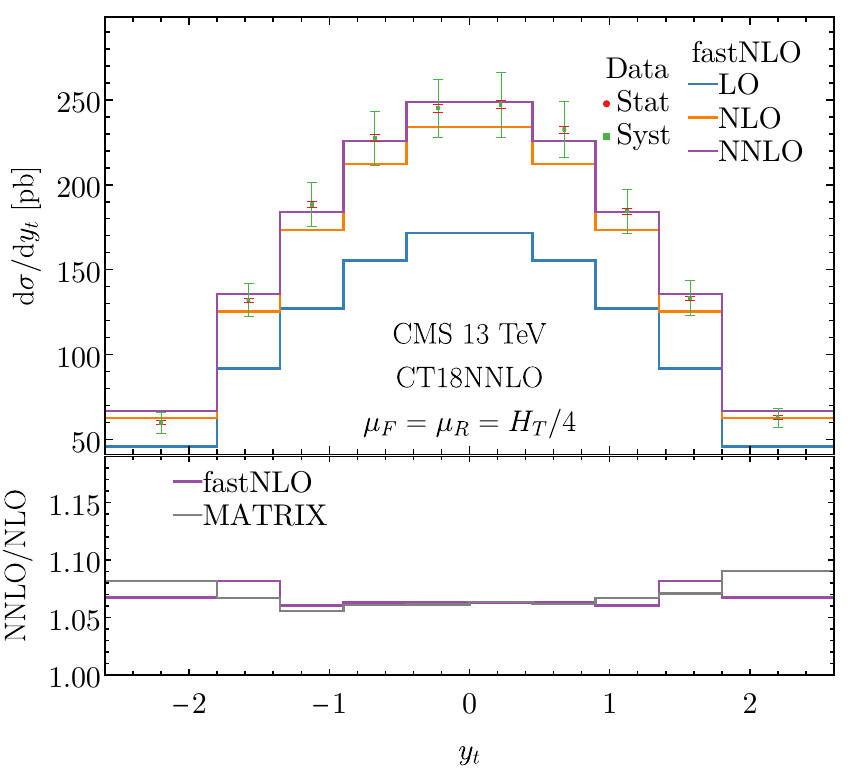}
\caption{Theoretical predictions for the $m_{t\bar{t}}$, $y_{t\bar{t}}$, $y_{t}$, and $p_{T,t}$ distributions compared to the 13 TeV CMS measurements in the dilepton channel~\cite{CMS:2018adi}. 
The CT18 NNLO PDFs~\cite{Hou:2019efy}, scale choice $\mu_{F}=\mu_{R}=H_T/4$, and pole mass $m_t^{\rm(pole)}=172.5$ GeV are selected here as default parameters.}
\label{fig:CMS13NNLO}
\end{figure}

\begin{figure}[htbp]
\includegraphics[width=0.49\textwidth]{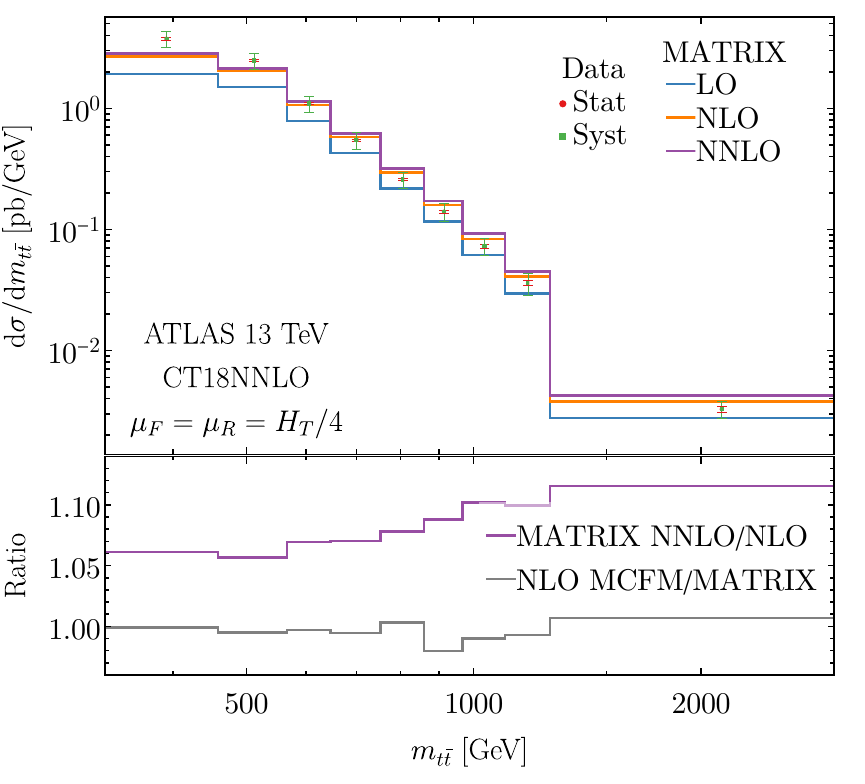}
\includegraphics[width=0.49\textwidth]{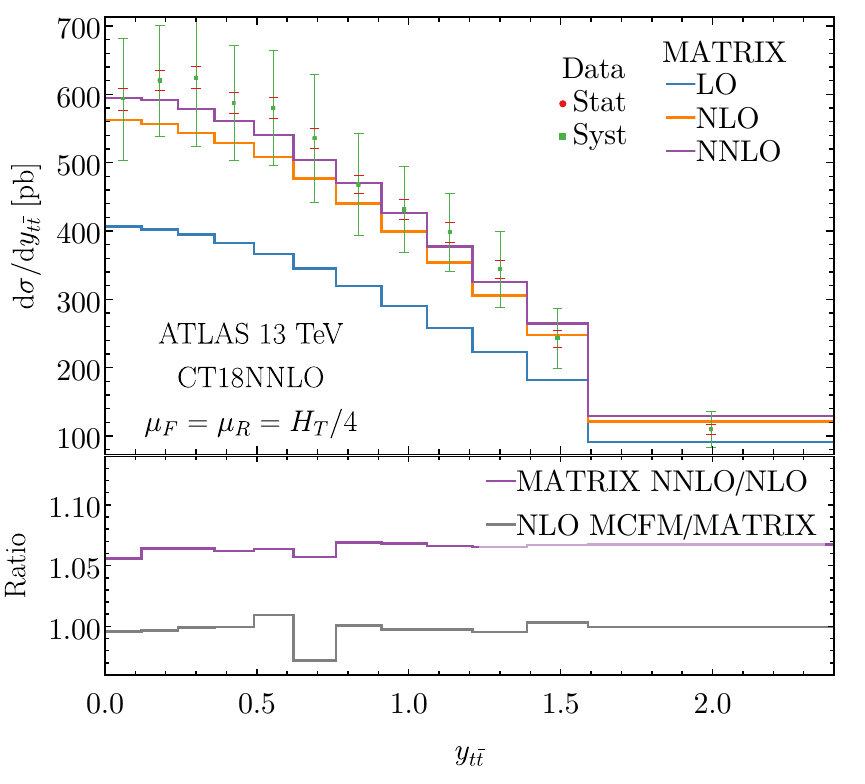}
\includegraphics[width=0.49\textwidth]{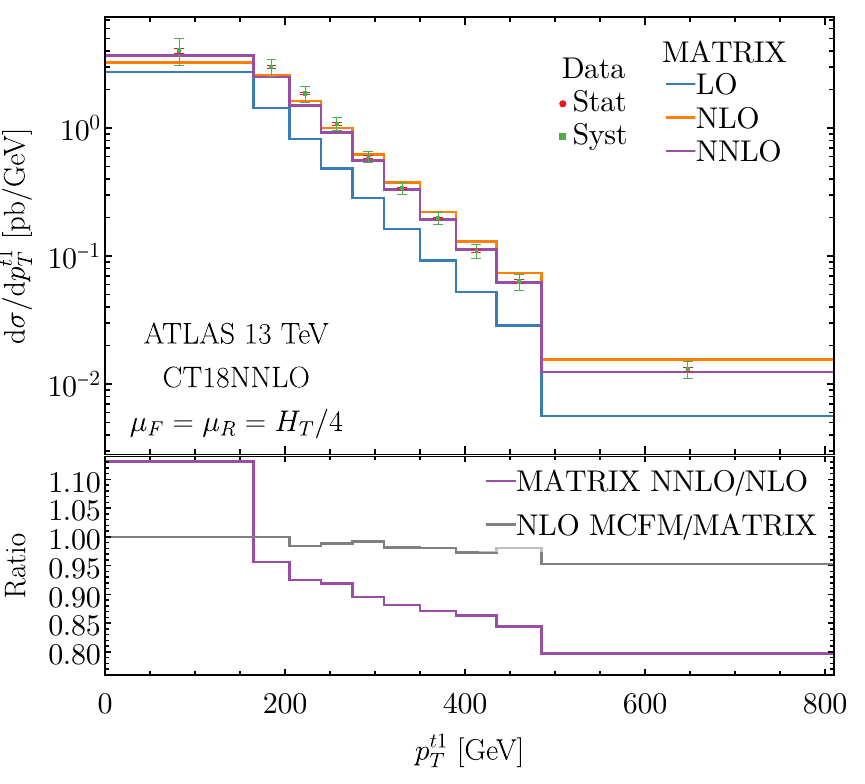}
\includegraphics[width=0.49\textwidth]{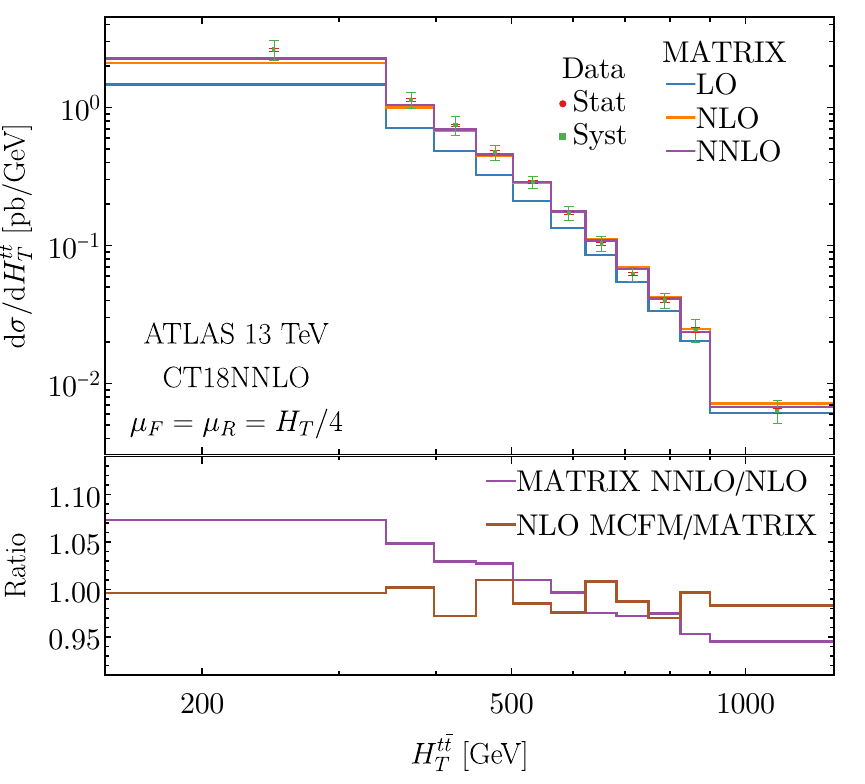}
\caption{Theoretical predictions for the $m_{t\bar{t}}$, $y_{t\bar{t}}$, $p_{T,t_1}$, and $H_T^{t\bar{t}}$ distributions compared to the 13 TeV ATLAS measurements in the all-hadronic channel~\cite{ATLAS:2020ccu}. Here, default parameters are used as in Fig.~\ref{fig:CMS13NNLO}.}
\label{fig:ATL13NNLO}
\end{figure}

Comparisons between these two calculations are shown in Figs.~\ref{fig:CMS13NNLO} and Fig.~\ref{fig:ATL13NNLO}, where the theory is compared to the $m_{t\bar{t}}$, $y_{t\bar{t}}$, $y_{t}$, and $p_{T,t}$ distributions measured at CMS at 13 TeV in the dilepton channel~\cite{CMS:2018adi}, and to the $m_{t\bar{t}}$, $y_{t\bar {t}}$, $p_{T,t_1}$, and $H_T^{t\bar{t}}$ distributions measured at ATLAS at 13 TeV in the all-hadronic channel~\cite{ATLAS:2020ccu}, respectively. 

For this case study, we use CT18 NNLO PDFs~\cite{Hou:2019efy}, the central scale is set to $\mu_{F}=\mu_{R}=H_T/4$, and the top-quark mass is $m_t^{\rm(pole)}=172.5$ GeV. These are our default parameters. 
In general, we find agreement between the two calculations within $1\%$ accuracy, which is sufficient for all of the analyses in this work.

In the upper insets of each panel of Fig.~\ref{fig:CMS13NNLO}, 
theory predictions obtained with \texttt{fastNLO} at LO, NLO, and NNLO perturbative orders for the absolute differential distributions are compared to the CMS measurements. Statistical and systematical uncertainties are shown separately using error bars with different colors. 
In the lower insets, we show NNLO/NLO $K$-factors for the two calculations. 
We note that for the $y_{t}$, $y_{t\bar{t}}$ and $p_{T,t}$ distributions the overall NNLO correction is about 7\%, except for the $m_{t\bar{t}}$ where the $K$-factor has larger variation ranging from 5\% to 12\%. We observe agreement between \texttt{MATRIX} and \texttt{fastNLO} at the percent level for all distributions, though they agree better in the $y_{t\bar t}$ than in $m_{t\bar t}$ distributions.  

In the upper insets of each panel in Fig.~\ref{fig:ATL13NNLO}, the theory predictions that are compared to the ATLAS measurements are obtained with \texttt{MATRIX} using default parameters. Statistical and systematical uncertainties are displayed as in Fig.~\ref{fig:CMS13NNLO}. 
In the lower insets, we show the \texttt{MATRIX} NNLO/NLO $K$-factor as well as the ratio between the \texttt{AppGrid} theory prediction from \texttt{MCFM} and \texttt{MATRIX} both computed at NLO. These NLO calculations agree within $1\%$ accuracy. 
In addition, we note that the leading transverse momentum $p_{T,t_1}$ and $H_T^{t\bar{t}}$ distributions are affected by large QCD perturbative corrections as their $K$-factors produce large variations.

\subsection{NLO electroweak corrections \label{app:EWcorr}}

We explore the impact of the electroweak (EW) corrections on the $t\bar t$ differential distributions at 13 TeV and discuss their impact in our global PDF analyses.

EW corrections are computed as $K$-factors using the multiplicative scheme as described in Ref.~\cite{Czakon:2017wor}, which are available in the repository~\cite{repo}. 
These EW corrections include contributions of order ${\cal O}(\alpha_s^2\alpha)$ as well as subleading ones, of order ${\cal O}(\alpha_s\alpha^2)$ and ${\cal O}(\alpha^3)$. Meanwhile, EW corrections are also incorporated in \texttt{MadGraph5\_aMC@NLO}~\cite{Alwall:2014hca,Pagani:2016caq} which performs differential cross section calculations in an automated fashion, up to NLO in both couplings. Recently, \texttt{MadGraph5\_aMC@NLO} has been interfaced with the \texttt{PineAPPL} library~\cite{Carrazza:2020gss} to obtain fast interpolation grids which include EW corrections up to ${\cal O}(\alpha_s^2\alpha)$. In addition, EW corrections for top-quark pair production at hadron colliders are also included in \texttt{MCFM}~\cite{Campbell:2016dks}.

\begin{figure}
\centering
\includegraphics[width=0.49\textwidth]{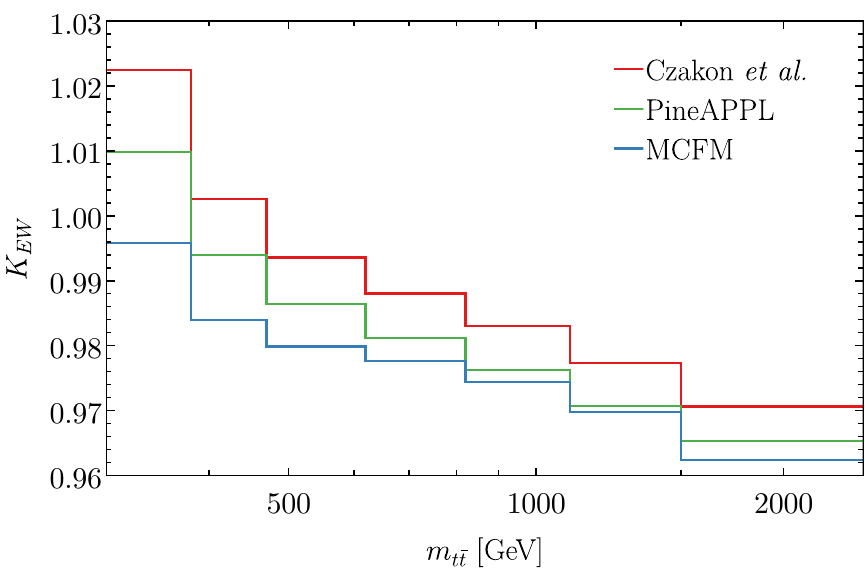}
\includegraphics[width=0.49\textwidth]{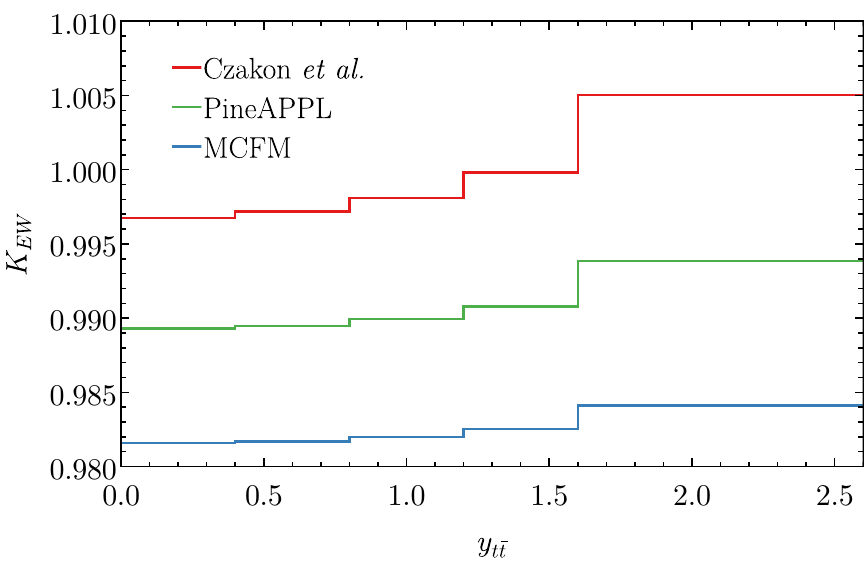}
\includegraphics[width=0.49\textwidth]{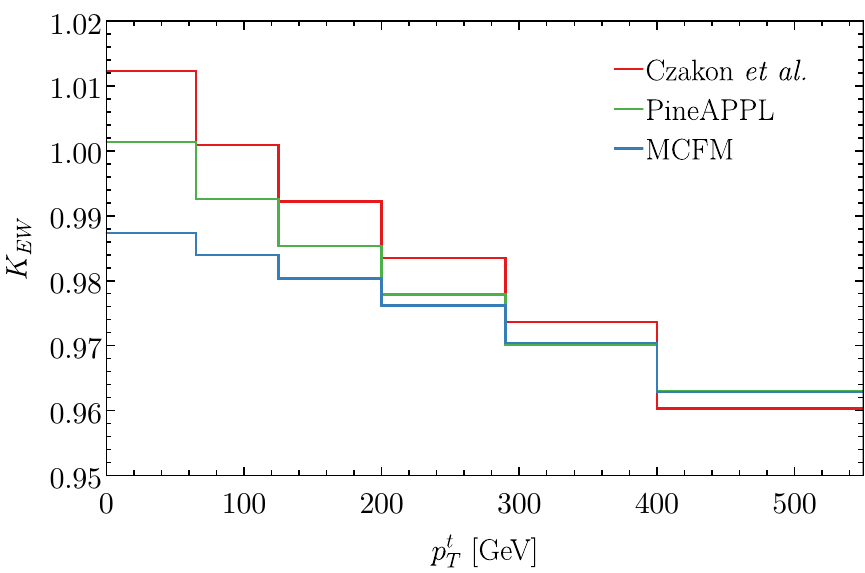}
\includegraphics[width=0.49\textwidth]{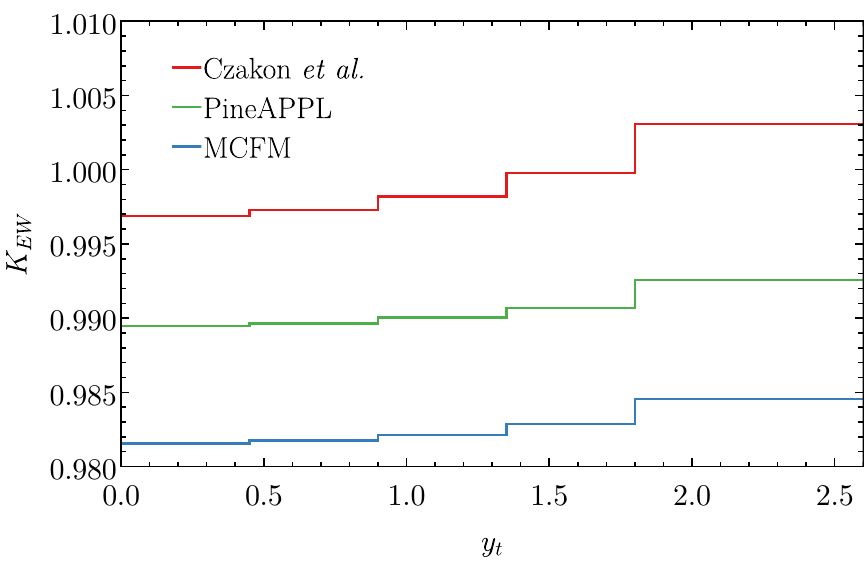}
\caption{Comparison of the EW corrections from the Czakon \emph{et al.}~\cite{Czakon:2017wor}, \texttt{PineAPPL}\cite{Carrazza:2020gss}, and \texttt{MCFM}~\cite{Campbell:2016dks}
for the $t\bar{t}$ production at the LHC 13 TeV measured by the ATLAS collaboration in the lepton+jets channel~\cite{ATLAS:2019hxz} and CMS in the dilepton one~\cite{CMS:2018adi}.}
\label{fig:EW}
\end{figure}

In Fig.~\ref{fig:EW}, we illustrate EW corrections from Czakon \emph{et~al.}~\cite{Czakon:2017wor,repo}, \texttt{PineAPPL}~\cite{Pagani:2016caq,Carrazza:2020gss}, and \texttt{MCFM}~\cite{Campbell:2016dks} which are defined as multiplicative $K$-factors $K_{EW}=\textrm{QCD}\times\textrm{EW/QCD}$. 
The 13 TeV $t\bar t$ theory predictions for the differential distributions are calculated using the CT18NNLO PDFs, and use the same bins shared by the ATLAS distributions in the lepton+jets~\cite{ATLAS:2019hxz} channel and the CMS distributions in the dilepton channel~\cite{CMS:2018adi}. 

While the two calculations produce almost identical shapes in the various distributions, differences within 1\% or smaller are found, 
mainly driven by the subleading EW corrections.
In addition, we observe that the rapidity distributions $y_t$ and $y_{t\bar t}$ appear to be more stable against EW corrections as compared to $m_{t\bar t}$ and $p_{T,t}$, with most of the impact affecting bins at large rapidity. 

EW corrections calculated using multiplicative EW $K$-factors are expolored in our global PDF analyses. 
We essentially observe no sizable impact on either the central-value PDFs or their uncertainties. The impact of EW corrections in our PDF analyses is negligible given the current size of the experimental errors and other theoretical uncertainties affecting the calculation in the global fit.

\section{CMSlj13 $m_{t\bar t}$ and $y_{t\bar t}$ with CT18+nTT1 and CT18+nTT2 PDFs.}
\label{CMS13lj-postfit}

\begin{figure}
\includegraphics[width=0.49\linewidth]{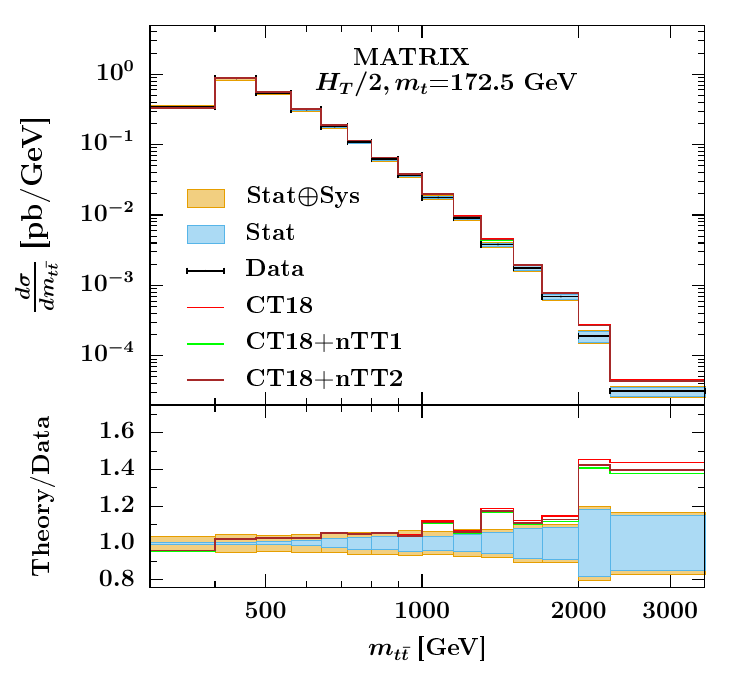}
\includegraphics[width=0.49\linewidth]{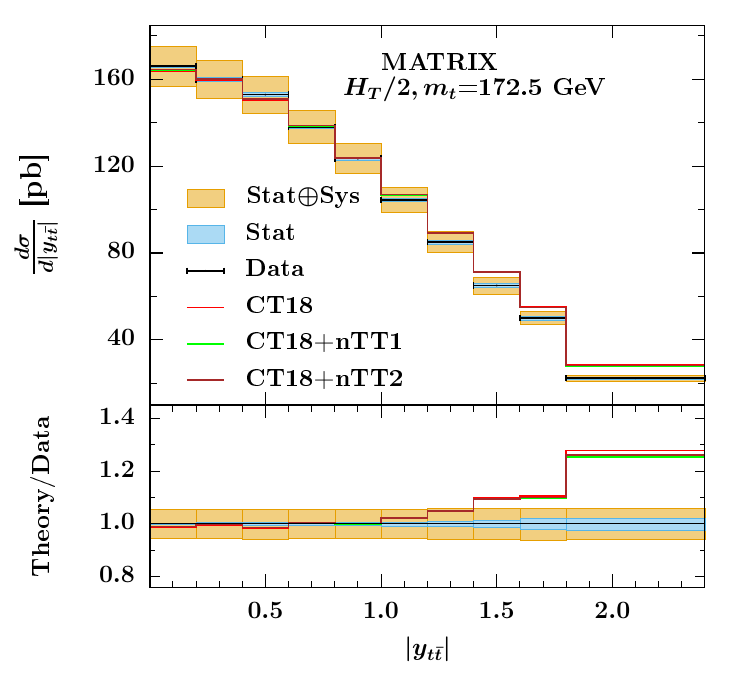}
\caption{Theory predictions for the $m_{t\bar{t}}$(left) and $|y_{t\bar t}|$(right) distributions obtained with CT18 PDF, and with CT18+nTT1 and CT18+nTT2 PDFs after the global fit. The theory is compared to the CMS13lj measurements.}
\label{CMS13lj-distr-postfit}
\end{figure}

In this section, we illustrate a data vs theory comparison for the $m_{t\bar t}$ and $y_{t\bar t}$ distributions at CMS 13 TeV in the lepton+jet channel (CMS13lj) which we show in Fig.~\ref{CMS13lj-distr-postfit}. Here, the theory predictions are obtained with the CT18,  CT18+nTT1 and CT18+nTT2 PDFs after the global fit. In both cases we observe a mild improvement when using the updated PDFs, as compared to Fig.~\ref{fig:mtt-ytt-HTO24-CMS137} As discussed in this work, cf. Tab.~\ref{global-fit-opt}, the $\chi^2/N_{pt}$ for the $m_{t\bar t}$ distribution in the CT18+nTT1 and CT18+nTT2 fits goes from  1.4 down to 1.1 for the scale chosen to be $H_T/2$. 
This is mostly due to an improvement in bin 8, 10, 12, and 13. However, the theory for both distributions continues to overestimate the data in the higher bins. As discussed in the main text, the $y_{t\bar t}$ distribution cannot produce a good fit and therefore is not included in this analysis. 

\section{\texttt{ePump} vs global fit: gluon PDF from the CT18+CMSlj13-$y_{t\bar t}$ baseline fit.}
\label{CMS13lj-ytt-gluon-postfit}

As pointed out in the original papers~\cite{Schmidt:2018hvu,Hou:2019gfw}, the \texttt{ePump} profiling method may not be reliable
when the updated central PDF deviates significantly from the baseline. In this work, we have only
used \texttt{ePump} as an auxiliary tool for the selection of the optimal combinations of the data sets to be analyzed via global fits. Therefore, in this work we do not report the full updated results of  \texttt{ePump} as compared to a true global fit. The global fit is the main result of our PDF global analysis. With that said, we have checked that the \texttt{ePump} updated results agree well with the global fit for the optimal combinations of the data sets. 

\begin{figure}
\includegraphics[width=0.49\linewidth]{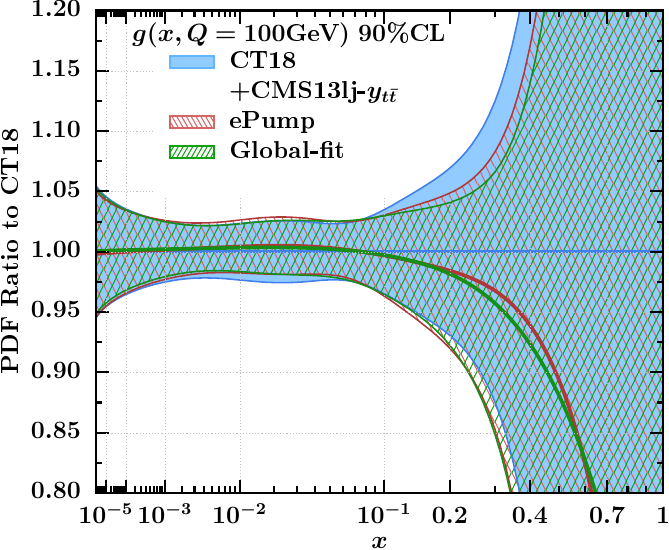}
\includegraphics[width=0.49\linewidth]{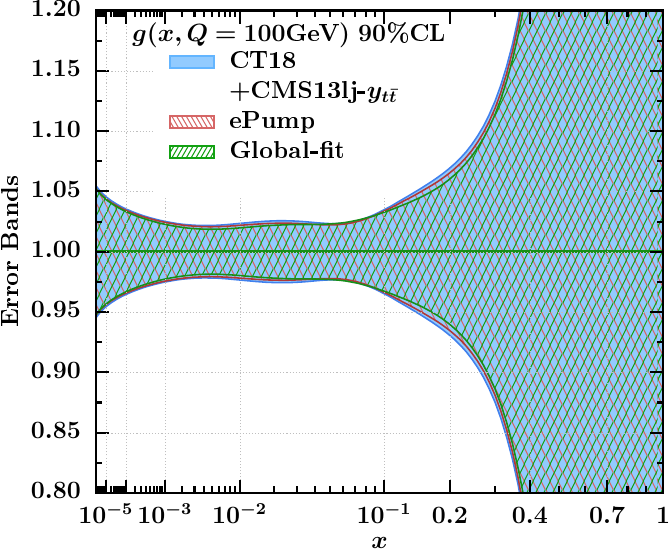}
\caption{Post-CT18 gluon PDF and its uncertainty in the CT18+CMS13lj-$y_{t\bar t}$ fit. Bands with different hatching represent the \texttt{ePump} update(red), and global fit results(green). The blue band is CT18 at 90\% CL.}
\label{CMS13lj-gluon-postfit}
\end{figure}

As shown in Table~\ref{global-fit-res},  the CT18+CMSlj13-$y_{t\bar t}$ data cannot be fit well either by \texttt{ePump}-updating or a global fit, which yields a large value of $\chi^2/N_{pt}$ for describing the CMSlj13-$y_{t\bar t}$ data.  
Hence, the CMSlj13-$y_{t\bar t}$ data set is not included in the optimal combinations nTT1 and nTT2, as introduced in the main text. Here, we update the CT18 PDFs by including this data to the original CT18 data set, as an example, to  compare the post-CT18 gluon PDF obtained from \texttt{ePump}-updating to that from a global fit.  In Fig.~\ref{CMS13lj-gluon-postfit}, we show the impact of the CMSlj13-$y_{t\bar t}$ measurements on both the central value and error band, and we also show the nominal CT18 results. The \texttt{ePump} result for the gluon central value in Fig.~\ref{CMS13lj-gluon-postfit}(left) is very close to the result from the global fit and the same is true for the error band in Fig.~\ref{CMS13lj-gluon-postfit}(right). 
 Namely, even for this data set, with its $\chi^2/N_{pt}$ value much larger than 1 the \texttt{ePump} update is still found to be in good agreement with the result from  a global fit. A similar (or better) level of agreement is found for the other data sets in the new nTT1 and nTT2 combinations.  

\section{Treatment of experimental uncertainties}
\label{exp-unc-treatment}
In global QCD analyses, the agreement between theory and experiment can be investigated by quantifying the agreement between individual data points and shifted residuals (which will be defined later) with the corresponding nuisance parameters. Under the assumption that both these quantities are distributed according to their normal distribution, which we refer to as $\mathcal{N}(0,1)$, this can be used to test the goodness-of-fit (see refs.~\cite{Kovarik:2019xvh} and references therein for more details). The experimental collaborations at the LHC often present experimental uncertainties using the covariance matrix representation (see the discussion in Sec.~\ref{sec:13TeVdata}). Therefore, to examine the normal distribution of shifted residuals a conversion to the nuisance parameters representation is required. This conversion is generally not unique, especially when the statistical and uncorrelated systematical errors are not fully specified, and when correlated systematic uncertainties and their sources are not explicitly known. 
In this case, a question arises about finding the optimal strategy to factor the covariance matrix and perform the conversion to nuisance parameters, and select the optimal number $N_{corr}$ of correlated sources that captures most of the features of the true correlated sources of uncertainty and does not introduce artificial fluctuations in the $\chi^2$ calculation (see the discussion in ref.~\cite{Kassabov:2022pps}). 

In this work, when the information on the experimental uncertainties is not fully provided with the measurements used in our analysis ({\it e.g.}, the CMS13ll, CMS13lj measurements), we express the covariance matrix in terms of nuisance parameters representation by using a version the $\Sigma+K$ decomposition, which is an iterative procedure introduced in the CT18 study~\cite{Hou:2019efy} to obtain the nuisance parameter representation from the covariance matrix.
This allows us to numerically match the $\chi^2$ in the two representations and obtain identical values. 
An extended discussion of this problem, in which we consider alternative methods of performing the conversion from the covariance matrix to the nuisance parameter representation, and study the problem of finding an optimal representation for the independent experimental errors that limits artificial fluctuations in the calculation of the $\chi^2$ function, will be addressed in a separate work. 
Independent discussions on the treatment of the experimental uncertainty correlations in global PDF analyses can also be found in refs.~\cite{Kovarik:2019xvh,Kassabov:2022pps}.

\bibliographystyle{utphys}

\end{document}